\documentclass[aps,pra,twocolumn,superscriptaddress]{revtex4-2}
\usepackage[utf8]{inputenc}
\usepackage{amsmath,amsfonts, amsthm} 
\usepackage{dcolumn}
\usepackage{subcaption} 
\usepackage{comment}
\usepackage{cancel}
\usepackage{bm}
\usepackage{multirow}
\usepackage{mathdots}
\usepackage{amssymb}
\usepackage{hyperref}
\usepackage{xcolor}
\usepackage{soul}
\usepackage{MnSymbol}
\usepackage{float}
\usepackage{mathtools}
\usepackage{dsfont}
\usepackage{slashed}
\usepackage{tikz}
\usepackage{bigints}

\usepackage{graphicx}
\usepackage{cleveref} 
\crefname{figure}{Fig.}{Fig.}
\crefname{equation}{}{}
\usepackage{todonotes}
\usepackage{orcidlink}

\usepackage{algpseudocode}

\newcommand{\inner}[2]{\ensuremath{\langle{#1} \,,{#2}\rangle}}
\def\id{\mathrm{id}}

\def\ra{\rangle}
\def\la{\langle}

\newcommand{\ket}[1]{\ensuremath{|#1\rangle}}
\newcommand{\bra}[1]{\ensuremath{\langle#1|}}
\newcommand{\ketbra}[2]{\ensuremath{\ket{#1} \! \bra{#2}}}
\newcommand{\proj}[1]{\ensuremath{\ketbra{#1}{#1}}}
\DeclareMathOperator{\tr}{tr}
\newcommand{\Id}{{\rm 1\hspace{-0.9mm}l}}
\def\Tr{\text{Tr}}

\newcounter{algorithm}

\newenvironment{algorithmfigure}[1][h]
{%
  \refstepcounter{algorithm}%
  \begingroup
  \begin{figure}[#1]
}
{%
  \end{figure}
  \endgroup
}

\newcommand{\bl}[1]{{#1}}

\newcommand{\rd}[1]{{#1}}

\newcommand{\blu}[1]{{#1}}

\begin{document}

\title{Benchmarking quantum devices beyond classical capabilities}

\author{Rafa{\l} Bistro\'n \orcidlink{0000-0002-0837-8644}}
\affiliation{Faculty of Physics, Astronomy and Applied Computer Science, Jagiellonian University, ul. \L{}ojasiewicza 11, 30-348 Krak{\'o}w, Poland}
\affiliation{Doctoral School of Exact and Natural Sciences, Jagiellonian University, ul. \L{}ojasiewicza 11, 30-348 Krak{\'o}w, Poland}

\author{Marcin Rudzi\'nski \orcidlink{0000-0002-6638-3978}}
\affiliation{Faculty of Physics, Astronomy and Applied Computer Science, Jagiellonian University, ul. \L{}ojasiewicza 11, 30-348 Krak{\'o}w, Poland}
\affiliation{Doctoral School of Exact and Natural Sciences, Jagiellonian University, ul. \L{}ojasiewicza 11, 30-348 Krak{\'o}w, Poland}

\author{Ryszard Kukulski \orcidlink{0000-0002-9171-1734}}
\affiliation{Faculty of Physics, Astronomy and Applied Computer Science, Jagiellonian University, ul. \L{}ojasiewicza 11, 30-348 Krak{\'o}w, Poland}
\affiliation{IT4Innovations, VSB~-~Technical University of Ostrava, 17.~listopadu 2172/15, 708 33 Ostrava, Czech Republic}

\author{Karol {\.Z}yczkowski \orcidlink{0000-0002-0653-3639}}
\affiliation{Faculty of Physics, Astronomy and Applied Computer Science, Jagiellonian University, ul. \L{}ojasiewicza 11, 30-348 Krak{\'o}w, Poland}
\affiliation{Center for Theoretical Physics, Polish Academy of Sciences,  Al. Lotnik{\'o}w 32/46, 02-668 Warszawa, Poland}

\date{April 27, 2026}

\begin{abstract}
Rapid development of quantum computing technology has led to a wide variety of sophisticated quantum devices. Benchmarking these systems becomes crucial for understanding their capabilities and paving the way for future advancements. The Quantum Volume (QV) test is one of the most widely used benchmarks for evaluating quantum computer performance due to its architecture independence. 
However, as the number of qubits in a quantum device grows, the test faces a significant limitation. 
It requires determining the subspace of the most probable outcomes, a task that is typically performed via classical simulation of the quantum circuit and therefore incurs an exponential computational cost. In this work, we propose modifications to the QV test, \blu{by adopting a carefully restricted circuit ensemble generated from a gate set that remains universal for quantum computation,} that allows for the direct determination of the heavy-output subspace. Crucially, the modified circuits remain capable of general quantum computation. This approach overcomes the scalability barrier of the Quantum Volume test beyond classical computational limits, while still probing the key features of universal quantum computing.
\end{abstract}

\maketitle

\section{Introduction}
\label{sec:intro}
Quantum processors have made extraordinary progress in recent years, evolving from small--scale physical experiments to devices capable of integrating hundreds of qubits \cite{Arute, Jurcevic, Pino, Honeywell, Bluvstein,Kim}, bringing us closer to implementing computations intractable for classical machines \cite{Shor1994, Grover1996, Abrams, Harrow}. The wide variety of physical realizations of quantum computers \cite{Pino, Kielpinski, Kok, Bruzewicz, Kjaergaard, Vandersypen, Majidy} raises a natural question about comparing their capabilities.

Effective benchmarks are crucial for quantifying the performance of quantum devices and guiding technological development \cite{Boixo, Corcoles, proctor2024benchmarking, McGeoch, Proctor}. Benchmarks should be well--motivated, precisely defined, architecture--agnostic, robust to variations in implementation, and efficient in terms of required computational resources \cite{Pelofske,reexamining_QV,proctor2024benchmarking}.

Inspired by the above requirements, the Quantum Volume (QV) test was introduced \cite{Bishop, Cross2019} as a measure of the performance of a quantum processor. It is given by the maximal size of a generic quantum circuit of width (number of qubits) equal to depth (number of layers), which is realized with considerable accuracy. Due to hardware independence, QV has become a widely used and extensively studied benchmark \cite{Bishop, Cross2019,Pelofske, Sundaresan, Chen, LaRose, Cornelissen, reexamining_QV}. Despite its effectiveness for medium--scale quantum devices, QV has a single important limitation \cite{reexamining_QV, proctor2024benchmarking}: the cost of the necessary classical simulation makes this test unsuitable already for systems containing circa 100 qubits, which are crucial for achieving quantum computational advantage.
State of the art quantum computers like Condor and Heron \cite{IBM2023, abughanem2024ibm}, Willow \cite{Google2024quantum}, or Zuchongzhi \cite{gao2024establishing} already operate near this regime; thus, benchmarking them requires numerous tricks, like dividing qubits into non-interacting groups \cite{gao2024establishing}.

To overcome this obstacle, in this work we propose a modification to the QV test -- the parity test, as well as an auxiliary double-parity test.
Both methods are
based on parity-preserving gates, which encompass matchgates and qubit swap, thus allowing for general computation \cite{jozsa2008matchgates}. \blu{The proposed tests enable} one 
to directly determine the subspace of the most probable outcomes (the heavy output subspace) without costly classical simulations. This approach provides a new, efficient, and scalable benchmark for both current and future quantum systems \bl{while retaining} the advantages of the QV test, such as being well--motivated and universal.
Furthermore, we discuss the general problem of simulating quantum circuits and the characteristics of the heavy output subspace.

\rd{This work is organized as follows: In Section~\ref{sec:benchmarking} we review the Quantum Volume benchmark and discuss its limitations arising from the need for expensive classical simulation. In Section~\ref{sec:parity} we introduce parity-preserving \bl{modifications} of the QV test and analyze their robustness to realistic noise mechanisms. Section~\ref{sec:case-study} presents a detailed case study of heavy-output frequency decay for the IBM Brisbane processor, illustrating the practical behavior of different benchmarks.} 
\bl{Section~\ref{sec:etimating} presents an alternative application of parity-preserving circuits as a tool for \blu{estimating the} heavy output frequency in the standard QV test for certain noise models. Next, in Section \ref{sec:bench_discuss}, we discuss the problem of quantum benchmarks and place our results in the context of alternative propositions.} 
\rd{Finally, in Sec.~\ref{sec:conclusions}, we summarize our findings and comment on the implications for benchmarking near-term quantum devices beyond the reach of classical simulation.}

\section{Quantum Volume test}
\label{sec:benchmarking}
\begin{figure}[h]
    \centering
    \includegraphics[width = 8.2 cm ]{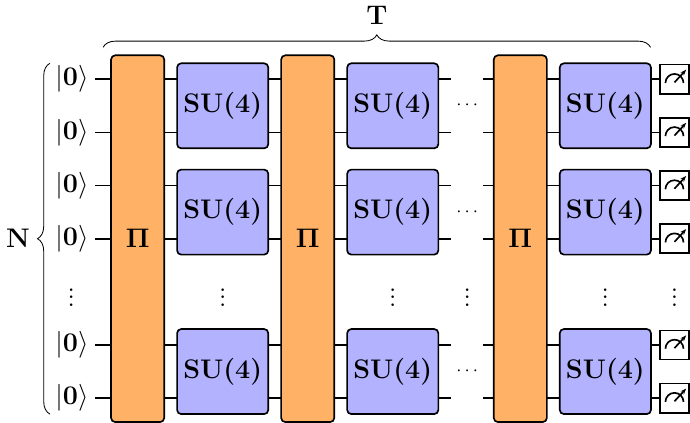}
    \caption{Quantum circuit used in the quantum volume test, consisting of $T$ layers and $N$ qubits, \blu{with} permutations $\Pi$ followed by two-qubit random gates SU$(4)$ \cite{Bishop, Cross2019}.}
    \label{fig:QV_circuit}
\end{figure}

\blu{The operational meaning of the Quantum Volume benchmark is the largest square circuit that can be executed effectively on the} tested device. The original construction of QV test utilized \rd{multiple random} quantum circuits, later \blu{referred to as} a QV circuit, consisting of $N$  qubits and $T$ layers, with the standard setting of a `square' circuit, $T = N$. In each layer the qubits are shuffled by a random permutation $\Pi$ and then they interact via random two-qubit gates $U \in SU(4)$ sampled with the Haar measure, see Fig. \ref{fig:QV_circuit}.

The test starts with simulation of a noiseless circuit \rd{to determine its \textit{heavy output probability}. For a given ideal QV circuit $U$ on $N$ qubits, let $p_x = |\langle x|U|0^{\otimes N}\rangle|^2$ denote the ideal output probabilities in the computational basis and let $m_U$ be the median of the set $\{p_x\}_{x\in\{0,1\}^N}$. The corresponding \textit{heavy output subspace} $H(U)$ consists of all bit strings $x$ with output probabilities above the median, $p_x \ge m_U$.  If a noisy implementation of $U$ produces an empirical output distribution $\{\hat{p}_x\}$, the \textit{heavy output probability for that circuit} is defined as}
\begin{equation}
  h_U \;=\; \sum_{x \in H(U)} \hat{p}_x.
\end{equation}
\rd{In the ideal, noiseless case one has, on average, $h_U = (1 + \ln{2})/2\approx 0.846$, whereas for a
completely depolarizing device $h_U = 1/2$ \cite{Math_monster_theoretic_fundations,reexamining_QV,Cross2019}. In practice, one estimates $h_U$ from a finite number of samples for
each circuit and then averages over an ensemble of random QV circuits of
the same size.  A device is said to pass the test for size $N$ if the
average observed heavy output probability over this ensemble
exceeds the threshold $2/3$ with two standard deviations confidence \cite{Cross2019}. The
quantum volume of the device is then reported as $V_Q = 2^N$ for the
largest such $N$. The threshold $2/3$ originates from Ref.\cite{Math_monster_theoretic_fundations} and corresponds to minimal success probability for algorithms in the BQP complexity class \cite{Bernstein1997}.}

\blu{The Quantum Volume test suffers from} one fundamental problem. To obtain heavy outputs, it relies on classical simulation of generic quantum circuit, which will soon become prohibitively expensive \cite{reexamining_QV, proctor2024benchmarking}. 
In order to overcome
this difficulty, we propose a redefinition of the
QV circuit so that the structure of heavy outputs is known without the need for expensive simulations, while deriving exact output distribution still requires general quantum computations.

\section{Parity-preserving benchmarks}
\label{sec:parity}
\rd{In this section we reformulate the Quantum Volume test in terms of parity-preserving gates, which restrict the dynamics to a well-structured heavy-output subspace while still allowing for universal quantum computation.}
\bl{We start by introducing our main \blu{proposal} --the parity preserving benchmark, later followed by an auxiliary double-parity test. Finally we discuss the extension of utilized circuits, so that the intermediate computations can be performed in entire Hilbert space, and mention potential case-specific modifications.}

\subsection{Single-parity benchmark}
\label{subsec:single-parity}

The first and main proposal is based on a general \textit{parity-preserving} two-qubit unitary of the form
\begin{equation}
\label{u_parity_1}
U = \begin{pmatrix} * & 0 & 0 & *\\ 0 & * & * & 0\\ 0 & 
    * & * & 0\\ * & 0 & 0 & *\end{pmatrix}.
\end{equation}
Since this construction encompasses both matchgates and swap gates, \bl{constituting a universal gate set \cite{jozsa2008matchgates}}, it allows for general quantum computations. 

\begin{figure}[t]
     \centering
     \vspace{-1cm}
    \subcaptionbox{\,\,\,\,\, \hspace{4.5cm} (b)}{
    \includegraphics[width=\columnwidth]{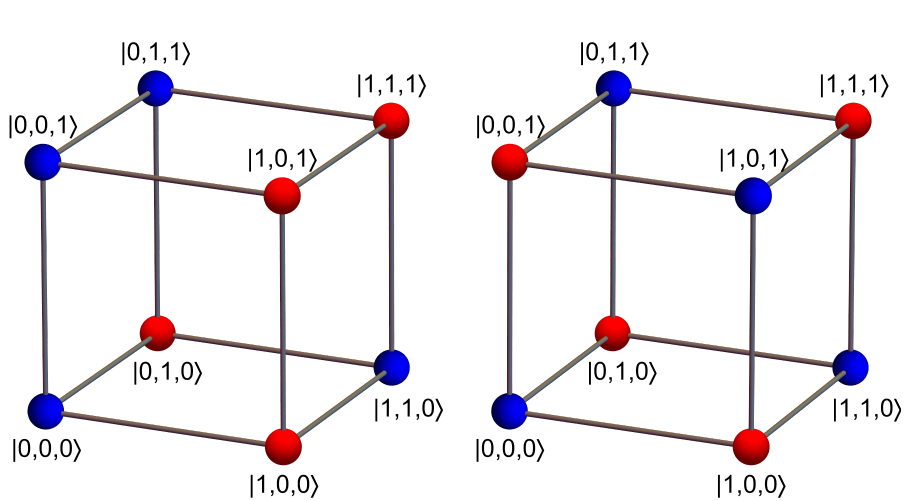}}
    \caption{Outputs \blu{of a} three-qubit QV circuit represented as vertices of a cube, each edge corresponds to one bit flip, 
    with heavy output subspace denoted in red: 
     (a) an exemplary
     realization and (b) the parity-preserving case - any bit flip sends a state out of the heavy-output subspace, as there are no edges connecting red vertices. 
     }
        \label{fig:heavy_outputs}
\end{figure}

We propose to replace all random 
two-qubit gates $U$ by parity-preserving ones. A circuit constructed from such matrices preserves
the global parity -- sum of all bits modulo $2$. 
The heavy-output probability for such a T--layer circuit can be written as
\begin{equation}
h_U = \sum_{P} |\la P|\prod_{j = 1}^T \overline{\mathcal{U}}_j ~|0^{\otimes N}\ra|^2.
\end{equation}
Here $P$ denotes bit strings of $N$ bits with an even number of $1$ and $\overline{\mathcal{U}}_j$ corresponds to averaged $j$-th layer of permutations and two-qubit gates. The difference between the original and the newly proposed heavy output subspace is depicted in Fig. \ref{fig:heavy_outputs}. 

From the standpoint of execution on quantum hardware, the introduced modifications are quite minor. Since each gate can be obtained from a parity--preserving one, combined with local pre- and post-processing \cite{Khaneja}, the number of elementary two-qubit gates 
does not change.

Finally, because the heavy-output subspace is {\sl a priori} known, and independent of drawn circuit, one may try to analytically study heavy output probability $h_U$ for the noise model of interest. 
In \blu{the simplest uniform model}, each two-qubit gate is \blu{perturbed} by $e^{i \alpha H}$, with $H$ being random Hamiltonian from the Gaussian Unitary Ensemble (GUE) and $\alpha$ the noise strength. 
\blu{The resulting formula for the heavy-output frequency} for a $T$--layer, $N$--qubit circuit reads
\begin{equation}
\label{hu_sol1}
h_U  = \frac{1}{2} \left(1 + \left(\frac{4f(\alpha)+1}{5} \right)^{\frac{N T}{2}} \right) \approx \frac{1}{2}\left(1 + e^{-2 \alpha^2 NT}\right), 
\end{equation}
where \blu{the function $f(\alpha)$ introduced in} \cite{Liu2018, Nadir2024fidelity} is related to the spectral form factor, see \rd{Appendix \ref{Section:I}} for details. In a perfect parity-preserving QV circuit, the heavy output frequency reaches $h_U = 1$, since no error "kicks" quantum states out of the heavy-output subspace.  

Furthermore, we extended error model into dissipative framework, considering interactions with an environment of dimension $d_E$ via a random GUE Hamiltonian $H$ -- see Appendix \ref{Section:I}. This extension effectively corresponds to increasing the noise level by a factor $\sqrt{d_E}$.
\rd{More explicitly, for a two-qubit gate coupled to an environment of dimension $d_E$ one finds}
\begin{equation}
\label{hu_sol1_en}
h_U  \approx \frac{1}{2}\left(1 + e^{-2 d_E \alpha^2 N T}\right)
\end{equation}
\rd{for the single-parity circuit. Dissipative noise effectively rescales the coherent noise parameter, without changing the functional dependence of the benchmarks on the circuit size and depth.}

In order to calculate the counterpart of the quantum volume using proposed circuit, one should, once again, look for the largest square circuit $T = N$ for which heavy output is above the threshold $h_U > 2/3$. We decided to utilize original threshold due to its strong anchoring in quantum computational complexity research \cite{Math_monster_theoretic_fundations}.

\subsection{Double-parity benchmark}
\label{subsec:double-parity}

\bl{In the introduced parity--preserving benchmark, the symmetry restricts the system to a specific heavy-output subspace. Thus errors within that subspace, i.e. the errors that do not affect parity, remain undetected. }
\rd{This includes, for example, single-qubit dephasing noise (Pauli-$Z$ errors) and certain multi-qubit bit-flip errors (e.g. Pauli-$XX$ on a pair of qubits), both of which leave the computational-basis parity unchanged.  Such parity-preserving processes are not exotic and can be relevant in some realistic devices.}
\rd{As a first solution} to this problem, we propose specialized quantum circuits, in which 
\rd{qubits are arbitrarily divided} into two sets of equal size. Its aim is to preserve parity both globally and within the two sets.
The circuit structure is \blu{similar to that of the} parity--preserving QV circuit, but with small modifications. 
\rd{Similarly to the QV test, one prepares a random circuit for the quantum computer to perform. At the entrance qubits are randomly divided into two sets of size $N/2$. 
In each layer we know which qubits belong to which subsets by tracking applied permutations. 
\bl{The difference from the QV and parity-preserving benchmarks is again the form of the 2-qubit gates. Here, two types of gates are used, depending on which qubits interact.}
Whenever two qubits from the same subset meet, we apply a generic parity-preserving two-qubit gate of the form \eqref{u_parity_1}, \bl{as in parity--preserving circuit}. When the interacting qubits belong to different subsets, we instead apply a diagonal interaction that does not mix the two parities,}
\begin{equation}
\label{u_parity_2}
u_{\text{diag}} = e^{ i a Z\otimes Z},
\end{equation}
\rd{where $a$ is a random phase drawn from a flat distribution, $a\in[0,2\pi)$, and $Z$ is the Pauli-$Z$ operator. In this way both the global parity and the parities of the red and blue subsets are conserved throughout the circuit. The heavy-output subspace is then defined as the set of bit strings that have even parity within each subset, so that only every fourth bit string is heavy.}

\rd{This explicit construction makes it clear that the double-parity benchmark probes error channels that conserve global parity but violate the finer parity structure imposed by the two subsets. Furthermore, since the division of qubits into two sets \blu{is random and therefore} usually different for each random circuit, this benchmark should, on average, detect simultaneous bit-flips which preserve global parity.}

To highlight the properties of the double-parity circuit we study its behavior not only in the presence of two-qubit GUE noise but also parity-preserving error models. We propose a noise \blu{in the} permutations, treated as an optimal composition of swap gates $S$. 
We assume that each swap was implemented with too short or too long pulse $S \to S^{\beta}$, where the exponent $\beta$ is drawn from a Gaussian distribution with mean $1$ and variance $\sigma^2$. Such an error model seems natural for computers with $\sqrt{S}$ as the native two-qubit gate. 
It turns out that this model is equivalent to the one with probabilistic swap omission with probability 
$p = \frac{1}{2}(1 - e^{- \frac{1}{2} \pi^2 \sigma^2})$ \cite{Nadir2024fidelity}.

As \rd{we demonstrate in Appendix \ref{Section:I}}, the final formula for heavy output frequency for such a model can be approximated as 
\begin{equation}
\label{hu_dub1}
h_U \approx \frac{1}{4}\left[2 e^{-\frac{3}{2} \alpha^2 N T \frac{(N-2/3)}{(N-1)}}  e^{-\frac{1}{2}p w(N)T } + e^{-2 \alpha^2 N T} + 1  \right] ~,
\end{equation}
where $w(N)$ is an architecture-dependent average number of swaps necessary to implement permutation of $N$ qubits. This decay is the sum of three \blu{distinct} terms. The background $1/4$, global parity decay $e^{-2 \alpha^2 N T}$ as in \eqref{hu_sol1} and double parity decay.
Due to the significantly different range of $h_U$ (decaying to $1/4$ instead of $1/2$), we apply a linear rescaling of the $2/3$ passing threshold to $(1+\log 2)/4\log 2\approx 0.61$.

\subsection{Implementation}\label{subsec:impl}
Operations performed by quantum processor are far from perfect unitaries. The computer can also be malicious and only pretend to perform quantum computation. 
To address these \bl{and other} issues, we consider extensions of above-discussed benchmarks for real-world applications, discussed in detail in \rd{Appendix \ref{app:appl}}.

\begin{figure}[h]
        \centering
        \includegraphics[width=0.98\columnwidth]{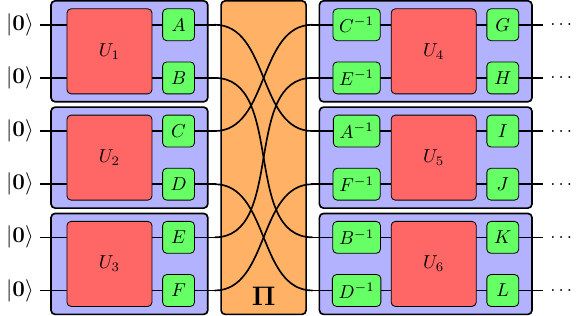}
    \caption{
    Diagram of the modified quantum volume circuit with hidden parity preservation (only first two layers presented for clarity). \bl{Each random 2-qubit gate (blue region) is constructed by surrounding parity-preserving gate $U$ with single-qubit gates, which are arranged in a way that they cancel each other after permutation $\Pi$, as explicitly drawn here to show the cancellation rule.} 
    } 
    \label{fig:circuit-modification}
\end{figure}

\bl{The main extension} addresses possible bias of quantum processor towards some even-parity bit strings \bl{and allows for intermediate computations in the entire Hilbert space}. To make gates in the proposed circuits more uniform, we append \blu{two single-qubit gates to each parity-preserving gate. Then we add} their inverses in the next layer, after permutation --see \cref{fig:circuit-modification}. In this way, the parity-preserving circuit is built from 2-qubit gates, with \blu{embedded} single-qubit pre- and post-processing, which no longer preserve parity on its own. This extension allows the test to detect parity-preserving errors as well.

\bl{As discussed in Appendix \ref{app:appl} this modification did not substantially change the results in our case study. However, it enables manipulation of quantum states outside the parity subspace and restores parity only at the last layer. Thus, for each circuit, the `preserved' subspaces in the intermediate calculations are different, hence we consider this extension substantially more robust and general. Thus, we include it in the proposed test. A step-by-step description of the entire test procedure, together with pseudocode, is presented in Appendix \ref{app:procedure}, and the implementation in the $\texttt{Python}$ language available in the GitHub repository \cite{bistronGithub_QV}.}

\bl{Additional modifications may aim} to detect malicious behavior by a quantum computer, which might calculate the more probable output parity and return random bit strings with that parity. 
A crude way to solve this issue is to randomly add a circuit with the opposite parity.
For example, this can be done by inserting a finite number of Pauli $X$ and hiding them in two-qubit gates.
Our more sophisticated approach is to sneak a few matchgate circuits in disguise, camouflaged as in the above paragraph. For such circuits, the testing party can calculate output probabilities efficiently and compare them with computer outcomes to detect \blu{such behavior}.

\section{Case study: Heavy-output frequency decay in IBM Brisbane processor}
\label{sec:case-study}

\begin{figure*}[ht!]
    \centering
    \begin{subfigure}{0.32\textwidth}
        \centering
        \includegraphics[width=\textwidth]{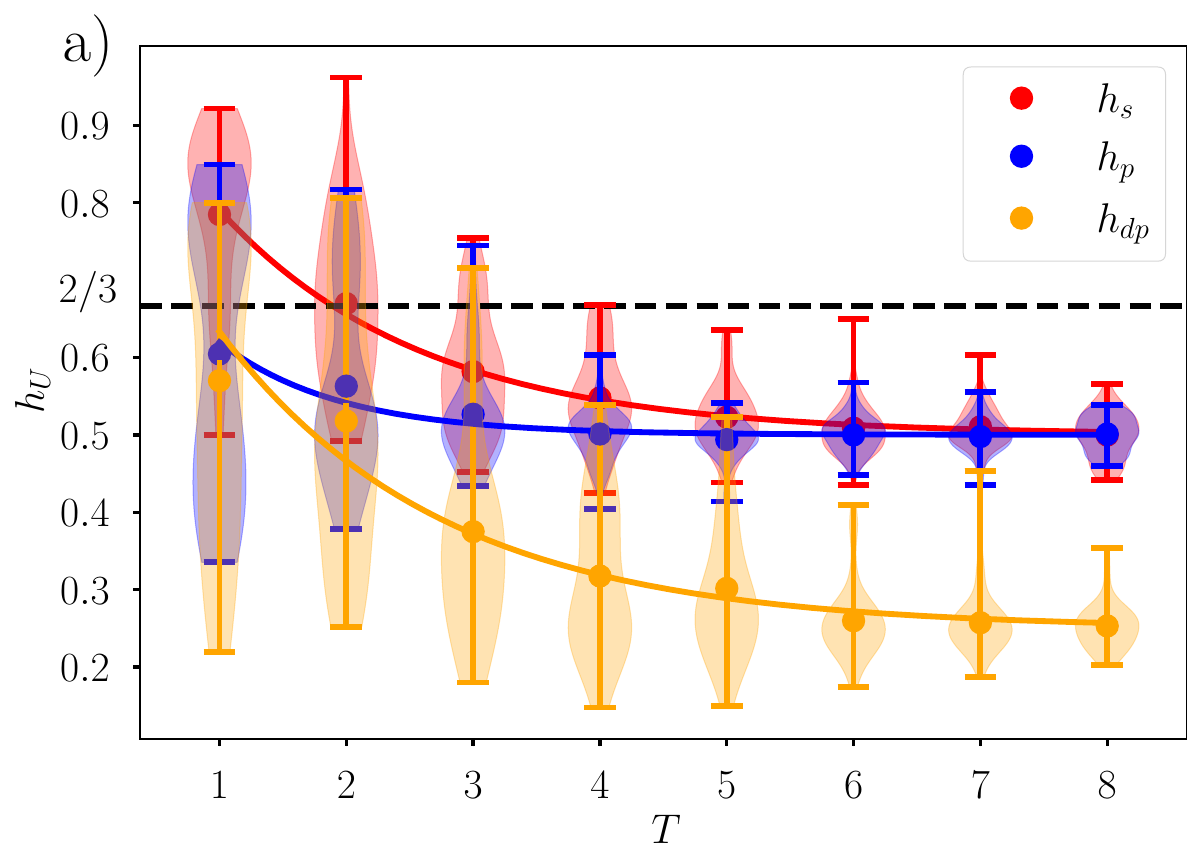}
        \label{subfig:ibm6}
    \end{subfigure}
    \hfill
    \begin{subfigure}{0.32\textwidth}
        \centering
        \includegraphics[width=\textwidth]{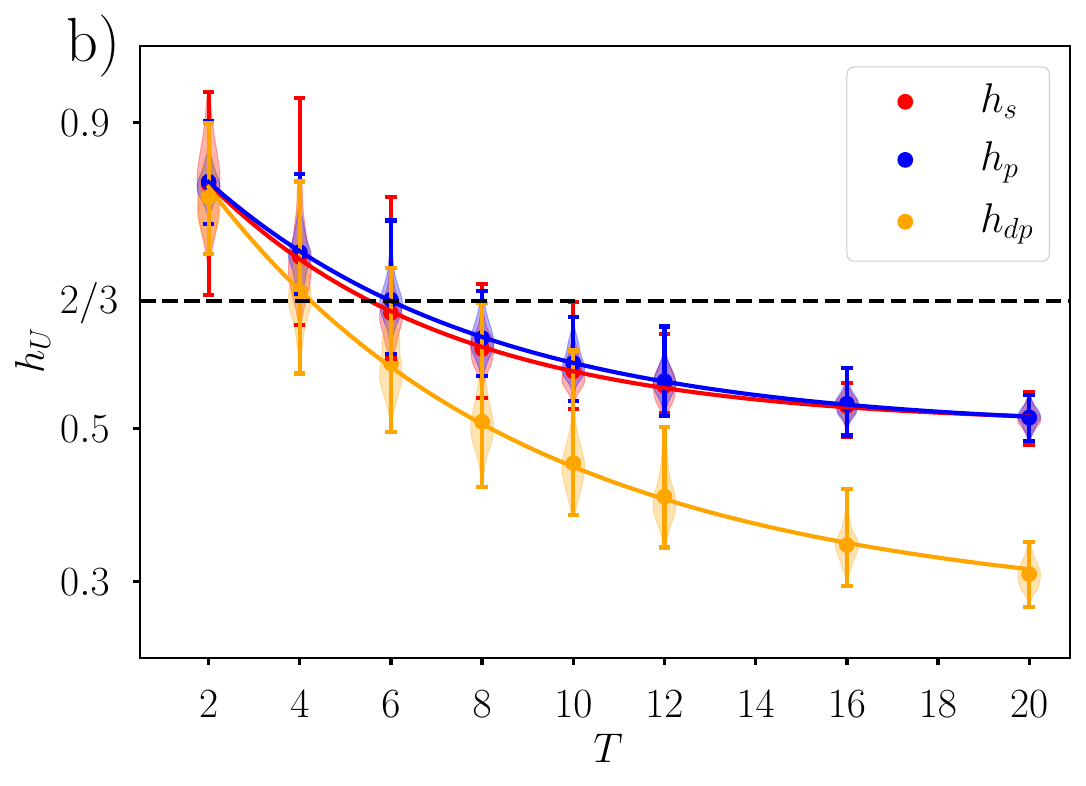}
        \label{subfig:ibm_simulation}
    \end{subfigure}
    \hfill
    \begin{subfigure}{0.315\textwidth}
        \centering
        \includegraphics[width=\textwidth]{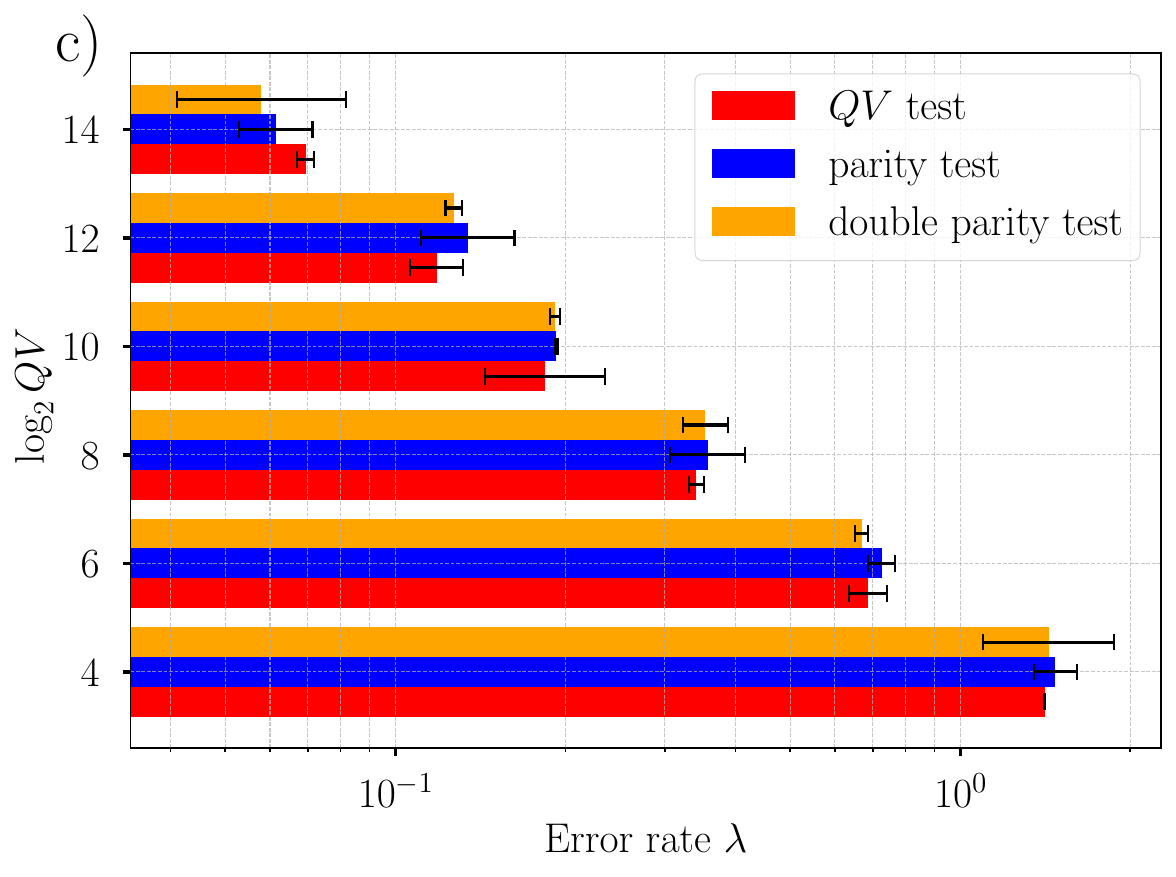}
        \label{subfig:binplot_qv}
    \end{subfigure}
    \caption{Comparison of single parity ($p$), double parity ($dp$) and standard ($s$) quantum volume ($QV$) tests. Heavy output probability $h_U$ as a function of number of layers $T$ for 6 qubits on a) the quantum device \textit{IBM Brisbane}, b) simulator of this computer with error parameters rescaled by a factor $\lambda = 0.5$, to obtain $\log_2(QV) = 6$. The dashed line indicates $h_U = 2/3$ threshold. c) The ranges of the scaling factor $\lambda$ for passing different tests -- for any noise level proposed benchmarks (parity and double parity) exhibit the same behavior as the original QV test.}
    \label{fig:IBM}
\end{figure*}

To \blu{demonstrate} the functionality of the proposed tests we compared them with the standard Quantum Volume \cite{Cross2019} on a real device. For this purpose, we executed Quantum Volume, single parity and double parity circuits on the \textit{IBM Brisbane} quantum computer -- see \cref{fig:IBM}(a). We also simulated their action using the Qiskit Aer simulator \cite{Qiskit} to check how the obtained results change with the scale of errors in the simulated circuits -- see \cref{fig:IBM}(b),(c). The code used for testing and simulating the quantum computer is available at \cite{bistronGithub_QV}.

We started by examining heavy output frequency decay in \textit{IBM Brisbane} in the standard quantum volume test $h_s$, which requires classical simulation to determine heavy output subspace, and in the parity-preserving $h_p$ as well as the double parity-preserving $h_{dp}$ test. In our experiment we sampled 6-qubit quantum circuits composed of a different number of layers $T$. In the QV test, the number of layers is equal to the number of qubits, $N=T$, but in order to compare the behavior of the investigated measures, circuits for $T = 1,\dots,8$ were considered. 

Next, we compared the considered benchmarks for reduced error levels, by performing Qiskit simulations of the
computer with all its error parameters scaled by a factor $\lambda$. 
Then, analogous simulations were performed for systems with different numbers of qubits and layers. An example of such simulation results is shown in \cref{fig:IBM}(b). These simulations were used to extract the $\lambda$ scale required to achieve Quantum Volume levels for the benchmarks of interest. The results are presented in \cref{fig:IBM}(c). For further details see \rd{Appendix \ref{App:case_study}}.

\section{Estimating Heavy Output Probability}\label{sec:etimating}

In this section we present another application of parity-preserving circuits. Namely, we show that parity-preserving circuits can also be used to estimate $h_U$ in the standard QV test, utilizing, in contrast to the original definition, \blu{only a} polynomial amount of resources.
\rd{We provide estimates for heavy output probability in standard and parity-preserving quantum volume circuits. Then, leveraging the obtained relations, we propose estimators for heavy output frequency in the standard Quantum Volume test for different scenarios \blu{that require only polynomial} classical and quantum resources.
Hereafter, we assume that the 
entire information about the noise occurring during the computation will be 
encoded into a sequence of arbitrary noise channels $(\Omega_z)_{z=1}^M$ that 
affect the 
circuit after each layer. That means, if the noiseless circuit returns a 
state $\ket{\psi} = U_{M} \cdots U_2U_1\ket{0}$, where $U_z$ is a composition 
of two-qubit gates and a permutation, then the noisy one returns 
$\Omega_M(U_M\ldots\Omega_1(U_1\proj{0}U_1^\dagger)\ldots U_M^\dagger)$. The 
channels $\Omega_z$ can 
depend on the number of qubits $N$ and the circuit depth $T$, as well as on the 
particular architecture type, connectivity, and single and two-qubit gate 
errors. However, the influence of the particular gate types used to implement 
the random state is negligible. Under these assumptions we get an estimator}
\begin{equation}
\label{hu_pred}
    \widetilde{h_{U}} = \frac{1}{2} + \left(p_* - \frac{1}{2}\right)\frac{2^N 
\mathbb{P}^{(M)}_0 - 1}{2^N 
- 1} \prod_{z=1}^{M-1} \frac{\la J_{\Omega_z} |{J_\Id}\ra-1}{4^{N}-1},
\end{equation}
where $p_* = (1 + \ln 2)/2$ \rd{is the average heavy output frequency for perfect implementation of random circuits and} $\mathbb{P}_0^{(M)}$ is the probability 
that uniformly sampled input basis state $\ket{i}$ is measured as $\ket{i}$ 
after the influence of $\Omega^{(M)}$ and $ 
\la J_{\Omega_z} |{J_\Id}\ra$ is the inner product between Choi-Jamio{\l}kowski 
isomorphism of channels $\Omega_z$ and $\Id$ (see derivation in Appendix \ref{app:hu_est_original}).

\begin{figure}[h]
    \centering
    \includegraphics[width=0.95\linewidth]{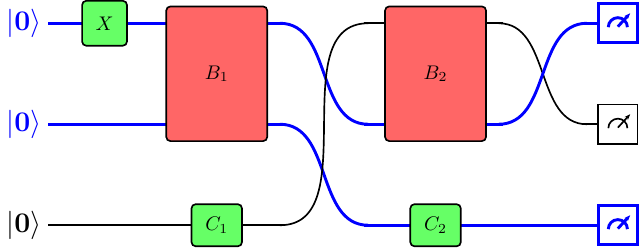}
    \caption{An example of parity-preserving circuit for $N=3$ 
    qubits and $T=2$ layers. We mark $m=2$ parity-preserving qubits \bl{with a 
    thick blue line}. Gate $X$ is applied on the 
    first qubit only. \blu{Since the} first two qubits preserve parity, $B_1$ is a random matrix 
    of the type $U_2^{\{00,11\}} \oplus U_2^{\{01, 10\}}$ and $C_1$ is arbitrary. \bl{After the first permutation,} matrix $B_2$ \bl{acts on one parity qubit, hence it has a form} 
    $U_2^{\{00,10\}} \oplus U_2^{\{01, 11\}}$ and $C_2$ is diagonal to preserve parity. The parity 
    subset at the end of the circuit equals $\mathcal{M}_2 = \{1,3\}$, hence, 
    if the circuit is noiseless, 
    we should measure one of the following bit-strings $100, 110, 001, 011$. }
    \label{fig:example}
\end{figure}

To estimate $h_U$ from experimental data we create family of QV circuits, preserving parity on $m = 1,\ldots,N$ qubits for a given $N, T$. 
At the beginning, we choose \blu{$m$ out of $N$ qubits at random on} which the test circuits will preserve 
parity and we keep track of them. Let us denote by $\mathcal{M}_0$ the 
initial subset 
of these qubits. Also, for 
the qubits from $\mathcal{M}_0$ we add randomly $X$ gates at the beginning 
of 
the circuit, 
for each qubit independently with equal probability. Each set-up 
consists of $T$ layers of two-qubit 
unitary gates and random 
permutations. Random permutations $\pi$ are arbitrary - the same as 
\blu{those} used in 
the original heavy-output problem. Since these permutations change the position of the qubits 
that keep parity, we update the position of the subset of $m$ qubits after 
each layer, $\mathcal{M}_t \xrightarrow{\pi} \mathcal{M}_{t+1}$.
Let $U_k^{\{b\}}$ be Haar-random unitary of size $k$ defined on 
subspace given by bit-strings $b$. For two-qubit gates we use the following 
three types of unitary matrices: 
\begin{itemize}
    \item if both qubits are not in the subset $\mathcal{M}_t$ then we choose generic random unitary matrix $U = U_4^{\{00,01,10,11\}}$.
    \item if the first qubit belongs to the set $\mathcal{M}_t$ and the second one is not, then we 
    choose a random unitary matrix represented as a direct sum, $U=U_2^{\{00,01\}} 
    \oplus 
    U_2^{\{10,11\}}$, which can be interpreted as random control operation with respect to parity-preserved qubit (analogously we define $U=U_2^{\{10,00\}} 
    \oplus 
    U_2^{\{01,11\}}$ if the second qubit is in $\mathcal{M}_t$, but the first is not).
    \item if both qubits are in the subset $\mathcal{M}_t$ then we choose 
    random unitary matrix of the type $U=U_2^{\{00,11\}} \oplus U_2^{\{01,10\}}$, as in our main proposition of parity-preserving circuit \eqref{u_parity_1}.
\end{itemize}
Additionally, if the number of qubits $N$ is odd then there is a qubit without \blu{a partner}. We apply a diagonal gate $U 
= U_1^{\{0\}} \oplus U_1^{\{1\}}$ for that qubit if it belongs to $\mathcal{M}_t$ 
and random single-qubit gate
$U = U_2^{\{0,1\}}$ otherwise. Finally, we 
perform a measurement and count how many bit-strings fall into the heavy subspace.
We show an example of parity-preserving circuit in Fig.~\ref{fig:example}.

\rd{For each $m$ we calculated an estimator for
the heavy-output probability $h_U^m$ of the proposed circuits \rd{(for details see Appendix \ref{app:hu_est_m})}
\begin{equation}\label{hum_est}
\begin{split}
\widetilde{h_U^m} = \frac12 \sum_{\substack{
p_z =\, 0,1 \\ p_0=p_M}}  \prod_{z=1}^M 
\frac{\la\Pi_{m,p_z}\otimes \Id|\Omega_z^{(s)}(\Pi_{m,p_{z-1}}\otimes 
\rho_*)\ra}{2^{m-1}},
\end{split}
\end{equation}
where $\Pi_{m,p_z}$ is the projector on the subspace in which chosen $m$ qubits have parity $p_z$ and $\Omega_z^{(s)}(\rho) = \frac{1}{N!} \sum_{\pi_z}  
\pi_z^\dagger \Omega_z(\pi_z 
\rho \pi_z^\dagger) \pi_z$ is the symmetrized channel $\Omega_z$ over all 
subsystem 
permutations $\pi_z$.}

By comparing heavy output probabilities \eqref{hu_pred} and \eqref{hum_est}, we see that parity-preserving circuit, when not extended by immersed single-qubit gates, can be insensitive to specific noise models. For example, if the noise
shifts the phase of basic states at the beginning of the circuit, then $\widetilde{h_U}$ can vary in $[1/2, p_*]$ while for the ``bare'' parity-preserving circuit
$\widetilde{h_U^m} = 1$ for all $m$ (see Appendix \ref{App:estimation-z} for derivation).
This observation leads to a very 
natural question: is it possible to use $\widetilde{h_U^m}$ to provide an upper bound 
for $\widetilde{h_U}$.
Indeed, direct calculations reveal $\widetilde{h_U} \le 
1/2+\left(p_*-1/2\right)
\prod_{z=1}^{M}\mathbb{P}_0^{(z)} + \mathcal{O}(4^{-N})
$ (see Appendix~\ref{App:estimation-upper}) and values $\mathbb{P}_0^{(z)}$ are captured by $\widetilde{h_U^m}$ test, which 
could support a positive answer to the \blu{aforementioned} question. However, determining an accurate upper bound 
remains an open problem.

For certain types of noises or particular models, one can estimate $\widetilde{h_U}$ directly from $\widetilde{h_U^m}$. As an example, we consider measurement errors  
$\Omega_z=\Id$ for $z<M$, 
where we only need to calculate 
\begin{equation*}
\mathbb{P}_0^{(M)} = \sum_{m = 1}^N \widetilde{h_U^m} \binom{N}{m}/2^{N-1} - \frac{2^{N-1} - 1}{2^{N-1}}
\end{equation*}
(derivation in Appendix~\ref{App:estimation-m}).

Another example \blu{is given} by single-qubit decoherence noise $\Phi_\epsilon(\rho) = (1-\epsilon)\rho + 
\epsilon \rho_*$ with a parameter $\epsilon \in [0,1]$ for which $\Omega_z = \Phi_\epsilon^{\otimes N}$ for $z=1,\ldots,M$. Here we get that 
\begin{equation}
\begin{split}
\widetilde{h_U} &= \frac12 + \left(p_* - \frac12\right)\frac{(2-\epsilon)^N-1}{2^N-1}
\left(\frac{(4-3\epsilon)^N-1}{4^{N}-1}\right)^{M-1},\\
\widetilde{h_U^m} & = \frac{1+(1-\epsilon)^{mM}}{2}
\end{split}
\end{equation}
(see derivation in Appendix~\ref{App:estimation-d}).
Thus the information obtained from multiple $\widetilde{h_U^m}$ is in fact redundant, and any of them can be used to estimate the value of $\widetilde{h_U}$.

\section{Benchmarking future quantum computers}\label{sec:bench_discuss}
A natural idea for characterizing quantum computers is to use some low level benchmarks, like error rates of elementary gates. This approach is currently used, for example, by IBM for publicly available quantum processing units \cite{IBM_Q}. Although informative, it is troublesome to translate it into quantum computer performance while executing real-life circuits \cite{proctor2024benchmarking}.

Alternatively one can employ computational problem benchmarks which describe performance of quantum computer for certain well-established problems, e.g. factoring large numbers \cite{gidney2019factor}. However, these types of tests fail to check universality of a quantum computer. 

Quantum Volume test was proposed as a remedy to this dilemma by considering multiple large random circuits at once \cite{Cross2019}. 
However, it suffers criticism due to its intrinsic non-scalability \cite{proctor2024benchmarking}.
Straightforward calculation of heavy output subspace corresponds to multiplication of $T$ matrices of size $2^N \times 2^N$, thus its complexity $c_{QV}$ scales as  
\begin{equation}
\label{complexity1}
c_{QV}=\mathcal{O}(T \; 2^{x N}),
\end{equation}
with $x = 3$ for standard matrix multiplication and $2<x<3$  for advanced algorithms   \cite{alman2024refined}.
For the {\sl parity-preserving} test the heavy output subspace is \textit{a priori} known, whereas for the double parity it is sufficient to track permutations of qubits, 
so in this case the complexity $c_{dp}$ scales as 
\begin{equation}
\label{complexity2}
c_{dp}=\mathcal{O}(T N).
\end{equation}

\rd{Beyond this issue of classical simulability, QV has other weaknesses. First, it can be relatively insensitive to certain coherent and correlated noise mechanisms, which can, in good approximation, preserve the heavy output subspace, while noticeably modifying intended unitary dynamics ~\cite{Hines2023,proctor2024benchmarking}.
Second, QV is usually employed only to probe square circuits of fixed width and depth, and thus provides a very limited information.
Volumetric benchmarking frameworks \cite{BlumeKohoutVolumetric} generalize the original QV idea to rectangular circuits. Our parity-preserving circuits can be adapted for similar analysis in a straightforward manner.
The goal of this work, however, is not to compete with such broad volumetric frameworks or with application-centric full-stack benchmarks \cite{proctor2024benchmarking}, but rather to construct the intuitive QV-style measure with an operational interpretation while removing its dependence on classically intractable simulations.}

One might try to restore QV test by using sophisticated classical algorithms for faster simulation of QV circuit. 
Natural candidates might be based on matchgates \cite{jozsa2008matchgates}, stabilizer states \cite{Aaronson2004}, or tensor networks \cite{Orus2019tensor}. Yet those approaches are doomed to failure due to the genericity of the QV circuit, which translates to large number of non-matchgates, e.g. swaps, non-Clifford gates and large bond dimension, respectively.

A more insightful method for simulating the QV circuit might try to leverage this universality. 
An approach based on Pauli shadows  \cite{bermejo2024quantum, angrisani2024classically} provides an algorithm to calculate any expectation value of quantum circuit outputs, for fixed precision, with  polynomial complexity. However, this approach also falls short, since, to classify outputs, one needs the errors to be smaller than the output probabilities $\approx 2^{-N}$. This requires exponential precision, resulting in exponential complexity \cite{angrisani2024classically}.
Thus, we infer that to overcome QV test scalability issue, one must modify its structure, as we propose in this work.

\rd{Mirror circuits provide an alternative scalable benchmark. Their core idea is to run a circuit followed by its (approximate, or recompiled) inverse and measure how well the overall transformation returns the
system to a reference state \cite{Proctor2022measuring, mayer2021theory, Hines2023}. 
In these schemes circuits are drawn from a prescribed ensemble. Then a corresponding mirror or approximate inverse is constructed, often using randomized compiling to suppress coherent-error cancellation and to ensure that the resulting error model is well behaved \cite{Hines2023}.
Some mirror-based benchmarks, such as mirror quantum volume \cite{Hines2023,Amico2023}, are related to the original QV construction but do so at the cost of introducing an inversion-based structure. Others, like mirror randomized benchmarking, focus on scalable characterization of Clifford layers \cite{Proctor2022}.
While these methods yield scalable estimates of gate or circuit fidelities, such metrics are not directly translatable to discrepancies in outcome probabilities, a distinction clearly illustrated by a phase-flip error model.
On the other hand, the QV test \blu{is concerned} with heavy outputs, so it examines directly the discrepancies of outcome probabilities for quantum processor.
Thus, it gives straightforward information about the accuracy of machine outcomes for a generic quantum circuit with given dimensions. 
Mirror circuit methods benchmark performance based on specifically structured mirror circuits, rather than preserving the original QV operational picture of implementing generic random circuits drawn from a universal ensemble, without any restrictions on circuit compilation and implementation, which is what our parity-based modifications aim to extend to larger system sizes.
Therefore, they can be considered as an alternative benchmarking method based on different motivation.}

\section{Concluding remarks}
\label{sec:conclusions}

We have introduced parity--preserving modifications of the Quantum Volume (QV) benchmark that address its scalability issue while preserving fundamental advantages, such as architecture independence and a clear operational meaning.
Unlike the standard QV test, which depends on expensive classical simulations to determine the heavy output subspace, our new benchmark directly identifies this subspace from the structure of the implemented gates. This eliminates the exponential cost associated with classical simulation and ensures that the test remains useful for an increasing number of qubits.
Because we managed to obtain this goal while still \blu{probing} the ability \blu{to perform} universal quantum computation and without significant simplification of applied quantum circuits, the features of the proposed benchmark coincide with those of the original QV test.

\bl{We also developed mechanisms to ensure that introduced parity-preserving structure does not compromise the benchmark’s diagnostic value. First, we proposed an auxiliary double-parity benchmark that serves as a systematic cross-check, tracking errors for which the parity-preserving benchmark could be partially insensitive. Second, we described practical strategies for hiding parity that could be implemented at the level of preparing of the logical circuit to be performed. They make underlying parity structure less transparent and therefore harder to exploit or mimic.}


Finally, \bl{as an additional contribution,} we introduce an efficient method for estimating the heavy-output probability in the standard QV test for certain types of noise using parity-preserving circuits. This approach requires only polynomial resources, both quantum and classical. The connection to the standard QV test is established via a simplified noise model and a structured approximation of the circuit, while preserving its essential properties.

\textit{Acknowledgement.} It is a pleasure to thank Jakub Czartowski for fruitful discussions and 
Nadir Samos de Buruaga for several useful comments.
Furthermore, we acknowledge Pedro Ribeiro and Rodrigo Pereira for a collaboration on an earlier project \cite{Nadir2024fidelity}, which offered valuable context for this work. 
We acknowledge funding by the European Union under ERC
Advanced Grant TAtypic, Project No. 101142236, and by
the National Science Centre, Poland, under Contract No.
2021/03/Y/ST2/00193 within the QuantERA II Programme,
which has received funding from the European Union’s Horizon 2020 research and innovation programme under Grant
Agreement No. 101017733.
RB acknowledges support by the National Science Centre, Poland, under the contract number 2023/50/E/ST2/00472.
RK acknowledges the financial support of the European Union under the REFRESH – Research Excellence For Region Sustainability and High-tech Industries project number CZ.10.03.01/00/22\_003/0000048 via the Operational Programme Just Transition.

\newpage
\bibliography{biblio}

@article{reexamining_QV,
  author    = {Charles H. Baldwin and Karl Mayer and Natalie C. Brown and Ciar{\'{a}}n Ryan{-}Anderson and David Hayes},
  title     = {Re-examining the quantum volume test: Ideal distributions, compiler optimizations, confidence intervals, and scalable resource estimations},
  journal   = {Quantum},
  volume    = {6},
  pages     = {707},
  year      = {2022},
  url       = {https://doi.org/10.22331/q-2022-05-09-707},
  doi       = {10.22331/Q-2022-05-09-707},
  bibsource = {dblp computer science bibliography, https://dblp.org}
}

@article{Math_monster_theoretic_fundations,
  title     = {Complexity-Theoretic Foundations of Quantum Supremacy Experiments},
  author    = {S. Aaronson and Lijie Chen},
  journal   = {Cybersecurity and Cyberforensics Conference},
  pages = {{Riga Latvia}},
  year      = {2016},
  url={https://api.semanticscholar.org/CorpusID:12591414},
}

@article{Bernstein1997,
  title = {Quantum Complexity Theory},
  volume = {26},
  ISSN = {1095-7111},
  url = {http://dx.doi.org/10.1137/S0097539796300921},
  DOI = {10.1137/s0097539796300921},
  number = {5},
  journal = {SIAM Journal on Computing},
  publisher = {Society for Industrial & Applied Mathematics (SIAM)},
  author = {Bernstein,  Ethan and Vazirani,  Umesh},
  year = {1997},
  month = oct,
  pages = {1411–1473}
}

@article{parameters_for_Haar_measure,
  title     = {Unitary quantum gates, perfect entanglers, and unistochastic maps},
  author    = {Musz, Marcin and Ku\'{s}, Marek and \ifmmode \dot{Z}\else \.{Z}\fi{}yczkowski, Karol},
  journal   = {Phys. Rev. A},
  volume    = {87},
  issue     = {2},
  pages     = {022111},
  numpages  = {12},
  year      = {2013},
  month     = {Feb},
  publisher = {American Physical Society},
  doi       = {10.1103/PhysRevA.87.022111},
  url       = {https://link.aps.org/doi/10.1103/PhysRevA.87.022111}
}

@article{Nadir2024fidelity,
	title={{Fidelity decay and error accumulation in random quantum circuits}},
	author={Nadir Samos Sáenz de Buruaga and Rafał Bistroń and Marcin Rudziński and Rodrigo M. C. Pereira and Karol Życzkowski and Pedro Ribeiro},
	journal={SciPost Phys.},
	volume={19},
	pages={013},
	year={2025},
	publisher={SciPost},
	doi={10.21468/SciPostPhys.19.1.013},
	url={https://scipost.org/10.21468/SciPostPhys.19.1.013},
}

@incollection{LAKSHMIVARAHAN1984295,
title = {Parallel Sorting Algorithms},
editor = {Marshall C. Yovits},
series = {Advances in Computers},
publisher = {Elsevier},
volume = {23},
pages = {295-354},
year = {1984},
issn = {0065-2458},
doi = {https://doi.org/10.1016/S0065-2458(08)60467-2},
url = {https://www.sciencedirect.com/science/article/pii/S0065245808604672},
author = {S. Lakshmivarahan and Sudarshan K. Dhall and Leslie L. Miller},
}

@article{CANFIELD2011109,
title = {The {Mahonian} probability distribution on words is asymptotically normal},
journal = {Advances in Applied Mathematics},
volume = {46},
number = {1},
pages = {109-124},
year = {2011},
note = {Special issue in honor of Dennis Stanton},
issn = {0196-8858},
doi = {https://doi.org/10.1016/j.aam.2009.10.001},
url = {https://www.sciencedirect.com/science/article/pii/S0196885810001090},
author = {E. Rodney Canfield and Svante Janson and Doron Zeilberger},
}

@article{proctor2024benchmarking,
  title   = {Benchmarking quantum computers},
  author  = {Timothy Proctor and Kevin Young and Andrew D. Baczewski and others},
  year    = {2025},
  journal = {Nat. Rev. Phys.},
  volume={7},
   pages={105–118},
   url={https://doi.org/10.1038/s42254-024-00796-z}
}

@article{gidney2019factor,
      journal   = {Quantum},
      volume    = {5},
      pages     = {433},
      year      = {2021},
      url       = {https://doi.org/10.22331/q-2021-04-15-433},
      title     = {How to factor 2048 bit {RSA} integers in 8 hours using 20 million noisy qubits},
    author    = {C. Gidney and Martin Ekerå},
}

@article{angrisani2024classically,
  title   = {Classically estimating observables of noiseless quantum circuits},
  author  = {Armando Angrisani and Alexander Schmidhuber and Manuel S. Rudolph and M. Cerezo and Zoë Holmes and Hsin-Yuan Huang},
  year    = {2024},
  journal = {preprint arXiv:2409.01706},
  url={https://doi.org/10.48550/arXiv.2409.01706}
}

@article{bermejo2024quantum,
  title   = {Quantum Convolutional Neural Networks are (Effectively) Classically Simulable},
  author  = {Pablo Bermejo and Paolo Braccia and Manuel S. Rudolph and Zoë Holmes and Lukasz Cincio and M. Cerezo},
  year    = {2024},
  journal = {preprint arXiv:2408.12739},
  url={https://doi.org/10.48550/arXiv.2408.12739}
}

@article{Liu2018,
  title = {Spectral form factors and late time quantum chaos},
  author = {Liu, Junyu},
  journal = {Phys. Rev. D},
  volume = {98},
  issue = {8},
  pages = {086026},
  numpages = {28},
  year = {2018},
  month = {Oct},
  publisher = {American Physical Society},
  doi = {10.1103/PhysRevD.98.086026},
  url = {https://link.aps.org/doi/10.1103/PhysRevD.98.086026}
}

@misc{IBM_Q,
  title = {{IBM Quantum Platform}},
  howpublished = {\url{https://quantum.ibm.com}},
  note = {Accessed: 2024-11-22}
}

@misc{Qiskit,
  title = {{Qiskit documentation}},
  howpublished = {\url{https://docs.quantum.ibm.com}},
}

@article{Cross2019,
  title = {Validating quantum computers using randomized model circuits},
  author = {Cross, Andrew W. and Bishop, Lev S. and Sheldon, Sarah and Nation, Paul D. and Gambetta, Jay M.},
  journal = {Phys. Rev. A},
  volume = {100},
  issue = {3},
  pages = {032328},
  numpages = {11},
  year = {2019},
  month = {Sep},
  publisher = {American Physical Society},
  doi = {10.1103/PhysRevA.100.032328},
  url = {https://link.aps.org/doi/10.1103/PhysRevA.100.032328}
}

@article{Aaronson2004,
  title = {Improved simulation of stabilizer circuits},
  author = {Aaronson, Scott and Gottesman, Daniel},
  journal = {Phys. Rev. A},
  volume = {70},
  issue = {5},
  pages = {052328},
  numpages = {14},
  year = {2004},
  month = {Nov},
  publisher = {American Physical Society},
  doi = {10.1103/PhysRevA.70.052328},
  url = {https://link.aps.org/doi/10.1103/PhysRevA.70.052328}
}

@article{Orus2019tensor,
  author  = {Román Orús},
  title   = {Tensor networks for complex quantum systems},
  journal = {Nat. Rev. Phys.},
  year    = {2019},
  volume={1},
   pages={538–550},
   url={https://doi.org/10.1038/s42254-019-0086-7}
}

@article{Abrams,
  title = {Quantum Algorithm Providing Exponential Speed Increase for Finding Eigenvalues and Eigenvectors},
  author = {Abrams, Daniel S. and Lloyd, Seth},
  journal = {Phys. Rev. Lett.},
  volume = {83},
  issue = {24},
  pages = {5162--5165},
  numpages = {0},
  year = {1999},
  month = {Dec},
  publisher = {American Physical Society},
  doi = {10.1103/PhysRevLett.83.5162},
  url = {https://link.aps.org/doi/10.1103/PhysRevLett.83.5162}
}

@article{Harrow,
  title = {Quantum Algorithm for Linear Systems of Equations},
  author = {Harrow, Aram W. and Hassidim, Avinatan and Lloyd, Seth},
  journal = {Phys. Rev. Lett.},
  volume = {103},
  issue = {15},
  pages = {150502},
  numpages = {4},
  year = {2009},
  month = {Oct},
  publisher = {American Physical Society},
  doi = {10.1103/PhysRevLett.103.150502},
  url = {https://link.aps.org/doi/10.1103/PhysRevLett.103.150502}
}

@INPROCEEDINGS{Shor1994,

  author={Shor, P.W.},

  booktitle={Proceedings 35th Annual Symposium on Foundations of Computer Science}, 

  title={Algorithms for quantum computation: discrete logarithms and factoring}, 

  year={1994},

  volume={},

  number={},

  pages={124-134},

  doi={10.1109/SFCS.1994.365700}}

@inproceedings{Grover1996,
author = {Grover, Lov K.},
title = {A fast quantum mechanical algorithm for database search},
year = {1996},
isbn = {0897917855},
url = {https://doi.org/10.1145/237814.237866},
doi = {10.1145/237814.237866},
booktitle = {Proceedings of the Twenty-Eighth Annual ACM Symposium on Theory of Computing},
pages = {212–219},
numpages = {8},
location = {Philadelphia, Pennsylvania, USA},
}

@ARTICLE{Corcoles,
  author={Córcoles, Antonio D. and others},
  journal={Proceedings of the IEEE}, 
  title={Challenges and Opportunities of Near-Term Quantum Computing Systems}, 
  year={2020},
  volume={108},
  number={8},
  pages={1338-1352},
  doi={10.1109/JPROC.2019.2954005}
  }

@ARTICLE{Pelofske,
  author={Pelofske, Elijah and Bärtschi, Andreas and Eidenbenz, Stephan},
  journal={IEEE Transactions on Quantum Engineering}, 
  title={Quantum Volume in Practice: What Users Can Expect From {NISQ} Devices}, 
  year={2022},
  volume={3},
  pages={1--19},
  doi={10.1109/TQE.2022.3184764}}

@article{Jurcevic,
doi = {10.1088/2058-9565/abe519},
url = {https://dx.doi.org/10.1088/2058-9565/abe519},
year = {2021},
month = {mar},
publisher = {IOP Publishing},
volume = {6},
number = {2},
pages = {025020},
author = {Petar Jurcevic and others},
title = {Demonstration of quantum volume 64 on a superconducting quantum computing system},
journal = {Quantum Sci. Technol.},
}

@article{Arute,
   title={Quantum supremacy using a programmable superconducting processor},
   volume={574},
   ISSN={1476-4687},
   url={http://dx.doi.org/10.1038/s41586-019-1666-5},
   DOI={10.1038/s41586-019-1666-5},
   number={7779},
   journal={Nature},
   publisher={Springer Science and Business Media LLC},
   author={Arute, Frank and Arya, Kunal and  Babbush, Ryan and others},
   year={2019},
   month=oct, pages={505–510} }

@article{Boixo,
   title={Characterizing quantum supremacy in near-term devices},
   volume={14},
   ISSN={1745-2481},
   url={http://dx.doi.org/10.1038/s41567-018-0124-x},
   DOI={10.1038/s41567-018-0124-x},
   number={6},
   journal={Nat. Phys.},
   publisher={Springer Science and Business Media LLC},
   author={Boixo, Sergio and Isakov, Sergei V. and Smelyanskiy, Vadim N. and others},
   year={2018},
   month=apr, pages={595–600} }

@article{Bishop,
  title={Quantum Volume},
  author={Lev S. Bishop and Sergey Bravyi and Andrew W. Cross and Jay M. Gambetta and John A. Smolin},
  year={2017},
  journal  = {Tech. Rep.},
  url={https://api.semanticscholar.org/CorpusID:13143757}
}

@article{Pino,
   title={Demonstration of the trapped-ion quantum CCD computer architecture},
   volume={592},
   ISSN={1476-4687},
   url={http://dx.doi.org/10.1038/s41586-021-03318-4},
   DOI={10.1038/s41586-021-03318-4},
   number={7853},
   journal={Nature},
   publisher={Springer Science and Business Media LLC},
   author={Pino, J. M. and Dreiling, J. M. and Figgatt, C. and others},
   year={2021},
   month=apr, pages={209–213} }

@online{Honeywell,
  title = {Honeywell {{Sets Another Record For Quantum Computing Performance}}},
  url = {https://www.honeywell.com/us/en/news/2021/07/honeywell-sets-another-record-for-quantum-computing-performance},
  urldate = {2024-11-22},
  note = {Internet announcement (2024)}
}

@article{McGeoch,
      title={How {NOT} to Fool the Masses When Giving Performance Results for Quantum Computers}, 
      author={Catherine McGeoch},
      year={2024},
      journal = {preprint arXiv:2411.08860},
      url={https://arxiv.org/abs/2411.08860}, 
}

@article{Sundaresan,
  title = {Reducing Unitary and Spectator Errors in Cross Resonance with Optimized Rotary Echoes},
  author = {Sundaresan, Neereja and Lauer, Isaac and Pritchett, Emily and Magesan, Easwar and Jurcevic, Petar and Gambetta, Jay M.},
  journal = {PRX Quantum},
  volume = {1},
  issue = {2},
  pages = {020318},
  numpages = {23},
  year = {2020},
  month = {Dec},
  publisher = {American Physical Society},
  doi = {10.1103/PRXQuantum.1.020318},
  url = {https://link.aps.org/doi/10.1103/PRXQuantum.1.020318}
}

@article{Kielpinski,
   title={Architecture for a large-scale ion-trap quantum computer},
   volume={417},
   url={https://doi.org/10.1038/nature00784},
   journal={Nature},
   author={Kielpinski, D. and Monroe, C. and Wineland, D.},
   year={2002},
    pages={709–711} 
}

@article{Kjaergaard,
   author = "Kjaergaard, Morten and Schwartz, Mollie E. and Braumüller, Jochen and Krantz, Philip and Wang, Joel I.-J. and Gustavsson, Simon and Oliver, William D.",
   title = "Superconducting Qubits: Current State of Play", 
   journal= "Annu. Rev. Condens. Matter Phys.",
   year = "2020",
   volume = "11",
   number = "Volume 11, 2020",
   pages = "369-395",
   doi = "https://doi.org/10.1146/annurev-conmatphys-031119-050605",
   url = "https://www.annualreviews.org/content/journals/10.1146/annurev-conmatphys-031119-050605",
   publisher = "Annual Reviews",
   issn = "1947-5462"
  }

@article{Bluvstein,
   title={Logical quantum processor based on reconfigurable atom arrays},
   volume={626},
   ISSN={1476-4687},
   url={http://dx.doi.org/10.1038/s41586-023-06927-3},
   DOI={10.1038/s41586-023-06927-3},
   number={7997},
   journal={Nature},
   publisher={Springer Science and Business Media LLC},
   author={Bluvstein, Dolev and Evered, Simon J. and Geim, Alexandra A. and others},
   year={2023},
   month=dec, pages={58–65} }

@article{Kim,
  title={Evidence for the utility of quantum computing before fault tolerance},
  author={Kim, Youngseok and Eddins, Andrew and Anand, Sajant and others},
  journal={Nature},
  volume={618},
  number={7965},
  pages={500--505},
  year={2023},
  url={https://doi.org/10.1038/s41586-023-06096-3}
}

@article{Chen,
     author    = {Jwo-Sy Chen and Erik Nielsen and Matthew Ebert and Volkan Inlek and Kenneth Wright and Vandiver Chaplin and Andrii Maksymov and Eduardo Páez and Amrit Poudel and Peter Maunz and John Gamble},
      journal   = {Quantum},
      volume    = {8},
      pages     = {1516},
      year      = {2024},
      url       = {https://doi.org/10.22331/q-2024-11-07-1516},
      title={Benchmarking a trapped-ion quantum computer with 30 qubits}, 
}

@article{Proctor,
   title={Measuring the capabilities of quantum computers},
   volume={18},
   ISSN={1745-2481},
   url={http://dx.doi.org/10.1038/s41567-021-01409-7},
   DOI={10.1038/s41567-021-01409-7},
   number={1},
   journal={Nat. Phys.},
   publisher={Springer Science and Business Media LLC},
   author={Proctor, Timothy and Rudinger, Kenneth and Young, Kevin and others},
   year={2021},
   month=dec, pages={75–79} }

@article{Bruzewicz,
    author = {Bruzewicz, Colin D. and Chiaverini, John and McConnell, Robert and Sage, Jeremy M.},
    title = {Trapped-ion quantum computing: Progress and challenges},
    journal = {Appl. Phys. Rev.},
    volume = {6},
    number = {2},
    pages = {021314},
    year = {2019},
    month = {05},
    issn = {1931-9401},
    doi = {10.1063/1.5088164},
    url = {https://doi.org/10.1063/1.5088164}
}

@article{Vandersypen,
  title = {{NMR} techniques for quantum control and computation},
  author = {Vandersypen, L. M. K. and Chuang, I. L.},
  journal = {Rev. Mod. Phys.},
  volume = {76},
  issue = {4},
  pages = {1037--1069},
  numpages = {0},
  year = {2005},
  month = {Jan},
  publisher = {American Physical Society},
  doi = {10.1103/RevModPhys.76.1037},
  url = {https://link.aps.org/doi/10.1103/RevModPhys.76.1037}
}

@book{Majidy, 
    place={Cambridge}, 
    title={Building Quantum Computers: A Practical Introduction}, 
    publisher={Cambridge University Press}, 
    author={Majidy, Shayan and Wilson, Christopher and Laflamme, Raymond}, 
    url={https://doi.org/10.1017/9781009417020},
    year={2024}
}

@article{Kok,
  title = {Linear optical quantum computing with photonic qubits},
  author = {Kok, Pieter and Munro, W. J. and Nemoto, Kae and Ralph, T. C. and Dowling, Jonathan P. and Milburn, G. J.},
  journal = {Rev. Mod. Phys.},
  volume = {79},
  issue = {1},
  pages = {135--174},
  numpages = {0},
  year = {2007},
  month = {Jan},
  publisher = {American Physical Society},
  doi = {10.1103/RevModPhys.79.135},
  url = {https://link.aps.org/doi/10.1103/RevModPhys.79.135}
}

@article{LaRose,
      title={Error mitigation increases the effective quantum volume of quantum computers}, 
      author={Ryan LaRose and Andrea Mari and Vincent Russo and Dan Strano and William J. Zeng},
      year={2022},
      journal = {preprint arXiv:2203.05489},
      url={https://arxiv.org/abs/2203.05489}, 
}

@article{Cornelissen,
      title={Scalable Benchmarks for Gate-Based Quantum Computers}, 
      author={Arjan Cornelissen and Johannes Bausch and András Gilyén},
      year={2021},
      journal = {preprint arXiv:2104.10698},
      url={https://arxiv.org/abs/2104.10698}, 
}

@article{alman2024refined,
    title      = {A Refined Laser Method and Faster Matrix Multiplication},
    author     = {Josh Alman and Virginia Vassilevska Williams},
    url        = {https://theoretics.episciences.org/11261},
    doi        = {10.46298/theoretics.24.21},
    journal    = {TheoretiCS},
    issn       = {2751-4838},
    volume     = {Volume 3},
    eid        = 21,
    year       = {2024},
    month      = {Sep},
}

@article{IBM2023,
  title = {{IBM} releases first-ever 1000-qubit quantum chip},
  volume = {624},
  ISSN = {1476-4687},
  url = {http://dx.doi.org/10.1038/d41586-023-03854-1},
  DOI = {10.1038/d41586-023-03854-1},
  number = {7991},
  journal = {Nature},
  publisher = {Springer Science and Business Media LLC},
  author = {Castelvecchi,  Davide},
  year = {2023},
  month = dec,
  pages = {238}
}

@article{abughanem2024ibm,
  title   = {{IBM} {Quantum Computers: Evolution, Performance, and Future Directions}},
  author  = {M. AbuGhanem},
  year    = {2024},
  journal = {preprint arXiv:2410.00916},
  url={https://doi.org/10.48550/arXiv.2410.00916}
}

@article{Google2024quantum,
  author  = {{Google Quantum AI and Collaborators}},
  title   = {Quantum error correction below the surface code threshold},
  journal = {Nature},
  year    = {2024},
  url     = {https://doi.org/10.1038/s41586-024-08449-y}
}

@article{gao2024establishing,
  title   = {Establishing a New Benchmark in Quantum Computational Advantage with 105-qubit {Zuchongzhi} 3.0 Processor},
  author  = {Dongxin Gao and Daojin Fan and Chen Zha and others},
  year    = {2024},
  journal = {preprint arXiv:2412.11924},
  url={https://doi.org/10.48550/arXiv.2412.11924}
}

@misc{bistronGithub_QV,
 author = {Bistro{\'n}, Rafa{\l}},
 title = {{{GitHub Repository Quantum Volume}}},
 url = {https://github.com/RafalBistron/Quantum_Volume},
}

@article{Khaneja,
  title = {Time optimal control in spin systems},
  author = {Khaneja, Navin and Brockett, Roger and Glaser, Steffen J.},
  journal = {Phys. Rev. A},
  volume = {63},
  issue = {3},
  pages = {032308},
  numpages = {13},
  year = {2001},
  url={https://doi.org/10.1103/PhysRevA.63.032308}
}

@article{Proctor2022measuring,
  author  = {Timothy Proctor and Kenneth Rudinger and Kevin Young and Erik Nielsen and Robin Blume-Kohout},
  title   = {Measuring the capabilities of quantum computers},
  journal = {Nat. Phys.},
  year    = {2022},
  volume={18},
   pages={75–79},
   url={https://doi.org/10.1038/s41567-021-01409-7}
}

@article{Proctor2022,
  title = {Scalable Randomized Benchmarking of Quantum Computers Using Mirror Circuits},
  author = {Proctor, Timothy and Seritan, Stefan and Rudinger, Kenneth and Nielsen, Erik and Blume-Kohout, Robin and Young, Kevin},
  journal = {Phys. Rev. Lett.},
  volume = {129},
  issue = {15},
  pages = {150502},
  numpages = {7},
  year = {2022},
  month = {Oct},
  publisher = {American Physical Society},
  doi = {10.1103/PhysRevLett.129.150502},
  url = {https://link.aps.org/doi/10.1103/PhysRevLett.129.150502}
}

@article{mayer2021theory,
  title   = {Theory of mirror benchmarking and demonstration on a quantum computer},
  author  = {Karl Mayer and Alex Hall and Thomas Gatterman and Si Khadir Halit and Kenny Lee and Justin Bohnet and Dan Gresh and Aaron Hankin and Kevin Gilmore and Justin Gerber and John Gaebler},
  year    = {2021},
  journal = {preprint arXiv:2108.10431},
  url={https://doi.org/10.48550/arXiv.2108.10431}
}

@inproceedings{Amico2023,
  title = {Defining Best Practices for Quantum Benchmarks},
  booktitle = {2023 IEEE International Conference on Quantum Computing and Engineering (QCE)},
  publisher = {IEEE},
  author = {Amico,  Mirko and Zhang,  Helena and Jurcevic,  Petar and Bishop,  Lev S. and Nation,  Paul and Wack,  Andrew and McKay,  David C.},
  year = {2023},
  month = sep,
  pages = {692–702},
  url={https://doi.org/10.1109/QCE57702.2023.00084}
}

@article{Hines2023,
  title = {Demonstrating Scalable Randomized Benchmarking of Universal Gate Sets},
  author = {Hines, Jordan and Lu, Marie and Naik, Ravi K. and Hashim, Akel and Ville, Jean-Loup and Mitchell, Brad and Kriekebaum, John Mark and Santiago, David I. and Seritan, Stefan and Nielsen, Erik and Blume-Kohout, Robin and Young, Kevin and Siddiqi, Irfan and Whaley, Birgitta and Proctor, Timothy},
  journal = {Phys. Rev. X},
  volume = {13},
  issue = {4},
  pages = {041030},
  numpages = {38},
  year = {2023},
  month = {Nov},
  publisher = {American Physical Society},
  doi = {10.1103/PhysRevX.13.041030},
  url = {https://link.aps.org/doi/10.1103/PhysRevX.13.041030}
}

@article{jozsa2008matchgates,
  title     = {Matchgates and classical simulation of quantum circuits},
  author    = {Jozsa, Richard and Miyake, Akimasa},
  journal   = {Proceedings of the Royal Society A: Mathematical, Physical and Engineering Sciences},
  volume    = {464},
  number    = {2100},
  pages     = {3089-3106},
  year      = {2008},
  publisher = {The Royal Society London}
}

@article{reardon-smith2023improved,
  title = {Improved simulation of quantum circuits dominated by free fermionic operations},
  volume = {8},
  ISSN = {2521-327X},
  url = {http://dx.doi.org/10.22331/q-2024-12-04-1549},
  DOI = {10.22331/q-2024-12-04-1549},
  journal = {Quantum},
  publisher = {Verein zur Forderung des Open Access Publizierens in den Quantenwissenschaften},
  author = {Reardon-Smith,  Oliver and Oszmaniec,  Michał and Korzekwa,  Kamil},
  year = {2024},
  month = dec,
  pages = {1549}
}

@article{Kraus01,
  title = {Optimal creation of entanglement using a two-qubit gate},
  author = {Kraus, B. and Cirac, J. I.},
  journal = {Phys. Rev. A},
  volume = {63},
  issue = {6},
  pages = {062309},
  numpages = {8},
  year = {2001},
  month = {May},
  publisher = {American Physical Society},
  doi = {10.1103/PhysRevA.63.062309},
  url = {https://link.aps.org/doi/10.1103/PhysRevA.63.062309}
}

@article{Vatan04,
  title = {Optimal quantum circuits for general two-qubit gates},
  author = {Vatan, Farrokh and Williams, Colin},
  journal = {Phys. Rev. A},
  volume = {69},
  issue = {3},
  pages = {032315},
  numpages = {5},
  year = {2004},
  month = {Mar},
  publisher = {American Physical Society},
  doi = {10.1103/PhysRevA.69.032315},
  url = {https://link.aps.org/doi/10.1103/PhysRevA.69.032315}
}

@article{BlumeKohoutVolumetric,
  doi = {10.22331/q-2020-11-15-362},
  url = {https://doi.org/10.22331/q-2020-11-15-362},
  title = {A volumetric framework for quantum computer benchmarks},
  author = {Blume-Kohout, Robin and Young, Kevin C.},
  journal = {{Quantum}},
  issn = {2521-327X},
  publisher = {{Verein zur F{\"{o}}rderung des Open Access Publizierens in den Quantenwissenschaften}},
  volume = {4},
  pages = {362},
  month = nov,
  year = {2020}
}

\onecolumngrid
\appendix


\section{Circuits with parity preservation, and their analytical counterparts}\label{Section:I}

In this \rd{Appendix}, we briefly reintroduce parity-preserving and double parity-preserving quantum volume circuits discussed in Sections \ref{subsec:single-parity}, \ref{subsec:double-parity}. \bl{Then we present slight theoretical modifications, that enable us} to derive an analytical formula for heavy output frequency for a reasonable choice of a noise model. 
\bl{Subsequently} we present the exact derivation of heavy output frequency for the modified circuits for different noise models including non-unitary dissipation errors \bl{(equations~(\ref{hu_sol1}, \ref{hu_sol1_en}, \ref{hu_dub1}))}. We finish the \rd{Appendix} by comparing results obtained from modified circuits with numerical simulations of their originals.

Let us start with a quantum volume circuit consisting of $N$ qubits and $T$ layers, each layer consisting of qubit permutation and a $\lfloor N/2 \rfloor$ two-qubit gates. In order to make a circuit action parity-preserving, we substitute arbitrary two-qubit gates with parity-preserving ones of a form
\begin{equation}
\label{u_parity_1a}
u = \begin{pmatrix} * & 0 & 0 & *\\ 0 & * & * & 0\\ 0 & 
    * & * & 0\\ * & 0 & 0 & *\end{pmatrix}~.
\end{equation}
where $*$ represents nonzero matrix entries. These gates are drawn randomly to constitute a QV circuit, with two uncoupled unitary blocks of size 2 (the inner and outer ones) taken independently from Haar measure on $U(2)$. Note that any two-qubit gate can be decomposed into parity-preserving interaction part and relatively simple single-qubit pre- and post-processing \cite{Kraus01, Vatan04, parameters_for_Haar_measure}. Thus our restriction does not substantially affect the implementation difficulty. Furthermore, parity-preserving gates \eqref{u_parity_1} encompass both matchgates and swap, enabling general quantum computations \cite{jozsa2008matchgates}.

Due to the block structure of gate \eqref{u_parity_1} a quantum state at any stage of the circuit is always a superposition of computational basis states with a conserved parity of the number of $1$s.
This is the case for permutations too. 
Thus, if the input state of the circuit is $|0\ra^{\otimes N}$, the heavy output subspace consists of the states with an even number of qubits in the state $|1\ra$. 

The downside of this approach, as mentioned in \rd{the main body of the work}, is that the heavy output subspace is not disturbed by any permutation - including the faulty one, so one cannot detect errors such as an omitted swap.
The second approach is to tackle this problem by randomly grouping qubits into two sets of $N/2$ qubits - let us call them red and blue.
In each layer, if two qubits from the same set meet, one applies the gate  \eqref{u_parity_1}. When a two-qubit gate has to act on qubits from different sets, the interaction is diagonal to not "mix colors" and takes the form
\begin{equation}
u_{diag} = e^{ i a Z\otimes Z},
\end{equation}
where $a$ is a random phase with a flat measure and $Z$ is the Pauli $Z$ gate, to introduce non-trivial and not-matchgate interaction, without spoiling the parities inside each subset. If the input state is $|0\ra^{\otimes N}$ and the device is noiseless, then only one-fourth of outputs are possible and we may choose them as a new "restricted" heavy output subspace.

From now on we will call the first circuit a \textit{parity quantum volume circuit}, and the second one a \textit{double-parity quantum volume circuit}. 
To investigate properties of the presented benchmarking methods we use following error model. We assume square root of swap $\sqrt{S}$ as a fundamental two-qubit gate, and we assume that each permutation is implemented as a combination of swaps $S$ between qubits. Since in such a scenario swap is relatively fast, we assume that the main error within permutations comes from imperfect swaps $S \to S^{\beta}$, where $\beta$ is some random variable sampled from Gaussian distribution with mean $1$ and variance $\sigma$. After simple integration over $\beta$  one may notice that this model is equivalent to the probabilistic application of swap gate with probability $p = \frac{1}{2}(1 - e^{- \frac{1}{2} \pi^2 \sigma^2})$. Therefore both permutations and their errors preserve parity.
In the case of two-qubit gates, we assume that each one is followed by a unitary $e^{i \alpha H}$, where $H$ is a random Hamiltonian form GUE and $\alpha$ is a noise strength parameter.

\subsection{Solvable counterpart of parity quantum volume circuit}

Now we are prepared to present a modification of the parity-preserving quantum volume circuit which does not substantially affect its behavior but enables us to obtain a close analytical formula for an average heavy output probability in the presence of introduced noise model. We note that some parts of the below calculations were inspired by \cite{Nadir2024fidelity}.

Let $\mathcal{U} = \Pi (\bigotimes_{i = 1}^{N/2} \tilde{u})$ denote one layer in circuit, where $\Pi$ is the permutation of qubits and $\tilde{u} = e^{i\alpha H} u$ is a non-perfect implementation of random two-qubit gate $u$.
Then the heavy output frequency might be written as

\begin{equation}
\label{h_U_formula}
h_U = \sum_{P} |\la P|\prod_{j = 1}^T \overline{\mathcal{U}} ~|0^{\otimes N}\ra|^2 = \sum_P \la PP| \prod_{j = 1}^T \overline{\mathcal{U}\otimes \mathcal{U}^*} ~|(0^{\otimes N})^{\otimes 2}\ra~,
\end{equation}
where $|P\ra$ (like parity) denotes a string of bits with an even number of $1$s, analogically $|N\ra$ (like non-parity) will denote the string of bits with an odd number of $1$s. It should not be confused with $N$, which denotes the number of qubits. The conjugation of second $\mathcal{U}$ comes from the absolute value, and hereafter overline represents an average over noise model and circuit realizations.

To obtain an average heavy output probability one has to calculate the average
$\overline{\mathcal{U}\otimes \mathcal{U}^*}$ and then raise it to proper power. 
To make it possible we introduce modification of each layer $\mathcal{U}$ by $2^{n-1}\times 2^{n-1}$ unitaries $\mathcal{R}_P, \mathcal{R}_N$ sampled with Haar measure
\begin{equation}
\label{modified_big_u_1}
\mathcal{U} = \Pi \left(\bigotimes_{i = 1}^{N/2} \tilde{u}\right) \to (\mathcal{R}_P \oplus \mathcal{R}_N) \Pi (\mathcal{R}_P \oplus \mathcal{R}_N) \left(\bigotimes_{i = 1}^{N/2} \tilde{u}\right) (\mathcal{R}_P \oplus \mathcal{R}_N),
\end{equation}
where $\mathcal{R}_P$ is a random unitary on a subspace spanned by all states with an even number of ones: $|P\ra$ and $\mathcal{R}_N$ is a random unitary on a complementary subspace, both of them unaffected by any noise, and sampled independently. Thus by averaging over random $\mathcal{R}_P$ and $\mathcal{R}_N$, we are effectively decoupling the permutations and two-qubit gates. The motivation for this modification stems from the intuition that if one considers the circuit layer by layer, after a few steps the input to the layer is "random enough" so the appearance of unitaries that mix two subspaces of interest separately does not affect the action of the next layer. For further arguments for this type of modification, one can consult \cite{Nadir2024fidelity}.
Thus we should calculate the average of each component in \eqref{modified_big_u_1}, multiplied by its complex conjugate according to \eqref{h_U_formula}. The average of big unitaries $\mathcal{R}$  are widely known
\begin{equation*}
\begin{aligned}
& \overline{\mathcal{R}_P \otimes \mathcal{R}_P^*} = \frac{1}{2^{N-1}}\left(\sum_P |P\ra^{\otimes 2} \right) \left(\sum_P \la P|^{\otimes 2} \right) = |+_P\ra\la +_P|~, \\
& \overline{\mathcal{R}_N \otimes \mathcal{R}_N^*} = \frac{1}{2^{N-1}}\left(\sum_N |N\ra^{\otimes 2} \right) \left(\sum_N \la N|^{\otimes 2} \right) = |+_N\ra\la +_N|~,
\end{aligned}
\end{equation*}
where we defined the Bell-like states $|+_P\ra$ nad $|+_N\ra$ on appropriate subspaces.
Moreover, the action of each permutation $\Pi$, no matter perfect or faulty (in the error model of omitted swaps), is a bijection in the set of bit-strings with an even (and odd) number of ones. Thus it leaves Bell-like states invariant and we may write
\begin{equation*}
\begin{aligned}
& \left(\overline{(\mathcal{R}_P \oplus \mathcal{R}_N) \otimes (\mathcal{R}_P^* \oplus \mathcal{R}_N^*)}\right) \overline{(\Pi \otimes \Pi)} \left(\overline{(\mathcal{R}_P \oplus \mathcal{R}_N) \otimes (\mathcal{R}_P^* \oplus \mathcal{R}_N^*)}\right) = \\
& = \left( |+_P\ra\la +_P| +  |+_N\ra\la +_N|\right) \overline{(\Pi \otimes \Pi)} \left( |+_P\ra\la +_P| +  |+_N\ra\la +_N|\right) = \\
& = \left( |+_P\ra\la +_P| +  |+_N\ra\la +_N|\right) \left( |+_P\ra\la +_P| +  |+_N\ra\la +_N|\right) = \\
& = \left( |+_P\ra\la +_P| +  |+_N\ra\la +_N|\right),
\end{aligned}
\end{equation*}
where $\Pi$ is not conjugated, since it is a real-entry matrix. 

Now let us discuss the action of two-qubit gates. If no errors are present, the average over single two-qubit parity-preserving gate is given by:

\begin{equation}
\label{u_parity_1_av}
\begin{aligned}
\overline{u \otimes u^*}  =\frac{1}{2}\Big{[}& (|00,00\ra + |11,11\ra )(\la 00,00| + \la 11,11| ) 
+ (|01,01\ra + |10,10\ra )(\la 01,01| + \la 10,10| )
\Big{]}.
\end{aligned}
\end{equation}
In this expression we separated the space of $u$ with the space of $u^*$ by commas.
Since the averaged gates $\bigotimes_{i = 1}^{N/2} \overline{u \otimes u^*}$ always follows unitaries  $\overline{(\mathcal{R}_P \oplus \mathcal{R}_N) \otimes (\mathcal{R}_P^* \oplus \mathcal{R}_N^*)} $, it is sufficient to consider its action on $|+_P\ra$ and $|+_N\ra$ states. Below we focus only on $|+_P\ra$, since the other case is analogous. 
For each pair of qubits, if for some state $|P\ra$ the corresponding qubit values are $00$ then there exists also a state with the same values on all qubits, except $11$ on two qubits of interest, thus the action of $\overline{u \otimes u^*}$ doesn't change the state $|+_P\ra$, the same holds as well for the scenario with $01$ and $10$. Finally by the same arguments, one may extend this scenario to the entire product $\bigotimes_{i = 1}^{N/2} \overline{u \otimes u^*}$. Thus we obtain
\begin{equation*}
\begin{aligned}
\overline{\mathcal{U} \otimes \mathcal{U}^*} = &  \left(\overline{(\mathcal{R}_P \oplus \mathcal{R}_N) \otimes (\mathcal{R}_P^* \oplus \mathcal{R}_N^*)}\right) \overline{(\Pi \otimes \Pi)} \left(\overline{(\mathcal{R}_P \oplus \mathcal{R}_N) \otimes (\mathcal{R}_P^* \oplus \mathcal{R}_N^*)}\right)\bigotimes_{i = 1}^{N/2} \overline{e^{i \alpha H}  \otimes e^{-i \alpha H^*}} \bigotimes_{i = 1}^{N/2} \overline{u \otimes u^*} \\
&  \left(\overline{(\mathcal{R}_P \oplus \mathcal{R}_N) \otimes (\mathcal{R}_P^* \oplus \mathcal{R}_N^*)}\right) = \\
& =  \left( |+_P\ra\la +_P| +  |+_N\ra\la +_N|\right) \overline{(\Pi \otimes \Pi)} \left( |+_P\ra\la +_P| +  |+_N\ra\la +_N|\right) \bigotimes_{i = 1}^{N/2} \overline{e^{i \alpha H}  \otimes e^{-i \alpha H^*}} \bigotimes_{i = 1}^{N/2} \overline{u \otimes u^*} \left( |+_P\ra\la +_P| +  |+_N\ra\la +_N|\right) = \\
& = \left( |+_P\ra\la +_P| +  |+_N\ra\la +_N|\right)  \bigotimes_{i = 1}^{N/2} \overline{e^{i \alpha H}  \otimes e^{-i \alpha H^*}}  \left( |+_P\ra\la +_P| +  |+_N\ra\la +_N|\right),
\end{aligned}
\end{equation*}
where we added one more layer of $2^{n-1} \times 2^{n-1}$ unitaries $\mathcal{R}$, at the end since their average is equivalent to projections, thus can be added multiple times without affecting the result of products of $\overline{\mathcal{U} \otimes \mathcal{U}^*}$.

The final missing average is $\overline{e^{i \alpha H}  \otimes e^{-i \alpha H^*}}$, which is equal to \cite{Nadir2024fidelity}
\begin{equation}
\label{gauss_noise_average}
\overline{e^{i \alpha H}  \otimes e^{-i \alpha H^*}} = 
\frac{4f(\alpha)+1}{4+1} \id^{\otimes 2} + \frac{1-f(\alpha)}{4+1} |+\ra\la+| = a~ \id^{\otimes 2} + b~ |+\ra\la+|~,
\end{equation}
with $|+\ra = (|11\ra + |00\ra)$ being a two-qubit Bell state and
\begin{equation*}
f(\alpha) = \frac{1}{36} e^{-\alpha ^2} \left(-\alpha ^{10}+\frac{25 \alpha ^8}{2}-64 \alpha ^6+138 \alpha ^4-144 \alpha ^2+36\right) \approx e^{- (4+1) \alpha^2}~.
\end{equation*}

Therefore, before finishing the calculations we 
need to consider the action of a product \eqref{gauss_noise_average}  on $|+_P\ra$ state (the action on the state $|+_N\ra$ is analogous) 
\begin{equation}
\label{2q_gtate_noise_action}
\begin{aligned}
& \bigotimes_{i = 1}^{N/2} \overline{e^{i \alpha H}  \otimes e^{-i \alpha H^*}} |+_P\ra = \frac{1}{2^{N-1}} \bigotimes_{i = 1}^{N/2} (a~ \id^{\otimes 2} + b~ |+\ra\la+|) \sum_P |P\ra^{\otimes 2} = \frac{1}{2^{N-1}} \sum_{i = 0}^{N/2} \binom{N/2}{i} a^i b^{\frac{N}{2} - i} (\id^{\otimes 2})^{\otimes i} \otimes (|+\ra\la+|)^{\otimes (\frac{N}{2} -i) } \sum_P |P\ra^{\otimes 2},
\end{aligned}
\end{equation}
where we slightly abused the notation by combining together $\id^{\otimes 2}$ and $|+\ra\la+|$ from different subsystems.
First, let us extract the term with $ (\id^{\otimes 2})^{\otimes N/2}$ from the sum. Note that for each state $|P\ra$ one can decompose $\la +|$ on each pair of qubits into a pair with the same parity as $|P\ra^{\otimes 2}$ on those qubits (ex. $(00,00)$ and $(11,11)$) and the pair with opposite parity (ex. $(01,01)$ and $(10,10)$), breaking and expanding the sum. If in the expanded product are even number $2j$ states with "opposite parity" the parity of the entire bit-string does not change. However, if it happens an odd number of times $2j +1$ the state $|P\ra^{\otimes 2}$ is changed into a tensor product of some odd state $|N\ra^{\otimes 2}$. Together with the fact that $|+_P\ra$ consists of the sum of all $|PP\ra$ states we may thus write:
\begin{equation*}
\begin{aligned}
&\bigotimes_{i = 1}^{N/2} \overline{e^{i \alpha H}  \otimes e^{-i \alpha H^*}} |+_P\ra = \frac{1}{2^{N-1}} \Big\{ a^{\frac{N}{2}} \sum_P |P\ra^{\otimes 2}  + \sum_{i = 0}^{\frac{N}{2}-1} \binom{N/2}{i} a^i b^{\frac{N}{2} - i} \\
&~~~~~~ \left[ 2^{\frac{N}{2} - i}\sum_{j = 0}^{\lfloor (N/2-i)/2 \rfloor} \binom{\lfloor (N/2-i)/2 \rfloor}{2j} \sum_P |P\ra^{\otimes 2} +  2^{\frac{N}{2} - i}\sum_{j = 0}^{\lfloor (N/2-i)/2 \rfloor} \binom{\lfloor (N/2-i)/2 \rfloor}{2j+1} \sum_N |N\ra^{\otimes 2}  \right]  \Big\} = \\
&  = \frac{1}{2^{N-1}} \Big\{ a^{\frac{N}{2}} \sum_P |P\ra^{\otimes 2}  + \sum_{i = 0}^{\frac{N}{2}-1} \binom{N/2}{i} a^i b^{\frac{N}{2} - i} \left[ 2^{\frac{N}{2} - i} 2^{\frac{N}{2} - i - 1} \sum_P |P\ra^{\otimes 2} +  2^{\frac{N}{2} - i} 2^{\frac{N}{2} - i  - 1} \sum_N |N\ra^{\otimes 2}  \right]  \Big\} = 
\end{aligned}
\end{equation*}
\begin{equation*}
\begin{aligned} 
&=   a^{\frac{N}{2}}|+_P\ra  + \sum_{i = 0}^{\frac{N}{2}-1} \binom{N/2}{i} a^i b^{\frac{N}{2} - i} \left[ 2^{\frac{N}{2} - i} 2^{\frac{N}{2} - i - 1}  |+_P\ra +  2^{\frac{N}{2} - i} 2^{\frac{N}{2} - i  - 1} |+_N\ra  \right] = \\
& = a^{\frac{N}{2}}|+_P\ra  + \sum_{i = 0}^{\frac{N}{2}-1} \frac{1}{2}\binom{N/2}{i} a^i (4b)^{\frac{N}{2} - i} (  |+_P\ra + |+_N\ra  )
=a^{\frac{N}{2}}|+_P\ra  - \frac{a^{\frac{N}{2}}}{2}(  |+_P\ra + |+_N\ra  ) +\frac{1}{2} \sum_{i = 0}^{\frac{N}{2}} \binom{N/2}{i} a^i (4b)^{\frac{N}{2} - i} (  |+_P\ra + |+_N\ra  )  =\\
& = a^{\frac{N}{2}}|+_P\ra  - \frac{a^{\frac{N}{2}}}{2}(  |+_P\ra + |+_N\ra  ) +\frac{1}{2} (a + 4b)^{\frac{N}{2}} (  |+_P\ra + |+_N\ra  ) ~ = \frac{1}{2} \left((a + 4b)^{\frac{N}{2}} + a^{\frac{N}{2}} \right)  |+_P\ra +  \frac{1}{2} \left((a + 4b)^{\frac{N}{2}} - a^{\frac{N}{2}} \right)  |+_N\ra = \\
& = \frac{1}{2} \left(1 + \left(\frac{4f(\alpha)+1}{5} \right)^{\frac{N}{2}} \right)  |+_P\ra +  \frac{1}{2} \left(1 - \left(\frac{4f(\alpha)+1}{5} \right)^{\frac{N}{2}} \right)  |+_N\ra = A |+_P\ra + B |+_N\ra, \\
\end{aligned}
\end{equation*}
where in the first step we used the fact that the sum of every second binomial is half of the sum of all binomials, which is $2$ to the appropriate power, and we defined coefficients $A,B$ by the last equality.
Thus, we may finally write:
\begin{equation}
\overline{\mathcal{U} \otimes \mathcal{U}^*} = A|+_P\ra\la +_P| + B|+_N\ra\la +_P| + B|+_P\ra\la +_N| + A|+_N\ra\la +_N| ~.
\end{equation}
Which is a $2\times2$ matrix on the subspace spanned by $|+_P\ra$ and $|+_N\ra$.
So the formula for heavy output subspace reads
\begin{equation*}
\begin{aligned}
h_U &= \sum_P \la P|^{\otimes 2} \prod_{j = 1}^T \overline{\mathcal{U}\otimes \mathcal{U}^*} ~|(0^{\otimes N})^{\otimes 2}\ra  = 2^{N} \la +_P|\prod_{j = 1}^T \left( A|+_P\ra\la +_P| + B|+_N\ra\la +_P| + B|+_P\ra\la +_N| + A|+_N\ra\la +_N| \right) ~|(0^{\otimes N})^{\otimes 2}\ra \\
& = \frac{1}{2} (A+B)^T + \frac{1}{2} (A-B)^T, 
\end{aligned}
\end{equation*}
in which we applied the formula for $2\times2$ matrix power:
\begin{equation}
\label{matrix_pow}
\left(
\begin{array}{cc}
 A & B \\
 B & A \\
\end{array}
\right)^T = \left(
\begin{array}{cc}
 \frac{1}{2} (A-B)^T+\frac{1}{2} (A+B)^T & \frac{1}{2} (A+B)^T-\frac{1}{2} (A-B)^T \\
 \frac{1}{2} (A+B)^T-\frac{1}{2} (A-B)^T & \frac{1}{2} (A-B)^T+\frac{1}{2} (A+B)^T \\
\end{array}
\right)~,
\end{equation}
which can be easily proven by induction.
After substitution of all symbols, we arrive at
the final form
\begin{equation}
\label{eq_sinal_single_parity}
h_U  = \frac{1}{2} \left(1 + \left(\frac{4f(\alpha)+1}{5} \right)^{\frac{N T}{2}} \right) = \frac{1}{2}\left(1 + e^{-2 \alpha^2 NT}\right) 
+\mathcal{O}(\alpha^4)~.
\end{equation}

\subsection{Solvable counterpart of double-parity quantum volume circuit}

Now we consider the circuit with double-parity preservation - parity on two, complementary, randomly selected subsets of qubits. Its general formula for the heavy output frequency looks exactly like in the previous case \eqref{h_U_formula}. The only difference is the diagonal gates $u_{diag}$ \eqref{u_parity_2} appearing each time the two qubits from different sets meet in a two-qubit gate. Once again to provide an analytical formula for heavy output frequency we modify the circuit by introducing large random unitaries:
\begin{equation}
\label{modified_big_u_2}
\begin{aligned}
& \mathcal{U} = \tilde{\Pi} \left(\bigotimes_{i = 1}^{N/2} \tilde{u}\right) \to(\mathcal{R}_{PP} \oplus \mathcal{R}_{NN} \oplus \mathcal{R}_{NP} \oplus \mathcal{R}_{PN}) \tilde{\Pi} (\mathcal{R}_{PP} \oplus \mathcal{R}_{NN} \oplus \mathcal{R}_{NP} \oplus \mathcal{R}_{PN}) \left(\bigotimes_{i = 1}^{N/2} \tilde{u}\right) (\mathcal{R}_{PP} \oplus \mathcal{R}_{NN} \oplus \mathcal{R}_{NP} \oplus \mathcal{R}_{PN}).
\end{aligned}
\end{equation}
To simplify this expression  we omitted subscripts $u_{diag}$ on diagonal two-qubit gates and the large unitaries act on $4$ subspaces corresponding to parity (or its lack) on both sets of qubits.
The tilde mark corresponds to the imperfect implementation of a given gate, and the error model in $\tilde{\Pi}$ assumes the omission of (some) swaps in $\Pi$ implementation.

Now one needs to calculate the average of each component in Eq. \eqref{modified_big_u_2}, multiplied by its complex conjugate according to Eq. \eqref{h_U_formula}. The average of big unitaries $\mathcal{R}$ works previously as before:
\begin{equation*}
\begin{aligned}
& \overline{\mathcal{R}_{PP} \otimes \mathcal{R}_{PP}^*} = \frac{1}{2^{N-2}}\left(\sum_{P_R P_B} |P_R P_B\ra^{\otimes 2} \right) \left(\sum_{P_R P_B} \la P_R P_B|^{\otimes 2} \right) = |+_{PP}\ra\la +_{PP}|, \\
& \overline{\mathcal{R}_{NN}\otimes \mathcal{R}_{NN}^*} = |+_{NN}\ra\la +_{NN}|, \\
& \overline{\mathcal{R}_{PN}\otimes \mathcal{R}_{PN}^*} = |+_{PN}\ra\la +_{PN}|, \\
& \overline{\mathcal{R}_{NP}\otimes \mathcal{R}_{NP}^*} = |+_{NP}\ra\la +_{NP}|. \\
\end{aligned}
\end{equation*}
Here we defined the Bell-like states $|+_{PP}\ra, \ldots$ on appropriate subspaces of odd (even) number of qubits in state $|1\rangle$ in both sets.

\subsubsection{Averaging faulty permutations}

As before let us first consider the average of permutations between the (averaged) large unitaries, thus we consider the "first" part of the expression \eqref{modified_big_u_2}. We decompose each permutation into a sequence of swaps acting on two selected qubits.
As the error model for permutations, we use imperfect swaps $S\to S^{\beta}$ with $\beta$ from Gaussian distribution, which averages out to the probabilistic scenario of swap omission with probability $p =\frac{1}{2}(1 - e^{-\frac{1}{2} \pi ^2 \sigma^2})$, where $\sigma$ is a standard deviation of the Gaussian.

Since the permutations (even faulty) cannot change the "joint" parity, for each permutation the matrix $\tilde{\Pi} \otimes \tilde{\Pi}$ is block diagonal with first block in the subspace spanned by the vectors $|PP\ra^{\otimes 2}$ and $|NN\ra^{\otimes 2}$, whereas other, separate block is supported on the subspace spanned by $|PN\ra^{\otimes 2}$ and $|NP\ra^{\otimes 2}$. Below we will consider only the first block, the calculations in the second block are analogous. Thus while considering the product of permutations with the averaged large mixing unitaries, the only relevant term is
\begin{equation}
\begin{aligned}
& (|+_{PP}\ra\la +_{PP}| + |+_{NN}\ra\la +_{NN}|) \overline{\tilde{\Pi} \otimes \tilde{\Pi}} (|+_{PP}\ra\la +_{PP}| + |+_{NN}\ra\la +_{NN}|) = \\
& = |+_{PP}\ra\la +_{PP}| \overline{\tilde{\Pi} \otimes \tilde{\Pi}} |+_{PP}\ra\la +_{PP}| +
|+_{PP}\ra\la +_{PP}| \overline{\tilde{\Pi} \otimes \tilde{\Pi}} |+_{NN}\ra\la +_{NN}| + \\
& ~~ + |+_{NN}\ra\la +_{NN}| \overline{\tilde{\Pi} \otimes \tilde{\Pi}} |+_{PP}\ra\la +_{PP}| +
|+_{NN}\ra\la +_{NN}| \overline{\tilde{\Pi} \otimes \tilde{\Pi}} |+_{NN}\ra\la +_{NN}|  = \\
& = |+_{PP}\ra\la +_{PP}| ~ \frac{1}{2^{N-2}}\overline{\Tr[P_{PP} \tilde{\Pi} P_{PP} \tilde{\Pi}^{-1}] } + |+_{PP}\ra\la +_{NN}| ~ \frac{1}{2^{N-2}}\overline{\Tr[P_{PP} \tilde{\Pi} P_{NN} \tilde{\Pi}^{-1}] } + \\
& ~~ + |+_{NN}\ra\la +_{PP}| ~ \frac{1}{2^{N-2}}\overline{\Tr[P_{NN} \tilde{\Pi} P_{PP} \tilde{\Pi}^{-1}] } + |+_{NN}\ra\la +_{NN}| ~ \frac{1}{2^{N-2}}\overline{\Tr[P_{NN} \tilde{\Pi} P_{NN} \tilde{\Pi}^{-1}] }~. \\
\end{aligned}
\end{equation}
Here $P_{PP}$ and $P_{NN}$ denote projections on subspaces with appropriate parity. In the second equality we used the fact, that for the Bell-like states $|+\ra$ and any two operators $M,N$ one has
\begin{equation*}
\la+ |M\otimes N|+\ra = \frac{1}{d} \sum_{i,j} \la i,i| M \otimes N|j,j \ra  = \frac{1}{d} \sum_{i,j} M_{i,j} N_{i,j} = \frac{1}{d} \Tr[M N^T].
\end{equation*}
The scalar terms to average, ex. $\overline{\Tr[P_{PP} \tilde{\Pi} P_{PP} \tilde{\Pi}^{-1}] }$ are just the number of basis states with appropriate parities, that after the action of the permutation have the opposite parities. Then the projections do not affect them, and the inverse permutation undo the action of the first one giving the contribution to the trace. All other scenarios give zero contribution. 

To compute this expression assume first, that the faulty implementation $\tilde{\Pi}$ replaced $k$ qubits from the "red" subset with the qubits from the "blue" subset, compared to the action of ideal permutation $\Pi$. If $k = 0$ the implementation is ideal, or all errors happen inside a subspace, thus all basis states from the given subspace contribute so the average factor is equal is $2^{N-2}$. 
If $N/2 >k > 0$ the averaged term is equal $2^{N-3}$, since all qubits may be in any basis state, except one qubit in each set which fixes the parity, and one of $2k$ exchanged qubits which fixes the exchange of parity. Finally, if all qubits are exchanged, the averaged term is equal to $2^{N-2}$ once again. To summarize:

\begin{equation*}
\begin{aligned}
&\frac{1}{2^{N-2}}\Tr[P_{PP} \tilde{\Pi} P_{PP} \tilde{\Pi}^{-1}]  = 
\left\{
\begin{matrix}
1& \text{ if } k = 0 \text{ or } k = N/2 \\
\frac{1}{2}& \text{ otherwise }
\end{matrix}
\right. ~,~~~\frac{1}{2^{N-2}}\Tr[P_{PP} \tilde{\Pi} P_{NN} \tilde{\Pi}^{-1}]  = 
\left\{
\begin{matrix}
0& \text{ if } k = 0 \text{ or } k = N/2 \\
\frac{1}{2}& \text{ otherwise }
\end{matrix}
\right. 
\end{aligned}
\end{equation*}
and analogously with other pairs of terms.

To determine the distribution of the number of exchanged qubits $P(k)$, we assume that omitted swaps are so sparse $k \ll N$, that the chance for one omission to interfere with the others is negligible. Thus each omitted swap results in one permutation error. Moreover, because we are interested only in errors mixing two subsets of qubits, and those two subsets were chosen randomly at the beginning, we assume that each omission has $1/2$ chance to cause the important error, lowering the error probability from $p$ to $p/2$.
Thus, if the error in the swaps are independent, we may approximate:
\begin{equation}
\begin{aligned}
& \frac{1}{2^{N-2}} \left\la\Tr[P_{PP} \tilde{\Pi} P_{PP} \tilde{\Pi}^{-1}]\right\ra_{p} \approx  \left(1-\frac{p}{2}\right)^{w} + \frac{1}{2}\sum_{k = 1}^{w} \binom{w}{k} \left(1 - \frac{p}{2}\right)^{w - k}\left(\frac{p}{2}\right)^{k} =  \\
& = \left(1-\frac{p}{2}\right)^{w} - \frac{1}{2} \left(1-\frac{p}{2}\right)^{w}  + \frac{1}{2} \left(\left(1 - \frac{p}{2}\right) + \frac{p}{2}\right)^{w} = \frac{1}{2} \left(1 + \left(1-\frac{p}{2}\right)^{w} \right) \\
&\frac{1}{2^{N-2}} \left\la\Tr[P_{PP} \tilde{\Pi} P_{NN} \tilde{\Pi}^{-1}]\right\ra  \approx \frac{1}{2}\sum_{l = k}^{w} \binom{w}{k} \left(1 - \frac{p}{2}\right)^{w - k}\left(\frac{p}{2}\right)^{k} =  \\
& = - \frac{1}{2} \left(1-\frac{p}{2}\right)^{w}  + \frac{1}{2} \left(\left(1 - \frac{p}{2}\right) + \frac{p}{2}\right)^{w} = \frac{1}{2} \left(1 - \left(1-\frac{p}{2}\right)^{w} \right) ,
\end{aligned}
\end{equation}
where $w$ is a number of swaps in permutation $\Pi$, which we also have to average over all permutations. 
To do so, we use another Ansatz, that the averaged expression is, for small $p$, approximately of the form $\left\la\left(1-\frac{p}{2}\right)^{w} \right\ra_{\pi} \approx e^{- \alpha(N) p}$. Thus we may calculate
\begin{equation*}
\alpha(N) = -\frac{\partial \log\left( \left\la \left(1-\frac{p}{2}\right)^{w} \right\ra_{\pi}\right)}{\partial p}\Big|_{p = 0} =
\frac{\left\la \frac{w}{2} \left(1-\frac{p}{2}\right)^{w-1} \right\ra_{\pi} }{\left\la\left(1-\frac{p}{2}\right)^{w} \right\ra_{\pi}} \Big|_{p = 0}  = \frac{w(N)}{2},
\end{equation*}
where $w(N)$ represents
the average number of swap gates in the implementation of $N$ qubits on a given architecture. For example in $1$D case $w(N)_{1D}  \leq \frac{N(N-1)}{4}$ \cite{Nadir2024fidelity}.

Thus, summarizing this part of the derivation, the averaged permutations sandwiched by large unitaries give:

\begin{equation}
\begin{aligned}
&\overline{(\mathcal{R}_{PP} \oplus \mathcal{R}_{NN} \oplus \mathcal{R}_{NP} \oplus \mathcal{R}_{PN})^{\otimes2}} ~\overline{\tilde{\Pi} \otimes \tilde{\Pi}}~ \overline{(\mathcal{R}_{PP} \oplus \mathcal{R}_{NN} \oplus \mathcal{R}_{NP} \oplus \mathcal{R}_{PN})^{\otimes 2}} = 
\begin{pmatrix}
x & y & 0 & 0\\
y & x & 0 & 0\\
0 & 0 & x & y\\
0 & 0 & y & x\\
\end{pmatrix}_{+}
\end{aligned}
\end{equation}
with matrix presented in the basis$\{|+_{PP}\ra, |+_{NN}\ra, |+_{PN}\ra, |+_{NP}\ra\}$ and coefficients $x, y$ are given by
\begin{equation}
x = \frac{1}{2}\left(1 + e^{-\frac{1}{2}p w(N)}\right) ~,~~~ y = \frac{1}{2}\left(1 - e^{-\frac{1}{2}p w(N)}\right)~.
\end{equation}

Note that $x+y = 1$, and $x - y = e^{-\frac{1}{2}p w(N)}$.

\subsubsection{Averaging faulty two-qubit gates}

Next, we move to the second part of the layer - faulty implemented two-qubit gates. Note, that the error of "mismatched" qubits was already handled while discussing faulty permutations sandwiched by large random unitaries, which are applied perfectly.
Thus each two-qubit gate is placed on an appropriate pair of qubits i.e.  $u$ always acts on qubits from the same set, whereas $u_{diag}$  always acts on qubits from the different sets.
As in the circuit with single parity \eqref{u_parity_1_av}, we have 
\begin{equation*}
\begin{aligned}
\overline{u \otimes u^*}  =\frac{1}{2}\Big{[}& (|00,00\ra + |11,11\ra )(\la 00,00| + \la 11,11| )  
+ (|01,01\ra + |10,10\ra )(\la 01,01| + \la 10,10| )
\Big{]},
\end{aligned}
\end{equation*}
where we separated the space of $u$ with the space of $u^*$ by commas. Thus we can once again infer, that the average action of $\overline{u \otimes u^*}$ does not change any of the states $|+_{PP}\ra, |+_{NN}\ra, |+_{PN}\ra, |+_{NP}\ra$, since it always acts on the qubits from the same set. Moreover, the average
\begin{equation}
\label{u_parity_2_av}
\begin{aligned}
\overline{u_{diag} \otimes u_{diag}^*}  =& (|00,00\ra\la 00,00| + |11,11\ra\la 11,11|)  ~+~ (|00,11\ra\la 00,11| + |11,00\ra\la 11,00|) + \\
&  (|01,01\ra\la 01,01| + |10,10\ra\la 10,10|)   ~+~ (|01,10\ra\la 01,10| + |10,01\ra\la 10,01|)
\end{aligned}
\end{equation}
is diagonal, so the Bell-like states
are invariant with respect to 
the action of this operator.

Discussion of errors $e^{i \alpha H} \otimes e^{-i \alpha H^*}$ in this scenario is 
more complicated. Once again we focus on only one Bell-like state $|+_{PP}\ra$, since the calculations for all of them are alike. Let us name by $K$ a number of cases in which qubits from different subsets meet. Then the action of two-qubit gate errors on the state $|+_{PP}\ra$ is given by
\begin{equation}
\label{abcd_def}
\begin{aligned}
\overline{e^{i \alpha H} \otimes e^{-i \alpha H^*}}^{\otimes N/2} |+_{PP}\ra 
& = (a~ \id^{\otimes 2} + b~ |+\ra\la+|)^{\otimes \left(\left(\frac{N}{2}\right) - K\right)/2} 
(a~ \id^{\otimes 2} + b~ |+\ra\la+|)^{\otimes \left(\left(\frac{N}{2}\right) - K\right)/2} (a~ \id^{\otimes 2} + b~ |+\ra\la+|)^{\otimes K}|+_{PP}\ra = \\
& = c |+_{PP}\ra + d |+_{NN}\ra + e (|+_{PN}\ra + |+_{NP}\ra).  \\
\end{aligned}
\end{equation}
With a slight abuse of the notation
we grouped the gates acting within, or between, subsets, the coefficients $a,b$ are defined as in the previous subsection \eqref{gauss_noise_average}, and the
coefficients $c,d,e$ are to be determined. Let us start with the coefficient $c$. We once again decomposed Bell-like states $\la+|$ into vectors with the same and opposite parity as $|+PP\ra$ on appropriate pair of qubits, naming the latter one error-terms. Similarly as in the circuit with single parity one obtains 
\begin{equation*}
\resizebox{\hsize}{!}{$
\begin{aligned}
c &= \Big(a^{\frac{N}{2}}\Big) + \\
& + 2\Bigg\{a^{\left(\left(\frac{N}{2}\right) - K\right)/2}  a^K \sum_{i = 0}^{\left(\left(\frac{N}{2}\right) - K\right)/2-1} \binom{\left(\left(\frac{N}{2}\right) - K\right)/2}{i} a^{i} b^{\left(\left(\frac{N}{2}\right) - K\right)/2 - i} 2^{\left(\left(\frac{N}{2}\right) - K\right)/2 - i} \sum_{m = 0}^{\left\lfloor\left(\left(\left(\frac{N}{2}\right) - K\right)/2 - i\right)/2\right\rfloor} \binom{\left(\left(\frac{N}{2}\right) - K\right)/2 - i}{2m} \Bigg\} + \\
& +\Bigg\{ a^{\frac{N}{2} - K} \sum_{j = 0}^{K-1} \binom{K}{j} a^{j} b^{K-j} \left( \sum_{n = 0}^{\left\lfloor\frac{K-j}{2} \right\rfloor } \binom{\left\lfloor\frac{K-j}{2} \right\rfloor}{2n}\right)^2 \Bigg\} + \\
& + \Bigg\{ a^K \sum_{i = 0}^{\left(\left(\frac{N}{2}\right) - K\right)/2-1} \binom{\left(\left(\frac{N}{2}\right) - K\right)/2}{i} \sum_{j = 0}^{\left(\left(\frac{N}{2}\right) - K\right)/2-1} \binom{\left(\left(\frac{N}{2}\right) - K\right)/2}{i} a^{i + j} b^{\frac{N}{2} - K - i - j} 2^{^{\frac{N}{2} - K - i - j}} \\
& \hspace{1cm} \Bigg[\left(\sum_{m = 0}^{\left\lfloor\left(\left(\left(\frac{N}{2}\right) - K\right)/2 - i\right)/2\right\rfloor} \binom{\left(\left(\frac{N}{2}\right) - K\right)/2 - i}{2m} \right)\left(\sum_{m = 0}^{\left\lfloor\left(\left(\left(\frac{N}{2}\right) - K\right)/2 - j\right)/2\right\rfloor} \binom{\left(\left(\frac{N}{2}\right) - K\right)/2 - j}{2m} \right)\Bigg]\Bigg\} + \\
& + 2\Bigg\{  a^{\left(\left(\frac{N}{2}\right) - K\right)/2} \sum_{i = 0}^{\left(\left(\frac{N}{2}\right) - K\right)/2-1} \binom{\left(\left(\frac{N}{2}\right) - K\right)/2}{i} a^{i} b^{\left(\left(\frac{N}{2}\right) - K\right)/2 - i} 2^{\left(\left(\frac{N}{2}\right) - K\right)/2 - i}
\sum_{j = 0}^{K-1} \binom{K}{j} a^{j} b^{K-j} \\
& \hspace{1 cm}\left[\left( 
\sum_{m = 0}^{\left\lfloor\left(\left(\left(\frac{N}{2}\right) - K\right)/2 - i\right)/2\right\rfloor} \binom{\left(\left(\frac{N}{2}\right) - K\right)/2 - i}{2m} \left(\sum_{n = 0}^{\left\lfloor\frac{K-j}{2} \right\rfloor } \binom{K-j}{2n}\right)^2
\right) +
\left( 
\sum_{m = 0}^{\left\lfloor\left(\left(\left(\frac{N}{2}\right) - K\right)/2 - i\right)/2\right\rfloor} \binom{\left(\left(\frac{N}{2}\right) - K\right)/2 - i}{2m+1} \sum_{n = 0}^{\left\lfloor\frac{K-j}{2} \right\rfloor } \binom{K-j}{2n+1}\sum_{n = 0}^{\left\lfloor\frac{K-j}{2} \right\rfloor } \binom{\left\lfloor\frac{K-j}{2} \right\rfloor}{2n}
\right)\right]\Bigg\} + \\
& + \Bigg\{  \sum_{i = 0}^{\left(\left(\frac{N}{2}\right) - K\right)/2-1} \binom{\left(\left(\frac{N}{2}\right) - K\right)/2}{i} a^{i} b^{\left(\left(\frac{N}{2}\right) - K\right)/2 - i} 2^{\left(\left(\frac{N}{2}\right) - K\right)/2 - i}
\sum_{j = 0}^{\left(\left(\frac{N}{2}\right) - K\right)/2-1} \binom{\left(\left(\frac{N}{2}\right) - K\right)/2}{j} a^{j} b^{\left(\left(\frac{N}{2}\right) - K\right)/2 - j} 2^{\left(\left(\frac{N}{2}\right) - K\right)/2 - j}
\sum_{k = 0}^{K-1} \binom{K}{k} a^{k} b^{K-k} \\
& \hspace{1cm} \Bigg[\left(\left(\sum_{m = 0}^{\left\lfloor\left(\left(\left(\frac{N}{2}\right) - K\right)/2 - i\right)/2\right\rfloor} \binom{\left(\left(\frac{N}{2}\right) - K\right)/2 - i}{2m}\right)
\left(\sum_{m = 0}^{\left\lfloor\left(\left(\left(\frac{N}{2}\right) - K\right)/2 - j\right)/2\right\rfloor} \binom{\left(\left(\frac{N}{2}\right) - K\right)/2 - j}{2m}\right)  
\left(\sum_{n = 0}^{\left\lfloor\frac{K-k}{2} \right\rfloor } \binom{K-j}{2n}\right)^2   \right) + \\
& \hspace{2cm}+ 2*\left(\sum_{m = 0}^{\left\lfloor\left(\left(\left(\frac{N}{2}\right) - K\right)/2 - i\right)/2\right\rfloor} \binom{\left(\left(\frac{N}{2}\right) - K\right)/2 - i}{2m}
\sum_{m = 0}^{\left\lfloor\left(\left(\left(\frac{N}{2}\right) - K\right)/2 - i\right)/2\right\rfloor} \binom{\left(\left(\frac{N}{2}\right) - K\right)/2 - i}{2m+1}
\sum_{n = 0}^{\left\lfloor\frac{K-k}{2} \right\rfloor } \binom{K-k}{2n+1}\sum_{n = 0}^{\left\lfloor\frac{K-k}{2} \right\rfloor } \binom{K-k}{2n}\right) + \\
& \hspace{2cm} + \left(\left(\sum_{m = 0}^{\left\lfloor\left(\left(\left(\frac{N}{2}\right) - K\right)/2 - i\right)/2\right\rfloor} \binom{\left(\left(\frac{N}{2}\right) - K\right)/2 - i}{2m+1}\right)
\left(\sum_{m = 0}^{\left\lfloor\left(\left(\left(\frac{N}{2}\right) - K\right)/2 - j\right)/2\right\rfloor} \binom{\left(\left(\frac{N}{2}\right) - K\right)/2 - j}{2m+1}\right)  
\left(\sum_{n = 0}^{\left\lfloor\frac{K-k}{2} \right\rfloor } \binom{K-k}{2n+1}\right)^2   \right)\Bigg]\Bigg\}~,
\end{aligned}$}
\end{equation*}
where in the consecutive curvy brackets in each line we considered the cases of: no error terms, error terms only inside one subset, error terms only between subsets, error terms inside both subsets, error terms inside one subset, and between two subsets and error terms everywhere.
The general structure of each component is as follows. First, one sums over all \rd{possible combinations of the term proportional to $a$ or $b$ coefficients, introduced in equation} \eqref{gauss_noise_average}, appropriate number of times in the two-qubit gates inside each of subsets and/or in two-qubit gates mixing subspaces  -- sums over $i,j,k$. Then comes the inner sums corresponding to the expansion of Bell-like states into all important cases - sums over $m$ and $n$. 

While considering two-qubit gates inside each subset, there are only two cases - either action on two-qubit switches the parity in the set, which gives two options, or it does not, which also gives two options. Since we are interested in only the odd (or even) number of switches, thus we sum over only every second binomial. The situation is analogous to a single parity circuit, see \eqref{2q_gtate_noise_action} and further discussion. The powers of $2$ corresponding to $2$ basis vectors for each case on each pair of qubits are taken in front of the sums and next to $b$ coefficient for clarity.

On the other hand, while considering two-qubit gates acting on qubits from different subsets we have four possibilities for each gate - change or preservation of parity in red or blue qubits. To tackle this complexity we divide the sum into two. The first corresponds to the joint action on the first subset and the second corresponds to the joint action on the second subset. We have freedom of such separation because each Bell-like state between a pair of qubits and their copy is equal to the product of the Bell states on the first qubit with its copy and the second qubit with its copy. 
After such decomposition, we have two independent sums, in which we once again take into account only every second binomial, depending on the scenario we consider in each term.
After exhausting but straightforward calculations, using the formula for binomial summations and the property that $a + 4b = 1$, one may obtain following form of the coefficients from eq. \eqref{abcd_def}
\begin{equation}
\label{c_process_form2}
\begin{aligned}
& c = \frac{1}{4} a^{\frac{N}{2}} + \frac{1}{2} a^{\left(\frac{N}{2}+K\right)/2} + \frac{1}{4} ~,~~~ d = \frac{1}{4} a^{\frac{N}{2}} -\frac{1}{2} a^{\left(\frac{N}{2}+K\right)/2} + \frac{1}{4} ~,~~~ e = -\frac{1}{4} a^{\frac{N}{2}} + \frac{1}{4} .
\end{aligned}
\end{equation}

The last step of this part of derivation is the average of the above coefficients over all possible numbers of two-qubit gates mixing subsets $K$. Since the qubits were matched randomly to two subsets and along the way we averaged over all possible permutations of qubits, one may assume, that the assignment of qubits in two-qubit gates was completely random with equal probability for all the cases.
Hence the expression for the probability of obtaining exactly $K$ pairs of different colors out of $N$ qubits has a form
\begin{equation*}
P(K) = \frac{1}{N!} \binom{N/2}{K} \binom{N/2-K}{\left(N/2-K\right)/2}\; 2^K \left(\frac{N}{2}\right)!\left(\frac{N}{2}\right)! = \binom{N}{N/2}^{-1} \binom{N/2}{K} \binom{N/2-K}{\left(N/2-K\right)/2}\; 2^K~,
\end{equation*}
where the first term is normalization - all possible distributions of $N$ qubits, the second one corresponds to all possible choices of $K$ pairs, the third to the distribution of colors (assignment to subsets) in all other pairs, the fourth to all possible layouts "red/blue" within pairs and the two last terms to all possible distributions of qubits within each subset. It is important to note, that $0\leq K \leq N/2$ and $N/2 - K$ must be even.
The only part of $c,d$ and $e$ coefficients which depends on $K$ can be averaged as
\begin{equation}
\begin{aligned}
g(a,N) &:= \la a^{\left(\frac{N}{2}+K\right)/2} \ra_K   = \sum_{K} a^{\left(\frac{N}{2}+K\right)/2}\binom{N}{N/2}^{-1} \binom{N/2}{K} \binom{N/2-K}{\left(N/2-K\right)/2}\; 2^K =  \\
& = \left\{\begin{matrix}
& a^{\frac{N}{4}} \, _2F_1\left(-\frac{N}{4},-\frac{N}{4};\frac{1}{2};a\right) \frac{\left(\left(\frac{N}{2}\right)!\right)^3}{\left(\left(\frac{N}{2}\right)!\right)^2 N!} ~~\text{ if } N\text{ mod }4 = 0 \\
& a^{\frac{N+2}{4}} \,
_2F_1\left(-\frac{N-2}{4},-\frac{N-2}{4};\frac{3}{2};a\right)  \frac{\left(\frac{N}{2}+1\right)! \left(\frac{N}{2}!\right)^2 }{\left(\frac{N-2}{4}\right)!\left(\frac{N+2}{4}\right)!  N!} ~~\text{ if } N\text{ mod }4 = 2 \\
\end{matrix}\right. 
\end{aligned}
\end{equation}

To obtain more convenient form for this expression we may expand
 $g(a,N)$ into a power series around $a = 1$ ($\alpha^2  = 0$). Thus in this case of small errors, one obtains
\begin{equation*}
g(a,N) \approx 1 - (1-a)N\frac{(3 N-2)}{8 (N-1)} + \mathcal{O}\left((1-a)^2\right) ,
\end{equation*}
Combining in with expansion of $a = \frac{4f(\alpha)+1}{4+1}$ in the power series in $\alpha^2$ gives
\begin{equation*}
g(a,N) \approx 1- \alpha^2 N \frac{(3 N-2)}{2 (N-1)} + \mathcal{O}(\alpha^4)~.
\end{equation*}
Thus for appropriately small errors one may assume
\begin{equation}
\label{g_simp}
g(a,N) \approx e^{- \frac{3}{2} \alpha^2 N \frac{(N-2/3)}{(N-1)}}~.
\end{equation}

Summarizing this part of the derivation, the averaged permutations sandwiched by large unitaries give
\begin{equation}
\begin{aligned}
&\overline{(\mathcal{R}_{PP} \oplus \mathcal{R}_{NN} \oplus \mathcal{R}_{NP} \oplus \mathcal{R}_{PN})^{\otimes(2)}} ~\left(\bigotimes_{i = 1}^{N/2} \tilde{u}\right)~ \overline{(\mathcal{R}_{PP} \oplus \mathcal{R}_{NN} \oplus \mathcal{R}_{NP} \oplus \mathcal{R}_{PN})^{\otimes 2}} =
\begin{pmatrix}
c & d & e & e\\
d & c & e & e\\
e & e & c & d\\
e & e & d & c\\
\end{pmatrix}_{+}~.
\end{aligned}
\end{equation}
Here once again for convenience we omitted the subscript $_{diag}$ and the average over $K$ of the parameters $c,d,e$ to keep the formula clean. The matrix is presented in the basis $\{|+_{PP}\ra, |+_{NN}\ra, |+_{PN}\ra, |+_{NP}\ra\}$.
Note useful relations between $c,d$ and $e$ coefficients $c+d+2e = 1$, $c+d - 2e = a^{\frac{N}{2}}$ and $c - d = g(a)$.

\subsubsection{Final form}
By combining the results form the previous subsections we can finally write 
\begin{equation}
\overline{\mathcal{U} \otimes \mathcal{U}^*} = 
\begin{pmatrix}
x & y & 0 & 0\\
y & x & 0 & 0\\
0 & 0 & x & y\\
0 & 0 & y & x\\
\end{pmatrix}_{+}\begin{pmatrix}
c & d & e & e\\
d & c & e & e\\
e & e & c & d\\
e & e & d & c\\
\end{pmatrix}_{+} = 
\left(
\begin{array}{cccc}
 c x+d y & c y+d x & e & e \\
 c y+d x & c x+d y & e & e \\
 e & e & c x+d y & c y+d x \\
 e & e & c y+d x & c x+d y \\
\end{array}
\right)_{+}~,
\end{equation}
where we used the fact that $x+y = 1$.
So the formula for heavy output subspace reads
\begin{equation*}
\begin{aligned}
h_U &= \sum_{P_R P_B} \la P_R P_B|^{\otimes 2} \prod_{j = 1}^T \overline{\mathcal{U}\otimes \mathcal{U}^*} ~|(0^{\otimes N})^{\otimes 2}\ra = \la +_{PP}| \left(
\begin{array}{cccc}
 c x+d y & c y+d x & e & e \\
 c y+d x & c x+d y & e & e \\
 e & e & c x+d y & c y+d x \\
 e & e & c y+d x & c x+d y \\
\end{array}
\right)_{+} ^{T} |(0^{\otimes N})^{\otimes 2}\ra 2^{N-2} = \\
& = \frac{1}{4} (2 ((c - d) (x - y))^T + (-2 e + (c + d) (x + y))^T + (2 e + (c + d) (x + y))^T)~,
\end{aligned}
\end{equation*}
where $T$ denotes the number of layers. The above expression was obtained using twice formula \eqref{matrix_pow} of raising symmetric $2\times2$ block matrix to power $T$.
By substituting (almost) all symbols we obtain
\begin{equation}
\begin{aligned}
h_U &= \frac{1}{4}\left[2~ \left(g\left(\frac{4f(\alpha)+1}{4+1}, N\right)\right)^T e^{-\frac{1}{2}p w(N)T } + \left(\frac{4f(\alpha)+1}{4+1}\right)^{\frac{1}{2} N T}  + 1  \right]~. \\ 
\end{aligned}
\end{equation}
Which can be further simplified if we approximate the functions $f(\alpha)$ and $g(\cdots)$ by exponents \eqref{g_simp}:
\begin{equation}
\label{eq_final_fouble_parity}
h_U \approx \frac{1}{4}\left[2 e^{-\frac{3}{2} \alpha^2 N T \frac{(N-2/3)}{(N-1)}}  e^{-\frac{1}{2}p w(N)T } + e^{-2 \alpha^2 N T} + 1  \right]
+\mathcal{O}(\alpha^4)~,
\end{equation}

One can see that the decrease of the first term corresponds to the "leakage" from the "double parity" subspace into the global parity subspace, whereas the decrease of the second term to the "leakage" into the rest of Hilbert space.

Note also that if there are no errors in two-qubit gates, the formula simplifies to
\begin{equation}
\begin{aligned}
\label{eq_double_parity_simplyfied}
h_U &= \frac{1}{2}\left[1 + e^{-\frac{1}{2}p w(N)T } \right]~.
\end{aligned}
\end{equation}

\subsection{Dissipative noise model}

One may generalize the above calculations to encompass other sorts of errors in numerous ways.
The most insightful modification from our point of view is the uncontrolled interaction with the environment in the faulty realization of two-qubit gates.
The adjustments in calculations are quite minor. The new formula for heavy output frequency reads
\begin{equation}
\label{h_U_formula_rho}
h_U = \sum_P \la P|\prod_{j = 1}^T \overline{\mathcal{K}} ~|0^{\otimes N}\ra\la 0^{\otimes N} |\left( \prod_{j = 1}^T \overline{\mathcal{K}}\right)^{\dagger}|P\ra  = 
\sum_P \la PP| \prod_{j = 1}^T \overline{\mathcal{K}\otimes \mathcal{K}^*} ~|(0^{\otimes N})^{\otimes 2}\ra~,
\end{equation}
where $\mathcal{K}$ correspond to the action of a single layer, with a sum over Kraus operators.

The non-unitary noise is modeled in the following way. First we assume each two-qubit gate has its own $d_E$-dimensional environment, then we consider interaction with the environment by a random Hamiltonians from the GUE on $4\times d_E$ space. Finally we averaged over all (unknown) input states of the environment, and partial trace over the environment after the interactions took place.

According to \cite{Nadir2024fidelity} the average over interactions defined by $4 d_E$ random Hamiltonians from Gaussian unitary ensemble with noise strength $\alpha$ are given by
\begin{equation}
\label{gauss_noise_average_env}
\overline{e^{i \alpha H}  \otimes e^{-i \alpha H^*}} = 
\frac{4 d_E f_{4d_E}(\alpha)+1}{4 d_E+1} \id_{S,E}^{\otimes 2} + \frac{1-f_{4d_E}(\alpha)}{4 d_E+1} |+\ra\la+|_{S,E}~,
\end{equation}
with subscript $S$ corresponding to two-qubit gate subsystem, and subscript $E$ corresponding to the environment and  $ f_{4d_E}(\alpha) \approx e^{-(4d_E+1)\alpha^2 }$.
The average over input states give a maximally mixed state, which vectorization reads $\frac{1}{d_E}|+\ra_E$, whereas partial trace in vectorized notation corresponds to $\la +|_E$, thus averaging the environment $E$ we obtain an expression for the partial trace,
\begin{equation}
\text{Tr}_E[\overline{e^{i \alpha H}  \otimes e^{-i \alpha H^*}}] = 
\frac{4 d_E f_{4d_E}(\alpha)+1}{4 d_E+1} \id_{S}^{\otimes 2} +d_E \frac{1-f_{4d_E}(\alpha)}{4 d_E+1} |+\ra\la+|_S = a_E~ \id^{\otimes 2} + b_E~ |+\ra\la+|_{S}~.
\end{equation}

The modified coefficient $a_E$, $b_E$ still satisfies the property $a_E + 4 b_E = 1$, thus all the calculations presented in the previous sections are analogous. The formula for heavy output frequency in one parity circuit reads
\begin{equation}
h_U  = \frac{1}{2} \left(1 + \left(\frac{4 d_E f_{4d_E}(\alpha)+1}{4 d_E +1} \right)^{\frac{N T}{2}} \right) \approx \frac{1}{2}\left(1 + e^{-2 d_E \alpha^2 NT}\right) ~,
\end{equation}
whereas in the circuit with double parity one obtains
\begin{equation}
\begin{aligned}
h_U &= \frac{1}{4}\left[2~ \left(g\left(\frac{4 d_E f_{4 d_E}(\alpha)+1}{4 d_E+1}, N\right)\right)^T e^{-\frac{1}{2}p w(N)T } + \left(\frac{4 d_Ef_{4 d_E}(\alpha)+1}{4 d_E+1}\right)^{\frac{1}{2} N T}  + 1  \right] \\
& \approx \frac{1}{4}\left[2 e^{-\frac{3}{2}d_E \alpha^2 N T \frac{(N-2/3)}{(N-1)}}  e^{-\frac{1}{2}p w(N)T } + e^{-2 d_E \alpha^2 N T} + 1  \right]~.
\end{aligned}
\end{equation}
Thus, on the level of approximated formulas, we may define effective noise strength as the noise strength resealed by the square root of the environment dimension $\alpha_{eff} = \sqrt{d_E}\alpha$, and re-obtain previous formulas.  

\subsection{Comparison of analytical and numerical results}

We finish this \bl{Appendix} by illustrating how well the formulas for heavy output frequency derived using modified circuits fit the simulations of actual proposed circuits.
To make the presentation more apparent we investigate simplified, exponential formulas \eqref{eq_sinal_single_parity} for single parity circuit, and \eqref{eq_final_fouble_parity} for double parity circuit. However since the approximations used to obtain these formulas relied on small errors, we mostly limit ourselves to such scenarios in the following discussion.

We modeled errors identically as above. Namely, we spoiled each two-qubit gate by associating it with random noise $e^{i \alpha H}$, where $\alpha$ is a noise strength and $H$ is a random Hamiltonian form Gaussian unitary ensemble. The noise for permutations is modeled by first decomposing them (in the optimal possible way) into swaps, and then assuming that each swap $S$ was executed inaccurately $S\to S^\beta$, where $\beta$ is a random Gaussian variable with mean $1$ and variance $\sigma^2$. As mentioned before, this model is equivalent to probabilistic omission of swaps with probability $p = \frac{1}{2}(1 - e^{- \frac{1}{2} \pi^2 \sigma^2})$.

Last but not least the essential assumption behind modified circuits was that after a few random layers, quantum states are "mixed enough" for large random unitary to not affect their properties. In order to make this assumption plausible, i.e. to provide enough layers for mixing quantum states, we consider a square circuit where the number of layers is equal to the number of qubits $T = N$. This scenario is especially interesting since it is also used to determine quantum volume.

For both parity and double parity quantum volume circuits we've performed numerous trials summarized in Table \ref{tab1}. Note that for a larger number of qubits, we restricted the range of $\sigma$. We did so, to ensure that permutation errors do not interfere with each other in which case the derived formulas serve only as an upper bound.

\begin{table}[h]
\begin{tabular}{l|l|l|l}
Number of qubits $N$ & sample size & $\sigma$ values & $\alpha$ values \\ \hline
4                    & 2000        &$10$ values from $[0,~0.05]$      &$10$ values from $[0,~0.05]$     \\
6                    & 2000        &$10$ values from $[0,~0.05]$      &$10$ values from $[0,~0.05]$     \\
8                    & 500         &$10$ values from $[0,~0.04]$      &$10$ values from $[0,~0.05]$     \\
10                   & 50          &$10$ values from $[0,~0.02]$      &$10$ values from $[0,~0.05]$    
\end{tabular}
\caption{Summary of performed numerical experiments. The same simulations were performed for the parity circuit and double-parity circuit. Each experiment consisted of 'sample size' runs, with $\alpha$ and $\sigma$ independently taken from 10 equally distributed values from appropriate intervals.}
\label{tab1}
\end{table}

\subsubsection{Parity circuit}
\begin{figure}[h]
    \centering
    \includegraphics[width = 10 cm ]{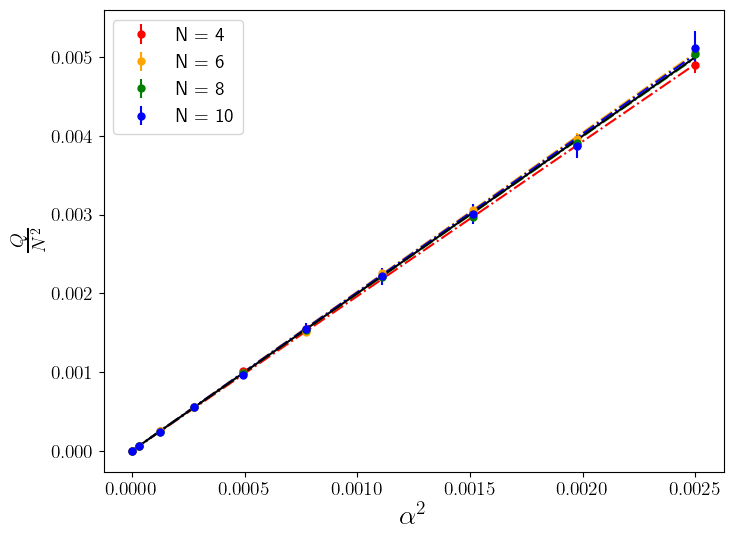}
    \caption{Exponent $Q$, normalized by the number of qubits and layers, as a function of squared noise parameter $\alpha^2$, together with linear fits. Predicted behavior \eqref{eq_sinal_single_parity} corresponds to the black line.}
    \label{fig:single_p_integrable_num1}
\end{figure}

Let us start with a single-parity quantum volume circuit. According to the formula for heavy output \eqref{eq_sinal_single_parity} we express heavy output frequency as
\begin{equation}
\label{sigle_p_fit}
h_U = \frac{1}{2} e^{-Q(N,T,\alpha)} + \frac{1}{2},
\end{equation}
where $Q$ is an unknown exponent dependent on number of qubits $N$, layers $T$ and noise strength $\alpha$. From the formula \eqref{eq_sinal_single_parity} we infer, that the expected value of $Q$ should be given by $Q =2 N T \alpha^2$. Therefore, for each test, we calculate the average value on $Q$, normalize it by the number of qubits and layers, and it plot as a function of noise strength squared $\alpha^2$.
The results are presented in the Figure \ref{fig:single_p_integrable_num1}. As one can see even for a small number of qubits, where the approximations were most crude, the obtained values are in line with theoretical predictions up to one standard deviation, with only exception of $N = 4$ which is extremely small given performed estimations. Additionally the intercept $b$ is negligible in all cases.
The linear fit of $Q/(N T)$ as a function of $\alpha^2$ yields:

\begin{equation*}
\begin{aligned}
&~\hspace{3 cm} \frac{Q}{NT} = a\; \alpha^2 + b ~~\text{ with}: \\
& a = 1.964 \pm 0.089 ~,~~~\; b = 1.8 \times 10^{-5} \pm 1.1 \times 10^{-5} ~~~\;\text{ for } N = 4, \\
& a = 2.021 \pm 0.022 ~,~~~\; b = -5.9 \times 10^{-6} \pm 9.7 \times 10^{-6} ~~\text{ for } N = 6, \\
& a = 1.996 \pm 0.010 ~,~~~~ b = -0.5 \times 10^{-5} \pm 1.2 \times 10^{-5} ~\;\text{ for } N = 8, \\
& a = 2.013 \pm 0.018 ~,~~~~ b = -1.3 \times 10^{-5} \pm 2.2 \times 10^{-5} ~~\text{ for } N = 10. \\
\end{aligned}
\end{equation*}

\begin{figure}[h]
    \centering
    \includegraphics[width = 10 cm ]{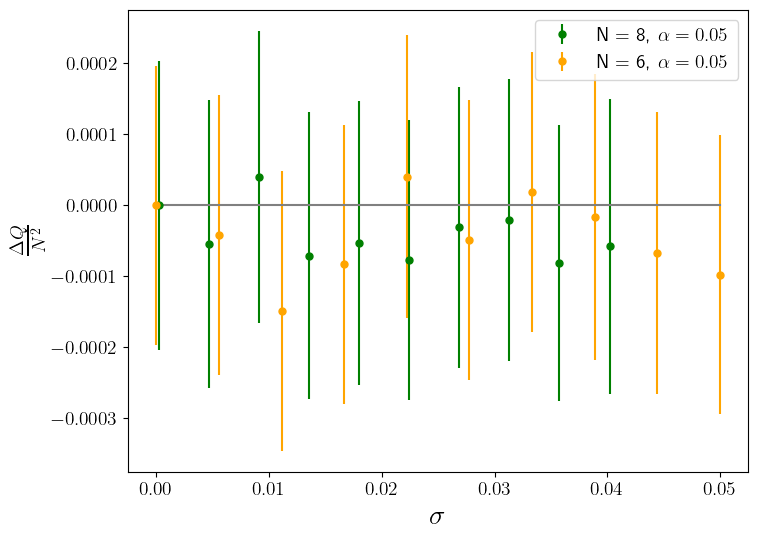}
    \caption{The difference between exponents $Q$, normalized by the number of two-qubit gates, between no swap errors scenario and swap errors strength $\sigma$. Two exemplary experiments with $N = 6$ and $N = 8$ are presented.}
    \label{fig:single_p_integrable_num2}
\end{figure}

Next we examined whether swap errors affected heavy output frequency for the single parity circuit. On the plot \ref{fig:single_p_integrable_num2} we present the difference between normalized exponents $Q$ for no swap noise and swap noise strength $\sigma$ for two exemplar experiments.
In this and every other experiment, swap errors did not affect the exponent $Q$, up to one standard deviation.

\subsubsection{Double parity circuit}

We start the discussion of the double parity circuit by considering the scenario with only errors within permutations. Following the equation \eqref{eq_double_parity_simplyfied} in such a case we express heavy output frequency as
\begin{equation}
\label{double_p_p_fit}
    h_U = \frac{1}{2} e^{-W(T,N,p)} + \frac{1}{2},
\end{equation}
where $W$ is an unknown exponent dependent on number of qubits $N$, layers $T$ and probability of swap omission $p$. From formula \eqref{eq_double_parity_simplyfied} we infer, that the expected value of exponent should be given by $W =\frac{1}{2} (T-1) w(N) p$, where $w(N)$ is an average number of swaps in implementation of $N$ qubit permutation. In our case, we studied linear architecture with permutations decomposed by brick-sort algorithm \cite{LAKSHMIVARAHAN1984295} which gives $w(N) = N(N-1)/4$ \cite{CANFIELD2011109}. Note also that in the formula for $W$ we may replace $T$ by $T-1$ since errors in the initial permutation of qubits, all in state $|0\ra$, are undetectable.

\begin{figure}[h]
    \centering
    \includegraphics[width = 10 cm ]{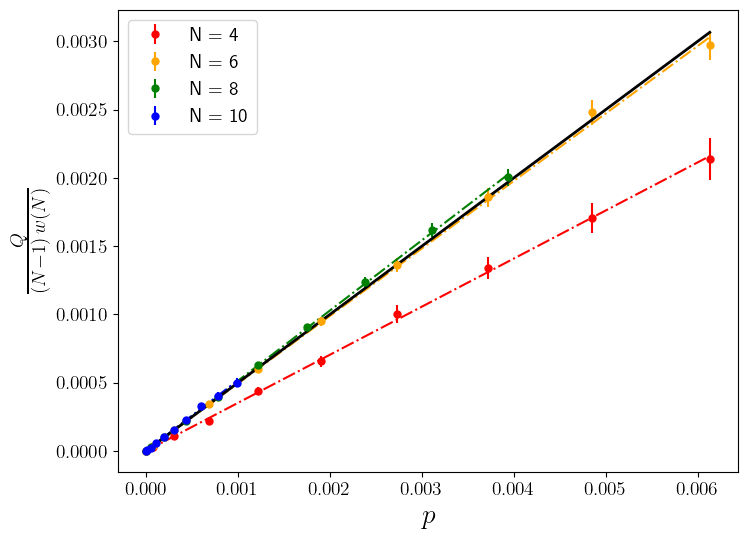}
    \caption{Exponent $W$, normalized by the number relevant layers $T-1$ and number of swaps within each permutation $w(N)$, as a function of noise parameter  $p$, together with linear fits. The predicted behavior \eqref{eq_double_parity_simplyfied} corresponds to the black line.}
    \label{fig:double_p_integrable_num1}
\end{figure}

As one can see in the figure \ref{fig:double_p_integrable_num1} the results from the experiments quickly converge to theoretical predictions as $N$ grows.
The linear fit of $W/((T-1)w(N))$ as a function of $p$ yields:
\begin{equation*}
\begin{aligned}
&~\hspace{2.5 cm} \frac{W}{(T-1) w(N)} = a\; p + b ~~\text{ with}: \\
& a = 0.3522 \pm 0.0033 ,~ b = 2.8 \times 10^{-6} \pm 9.8 \times 10^{-6} ~\;\text{ for } N = 4, \\
& a = 0.494 \pm 0.057 ~,~~~ b = 0.7 \times 10^{-5} \pm 1.7 \times 10^{-5} ~\;\text{ for } N = 6, \\
& a = 0.514 \pm 0.024 ~,~~~ b = 1.1 \times 10^{-6} \pm 4.5 \times 10^{-6} ~\;\text{ for } N = 8, \\
& a = 0.517 \pm 0.085 ~,~~~ b = 1.3 \times 10^{-6} \pm 4.1 \times 10^{-6} ~~\text{ for } N = 10. \\
\end{aligned}
\end{equation*}

Finally we examine the general formula for heavy output frequency $h_U$ decay in the double parity circuit in exponential form \eqref{eq_final_fouble_parity}. Although we performed a dozen numerical fits to study this scenario, here we present, in our opinion, the most convincing one. Note that the general formula \eqref{eq_final_fouble_parity} consists of three different exponents, thus we consider an expression
\begin{equation}
h_U = \frac{1}{4}(2 e^{-Q'(-N,T,\alpha)} e^{-W(N,T,p)} + e^{-Q(N,T,\alpha)}),
\end{equation}
Where the coefficients $Q(N,T,\alpha)$ and $W(N,T,p)$ are the same as in equations \eqref{sigle_p_fit} and \eqref{double_p_p_fit}. The new exponent depends on the number of qubits $N$, layers $T$ and noise strength $\alpha$. Its theoretical form should be given by $Q'(N,T,\alpha) = \frac{3}{2} \alpha^2 T q(N)$, where, for clarity, we define $q(N) := N \frac{N - 2/3}{N-1}$. In the presented experiments we first obtained values of $Q$ and $W$, utilising the entire range of $p$ and $\alpha$ values, as described above, and next used them to infer the value of $Q'$ from experimental data. 

\begin{figure}[h]
    \centering
    \includegraphics[width = 10 cm ]{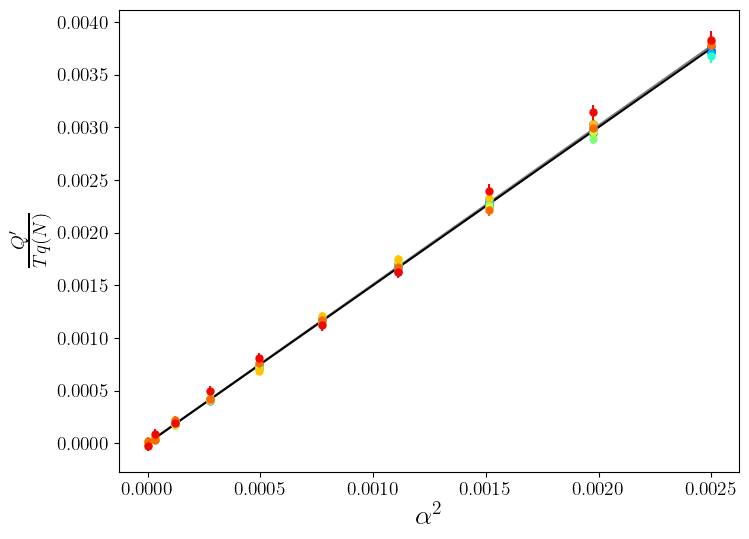}
    \caption{Exponent $Q'$, normalized by the number of layers $T$ and appropriate function of qubit's number $q(N)$, as a function of noise parameter squared $\alpha$. Data for $N = 8$ qubits.
    Each color of data points corresponds to different values of $p$ from $p = 0$ (blue) to $p = 0.04$ (red). The predicted behavior \eqref{eq_final_fouble_parity} corresponds to the gray line and is indistinguishable from linear fits.}
    \label{fig:double_p_integrable_num2}
\end{figure}

The exemple results for $N = 8$ are presented in \rd{Fig.} \ref{fig:double_p_integrable_num2} \rd{for six different values of $p$ equally distributed in the interval $[0,0.04]$}.
As one can see the experimental results are in line with theoretical predictions. Moreover the dispersion of data points for different values of $p$ is smaller than their dispersion for fixed values of $p$.
The linear fit of $Q'/(T q(N))$ as a function of $\alpha^2$ yields:
\begin{equation*}
\begin{aligned}
&~\hspace{3.5 cm} \frac{Q}{T q(N)} = a\; \alpha^2 + b ~~~~\text{ with }: \\
& a = 1.4932 \pm 0.0082 ~,~~~ b = 5.8 \times 10^{-6} \pm 5.9 \times 10^{-6} ~~~\;\text{ for } N = 4, \\
& a = 1.4956 \pm 0.0058 ~,~~~ b = -3.2 \times 10^{-6} \pm 5.8 \times 10^{-6} ~~\text{ for } N = 6, \\
& a = 1.5088 \pm 0.0095 ~,~~~ b = 1.0 \times 10^{-6} \pm 5.5 \times 10^{-6} ~~~\;\text{ for } N = 8, \\
& a = 1.5063 \pm 0.0079 ~,~~~ b = -3.7 \times 10^{-6} \pm 9.8 \times 10^{-6} ~~\text{ for } N = 10. \\
\end{aligned}
\end{equation*}

\section{Practical Applications and Test Modifications}\label{app:appl}

\bl{In this Appendix we extend the discussion of implementation for benchmarks proposed in Section \ref{subsec:impl}.}
In practice, the presented metrics can be used to test quantum computers both to evaluate hardware quality improvements and to compare diverse devices. The architecture-agnostic property enables direct benchmarking of systems with fundamentally distinct architectures.
Certain limitations in generality and inherent symmetries raise practical concerns regarding these benchmarks, which we address here. Test outcomes may be compromised independently of the user, either by device-specific properties, architectural features, or particular error types, or through  attempts to artificially increase performance metrics.

To maximize error detection coverage, we have already introduced both a parity test and a double parity test capable of identifying rare parity-preserving errors, while maintaining all error-detection capabilities of the classical Quantum Volume benchmark. However, these circuits can be further redesigned to make underlying symmetries less influential on the benchmark results.

\begin{figure}[h]
    \centering
    \begin{subfigure}[b]{0.49\textwidth}
        \centering
        \includegraphics[width=\textwidth]{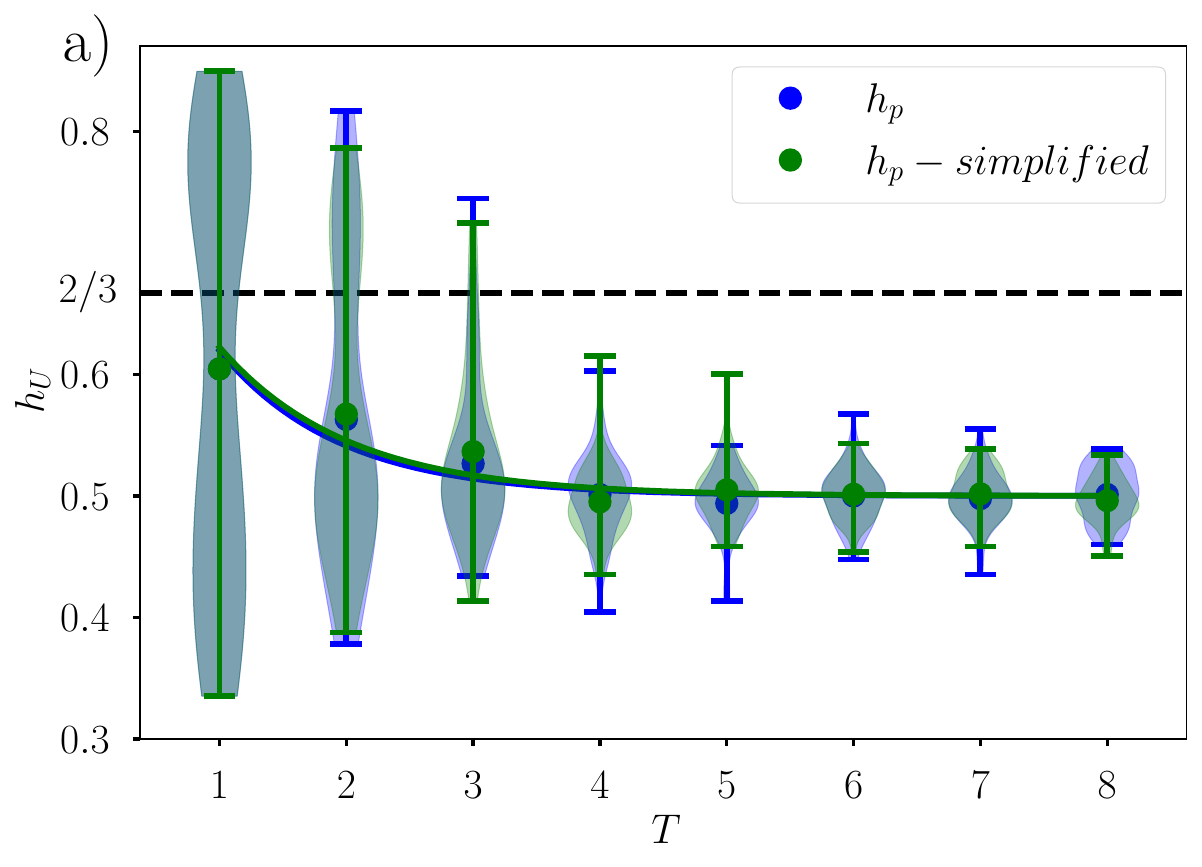}
    \end{subfigure}
    \hfill
    \begin{subfigure}[b]{0.49\textwidth}
        \centering
        \includegraphics[width=\textwidth]{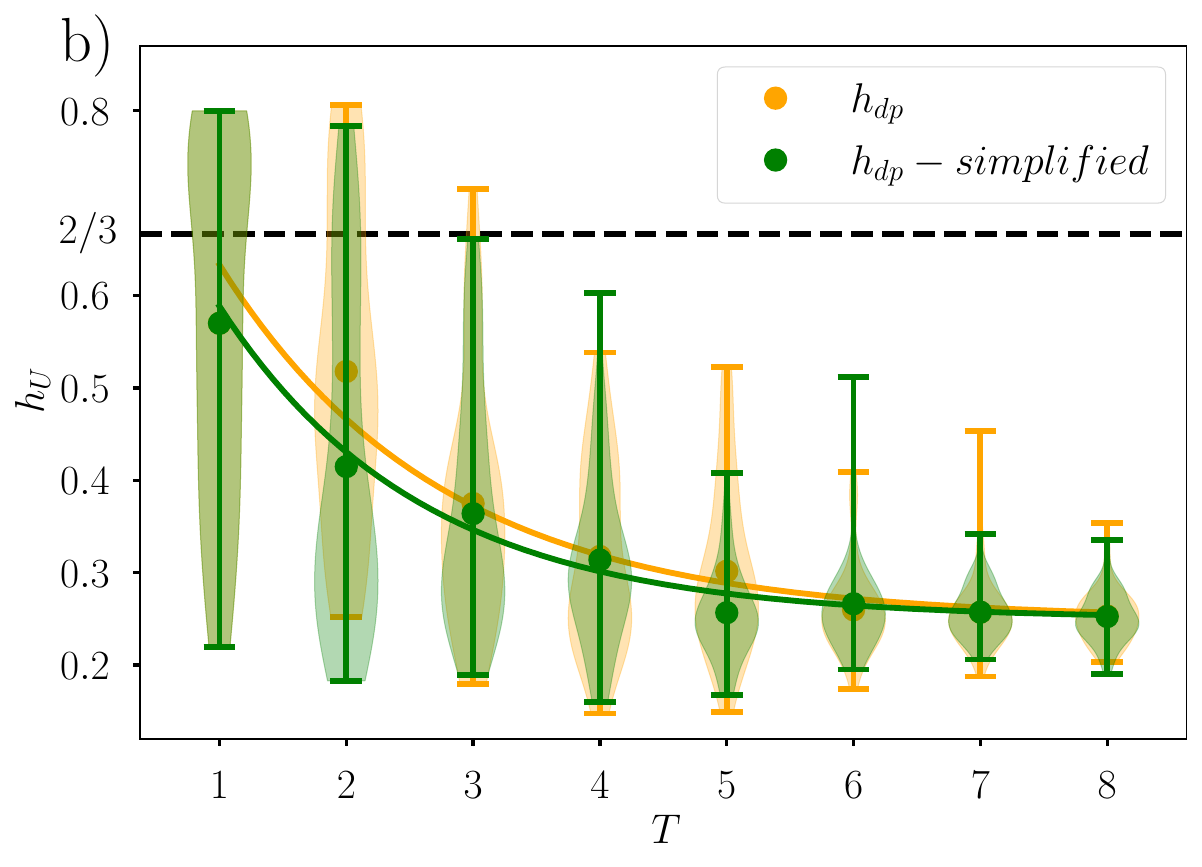}
    \end{subfigure}   
    \caption{Comparison of heavy output probability $h_U$ for $6$-qubit circuits ($N=6$) computed on \textit{IBM Brisbane} with gates of hidden parity preservation and with explicitly parity--preserving form \eqref{u_parity_1} denoted as \textit{simplified}. a) Single parity test comparison. b) Double parity test comparison.  }
    \label{fig:IBM-computation-comparison}
\end{figure}

To enhance circuit-layer generality, we append single-qubit operations before and after each parity-preserving two-qubit gate, combining them into one large two-qubit gate, while preserving global circuit parity. Crucially, every output single-qubit gate is counteracted by its inverse at the input of the next layer’s two-qubit gate following permutation --see Fig. \ref{fig:circuit-modification}. This ensures overall parity conservation despite non-preservation at individual layers. In practice, the test is executed by specifying lists of two-qubit gates and permutations, thereby hiding the circuit’s parity constraints and preventing gate simplification during decomposition to physical gates. Consequently, devices execute tests using generic gates, mitigating hardware-specific bias and making artificial improvements significantly harder.
Note that this modification naturally enables the test to detect parity-preserving errors as well.
For comparison of results for parity-preserving gates with results for gates with hidden parity see Fig. \ref{fig:IBM-computation-comparison}.

To prevent bias towards certain bit-stings, for example ground state represented as $|0^{\otimes N}\ra$, one may implement randomized Pauli $X$ gates embedded within two-qubit operations -- flipping parity when present in odd counts.

Finally, we address deliberate deception based on the estimation of correct output parity and returning a random bit-string with this parity. To tackle this problem, we propose to draw some random matchgate circuits supplemented with a few qubit swaps and non-match gates such that they are still quite efficiently simulable \cite{reardon-smith2023improved}. Then, disguise them by randomly permuting interacting pairs of qubits in each layer and adding single-qubit gates as discussed above and insert them into the test. Output probability distributions of these circuits can be benchmarked against classical predictions, exposing devices that return random valid-parity bit-strings to cheat the test.

\section{\bl{Procedures for modified quantum volume test}}\label{app:procedure}

\bl{In this Appendix we provide procedures for each step of proposed modifications of Quantum Volume test described in Section \ref{sec:parity}. The full implementation of the proposed benchmarks, as well as original quantum volume test in the \texttt{Python} language is available in GitHub repository \cite{bistronGithub_QV}.}

\bl{The procedure starts with generating random quantum circuits with a given number of qubits $N$ and layers $T$, with standard scenario $N=T$. The algorithm \ref{alg:gen_p_circ} describes the generation of random parity-preserving quantum volume circuits, whereas the Algorithm \ref{alg:gen_dp_circ} describes the generation of double parity circuits for auxiliary test.
In both of those pseudocodes there is a flag $\texttt{simplified}$. By default this flag is set to \texttt{True} which corresponds to ``dressing'' parity-preserving gates with random single-qubit gates and its inverses as described in Section \ref{subsec:impl}, and the Appendix \ref{app:appl}.}


\begin{algorithmfigure}[h]
\bl{\caption{\textbf{Generate Parity-Preserving Quantum Volume Circuit}}\label{alg:gen_p_circ}}
\centering
\begin{minipage}{0.95\columnwidth}
\hrule
\vspace{0.5ex}
\bl{
\begin{algorithmic}[1]
\Require $N$, $T$, flag \texttt{simplified}
\Ensure Quantum circuit $QC$
\State Initialize empty quantum circuit $QC$ on $N$ qubits
\State Initialize single-qubit gates table $S_q \gets \mathbb{I}$ for all qubits
\For{$i = 1$ to $T$}
    \State Sample a random permutation $\pi$
    \State Apply permutation gate $P(\pi)$ on qubits
    \For{$j = 0$ to $\lfloor N/2 \rfloor - 1$}
        \State $U \gets$ parity-preserving two-qubit unitary
        \If{not \texttt{simplified} and $i < T$}
            \State Sample random single-qubit unitaries for qubits $2j,2j+1$
            \State Combine $U$ with preprocessing by inverses of single-qubit unitaries from $S_q$ and postprocessing by new single qubit unitaries
            \State Update $S_q$ by new single qubit unitaries
        \EndIf
        \State Apply $U$ to qubits $(2j,2j+1)$
    \EndFor
    \If{$N$ \text{ is odd and not } \texttt{simplified}}
        \State Apply inverse of single-qubit unitary from $S_q$ to last qubit
    \EndIf
\EndFor
\State Measure all qubits
\Return $QC$
\end{algorithmic}}
\vspace{0.5ex}
\hrule
\end{minipage}
\end{algorithmfigure}

\begin{algorithmfigure}[h]
\bl{\caption{\textbf{Calculate Heavy Output Frequency for Parity-Preserving Circuit}}\label{alg:p_hu}}
\centering
\begin{minipage}{0.95\columnwidth}
\hrule
\vspace{0.5ex}
\bl{
\begin{algorithmic}[1]
\Require Measurement outcomes $\text{outputs}$
\Ensure Heavy output frequency $\mathrm{HU}$
\State Initialize $\mathrm{HU} \gets 0$, $\mathrm{Total} \gets 0$
\For{bitstrings $r$ with counts $c$ in $\text{outputs}$}
    \State $\mathrm{Total} \gets \mathrm{Total} + c$
    \State Compute parity $p \gets \sum_i r_i \bmod 2$
    \If{$p = 0$}
        \State $\mathrm{HU} \gets \mathrm{HU} + c$
    \EndIf
\EndFor
\Return $\mathrm{HU} / \mathrm{Total}$
\end{algorithmic}}
\vspace{0.5ex}
\hrule
\end{minipage}
\end{algorithmfigure}

\begin{algorithmfigure}[h]
\bl{\caption{\textbf{Generate Double-Parity Quantum Volume Circuit}}\label{alg:gen_dp_circ}}
\centering
\begin{minipage}{0.95\columnwidth}
\hrule
\vspace{0.5ex}
\bl{
\begin{algorithmic}[1]
\Require $N$, $T$, flag \texttt{simplified}
\Ensure Quantum circuit $QC$, parity assignment vector $\text{odds}$
\State Initialize empty quantum circuit $QC$
\State Initialize parity vector $\text{odds}$ with two parity sectors
\State Randomly permute $\text{odds}$
\State Initialize single-qubit table $S_q \gets \mathbb{I}$
\For{$i = 1$ to $T$}
    \State Sample random permutation $\pi$
    \State Apply permutation gate $P(\pi)$ on qubits
    \State Update $\text{odds} \gets \text{odds}[\pi]$
    \For{$j = 0$ to $\lfloor N/2 \rfloor - 1$}
        \If{$\text{odds}[2j] = \text{odds}[2j+1]$}
            \State $U \gets$ parity-preserving gate
        \Else
            \State $U \gets$ diagonal two-qubit gate
        \EndIf
        \If{not \texttt{simplified} and $i < T$}
            \State Sample random single-qubit unitaries for qubits $2j,2j+1$
            \State Combine $U$ with prepossessing by inverses of single-qubit unitaries from $S_q$ and postprocessing by new single qubit unitaries
            \State Update $S_q$ by new single qubit unitaries
            \State Update $S_q$ by new single qubit unitaries
        \EndIf
        \State Apply $U$ to qubits $(2j,2j+1)$
    \EndFor
    \If{$N$ \text{ is odd and not } \texttt{simplified}}
        \State Apply inverse of single-qubit unitary from $S_q$ to last qubit
    \EndIf
\EndFor
\State Measure all qubits
\Return $(QC, \text{odds})$
\end{algorithmic}}
\vspace{0.5ex}
\hrule
\end{minipage}
\end{algorithmfigure}

\bl{Then each quantum circuit is run multiple times to obtain estimates for outcome frequencies given by a number of counts for each measured bit-string. Algorithm \ref{alg:p_hu} provides the procedure to obtain heavy output frequency from outcome statistics in parity-preserving test, whereas the algorithm \ref{alg:dp_hu} corresponds to double parity test.}


\begin{algorithmfigure}[h]
\bl{\caption{\textbf{Calculate Heavy Output Frequency for Double-Parity-Preserving Circuit}}\label{alg:dp_hu}}
\centering
\begin{minipage}{0.95\columnwidth}
\hrule
\vspace{0.5ex}
\bl{
\begin{algorithmic}[1]
\Require Measurement outcomes $\text{outputs}$, parity assignment vector $\text{odds}$
\Ensure Heavy output frequency $\mathrm{HU}$
\State Define complementary parity vector $\text{even} \gets \mathbf{1} - \text{odds}$
\State Initialize $\mathrm{HU} \gets 0$, $\mathrm{Total} \gets 0$
\For{bitstrings $r$ with counts $c$ in $\text{outputs}$}
    \State $\mathrm{Total} \gets \mathrm{Total} + c$
    \State Convert $r$ to bit vector $\mathbf{b}$ ordered by qubit index
    \State Compute parity $p_1 \gets (\mathbf{b} \cdot \text{odds}) \bmod 2$
    \State Compute parity $p_2 \gets (\mathbf{b} \cdot \text{even}) \bmod 2$
    \If{$p_1 = 0$ and $p_2 = 0$}
        \State $\mathrm{HU} \gets \mathrm{HU} + c$
    \EndIf
\EndFor
\Return $\mathrm{HU} / \mathrm{Total}$
\end{algorithmic}}
\vspace{0.5ex}
\hrule
\end{minipage}
\end{algorithmfigure}

\bl{The quantum computer passes the test, if the heavy output frequency, averaged over all drawn circuits, exceeds the given threshold with $2\sigma$ confidence. For parity-preserving test we chose the same threshold as for standard quantum volume test, equal $2/3$. On the other hand, since in the double parity test the heavy output frequency can drop to $1/4$ instead of $1/2$, we propose rescaled threshold equal $(1+\log 2)/4\log 2\approx 0.61$. }

\newpage 
\newpage
\section{Case study: Present day quantum processor.}\label{App:case_study}

This Appendix provides \bl{detailed} descriptions of the computations and simulations of the \textit{IBM Brisbane} device \bl{discussed in Section \ref{sec:case-study}, whose} results are plotted in Fig. \ref{fig:IBM} in the main text.
All tasks were implemented using Qiskit \cite{Qiskit}. For a specified number of qubits $N$ and layers $T$, random circuits were constructed by sequentially applying random permutations and random two-qubit gates, as described in \rd{Appendix \ref{Section:I} and the main body of the work}. In the classical QV test two-qubit gates were sampled according to the Haar measure for $SU(4)$.  For the parity-preserving test, two-qubit gates were defined by \eqref{u_parity_1}. In the double parity-preserving case, one additional constraint was applied: qubits were divided into two groups, and any interaction between qubits from different groups used a random diagonal gate as specified in \eqref{u_parity_2}.

All circuits were optimized as much as possible using Qiskit with \textit{optimization{\_}level=3}. The {\sl layout} was trivial, which associates a physical qubit to each virtual qubit of the circuit in increasing order \cite{Qiskit} resulting in a one-dimensional qubit layout. The code used for testing and simulating quantum computers is available at \cite{bistronGithub_QV}.
 
\subsection{Device testing}
For performing the QV, single-parity, and double-parity tests we executed corresponding circuits on the \textit{IBM Brisbane} device. For $6$-qubit tests presented in the main text, 50 randomly generated circuits were executed 650 times. Heavy output subspaces for all circuits were determined through noiseless classical computation. The heavy output probabilities were then calculated based on the frequency of bit-strings within the heavy output subspace.

\subsection{Device simulation}
Simulations with rescaled errors on the IBM simulator were performed to test the performance of the proposed tests for different noise scales and to compare their results with those of the Quantum Volume test.

\begin{figure}[h]
    \centering
    \begin{subfigure}[b]{0.49\textwidth}
        \centering
        \includegraphics[width=\textwidth]{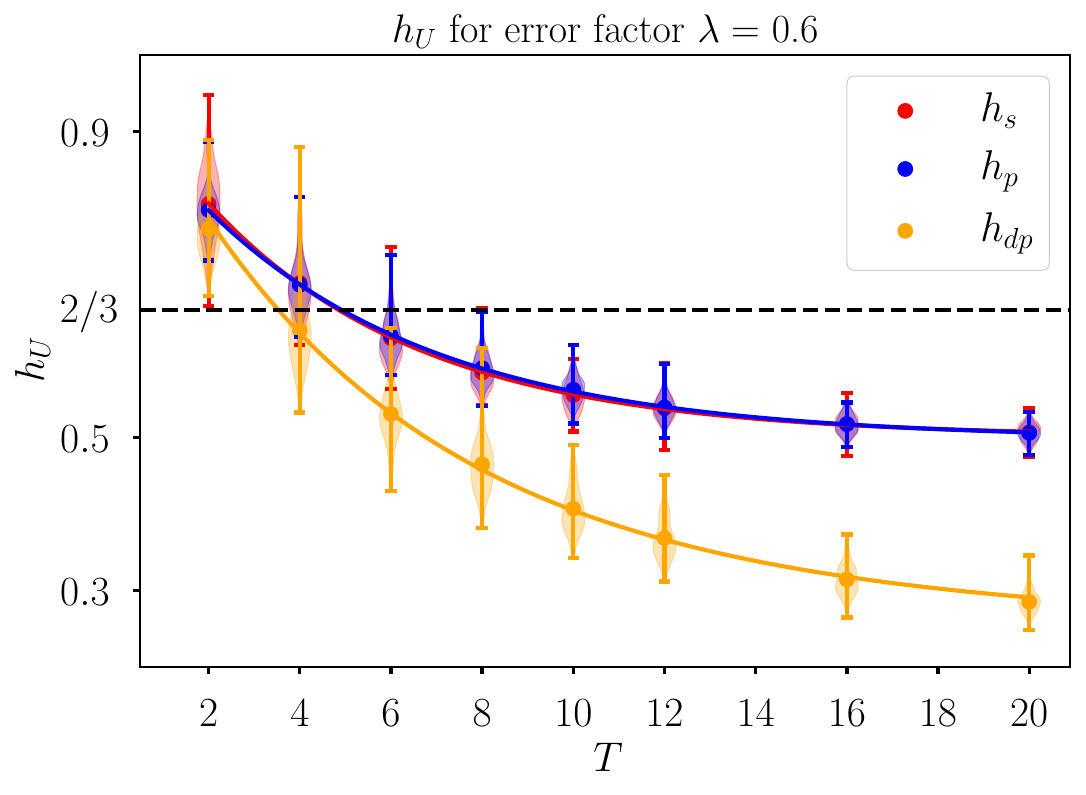}
    \end{subfigure}
    \hfill
    \begin{subfigure}[b]{0.49\textwidth}
        \centering
        \includegraphics[width=\textwidth]{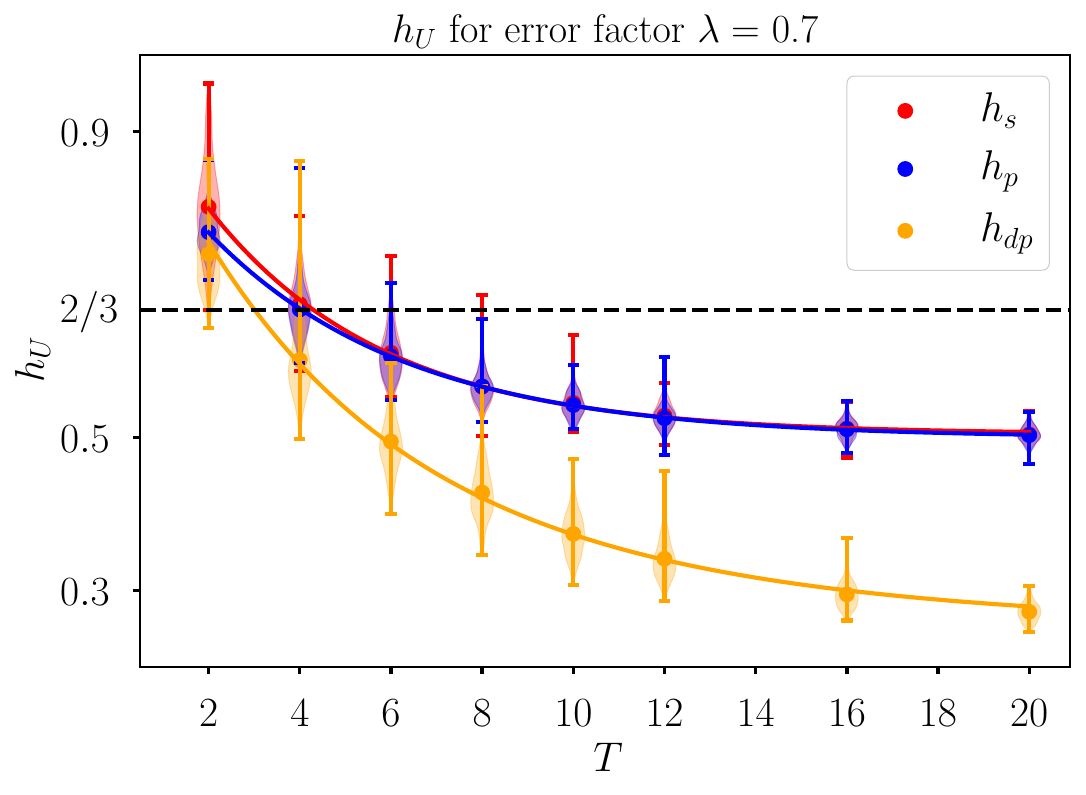}
    \end{subfigure}
    
    \vspace{0.5cm}
    
    \begin{subfigure}[b]{0.49\textwidth}
        \centering
        \includegraphics[width=\textwidth]{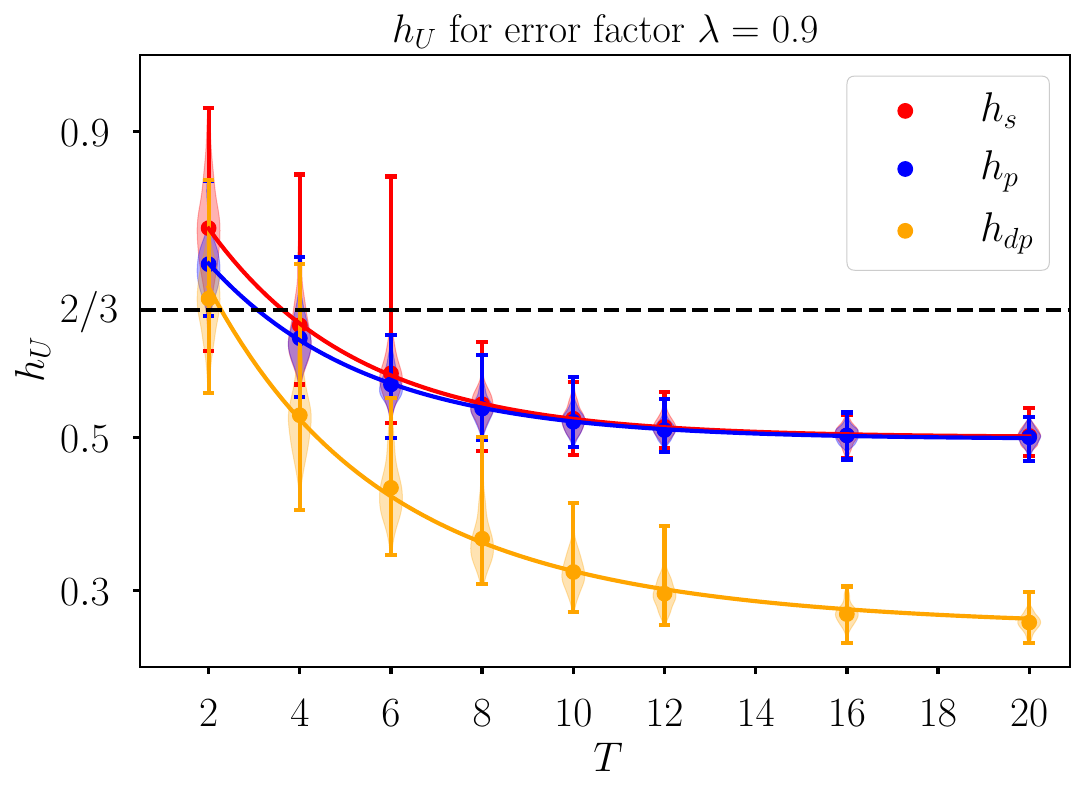}
    \end{subfigure}
    \hfill
    \begin{subfigure}[b]{0.49\textwidth}
        \centering
        \includegraphics[width=\textwidth]{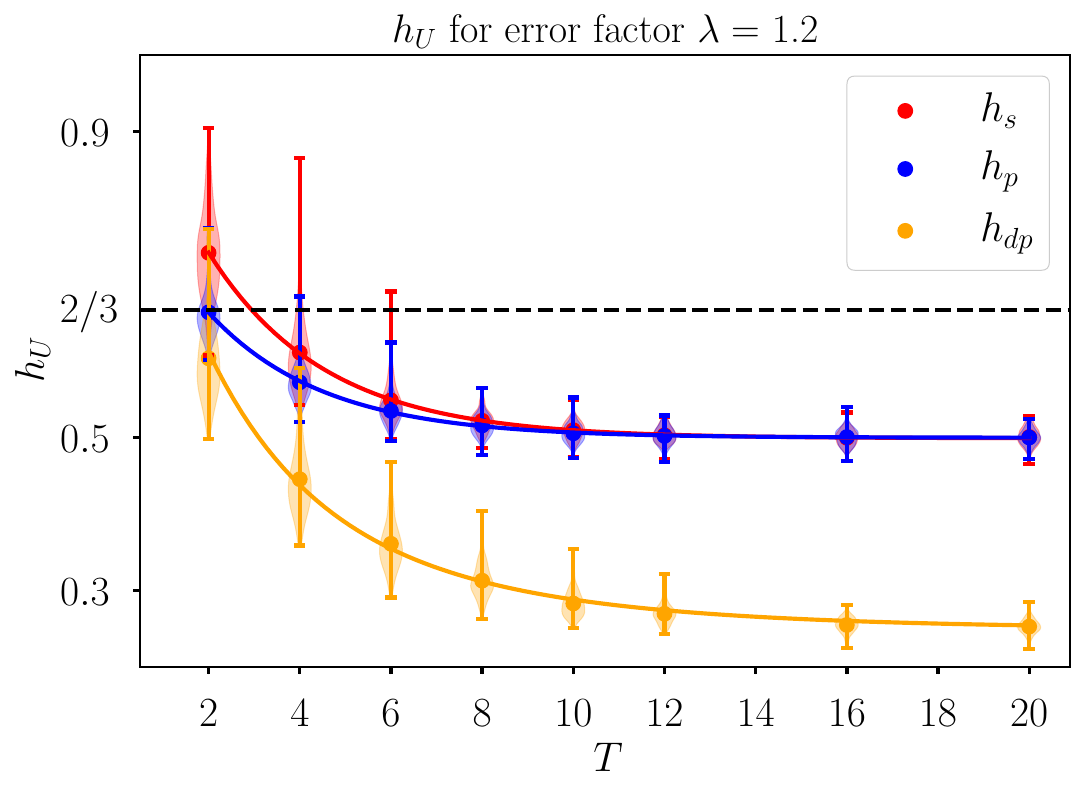}
    \end{subfigure}
    
    \caption{Heavy output probability $h_U$ for simulated $6$-qubit circuits ($N=6$) on \textit{IBM Brisbane} with several values of error factor $\lambda$. Note that for error factor $\lambda = 0.6$, for which the computer is close to pass QV test for $6$ qubits, heavy output frequencies of standard $h_{s}$ and parity-preserving $h_p$ QV circuits practically do coincide. Deviation between heavy output frequency in double parity test $h_{dp}$ from $h_s$ and $h_p$ originates from different minimal value $h_{dp}\geq 1/4$, and can be compensated by appropriate rescaling.}
    \label{fig:IBM-simulation}
\end{figure}

\begin{figure}[h]
        \centering
        \includegraphics[width=0.6\textwidth]{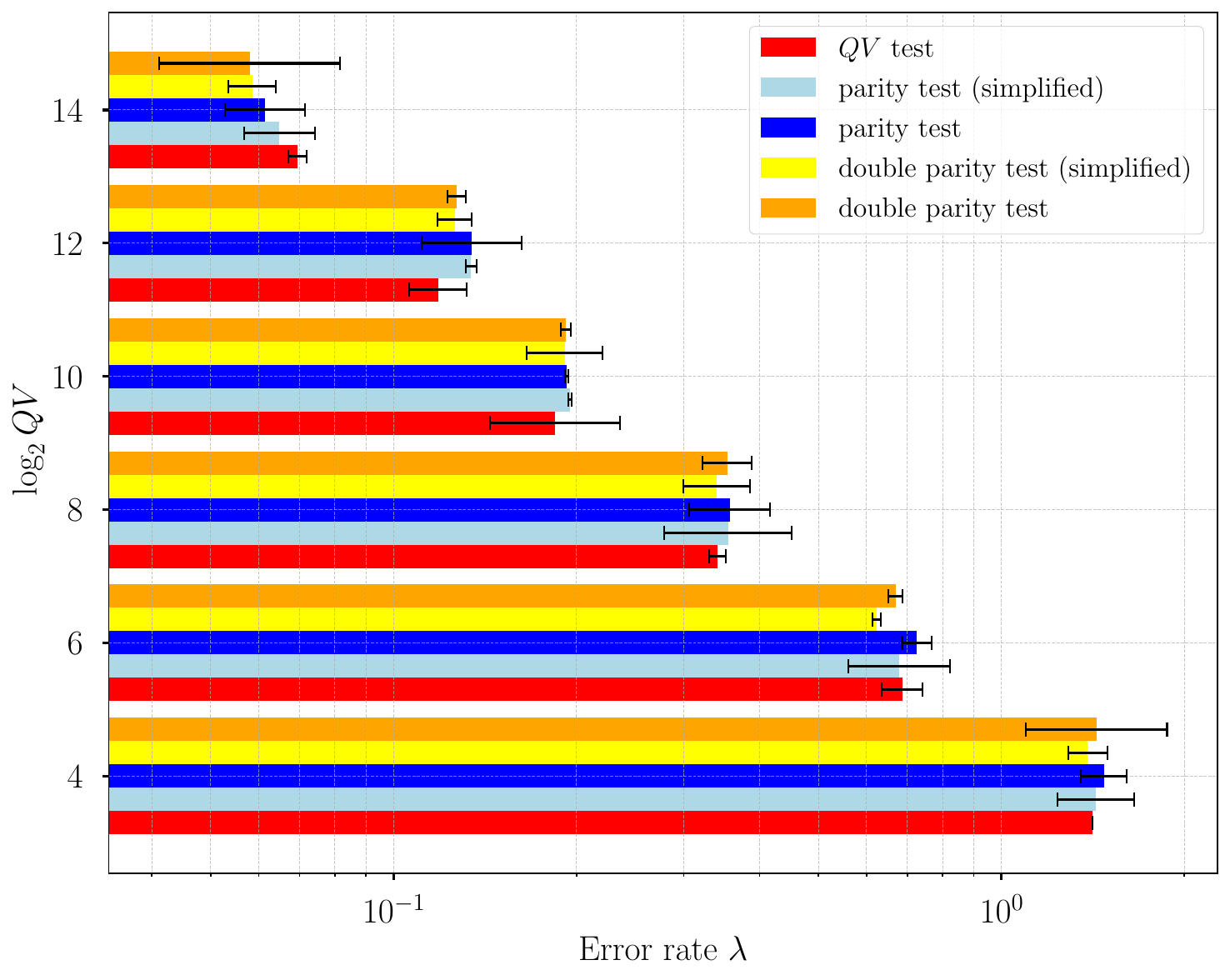}
    \caption{Ranges of the scaling factor $\lambda$ for passing the single parity, double parity, and quantum volume tests. Tests with explicitly parity-preserving 2-qubit gates \eqref{u_parity_1} are presented as \textit{simplified}, whereas the one remaining one are extended as discussed in previous Appendix. All four proposed benchmarks exhibit the same behavior as the original QV test for any noise strength, \bl{but the benchmarks proposed in this work does not require classical simulations.}} 
    \label{fig:IBM-Ranges-general}
\end{figure}

Simulation of the IBM device was performed with \textit{AerSimulator}. Using a fake backend, we modified the error parameters i.e.  \textit{readout{\_}error} for all qubits, \textit{gate{\_}error} for all single-qubit and all two-qubit gates. We multiplied them by the common error factor $\lambda >0$. Relaxation time $t_1$ and dephasing time $t_2$ parameters were multiplied by $1/\lambda$ error factor.

For each simulated circuit, heavy output subspace was determined by classical noiseless computation. Simulations were performed for varying values of the error factor, allowing the determination of its maximum value for which a given Quantum Volume was achieved for each test. The number of sampled circuits and performed simulations for each of them is presented in the table \ref{tab2}.

The range of error factors for which a test was passed--see Fig. \ref{fig:IBM-Ranges-general}, was determined as follows. For each data set obtained with different error factors $\lambda$, we estimated the value of $T$ at which the threshold line $h_U = 2/3$ was crossed. This threshold is denoted by $T_{\text{cros}}$. We then plotted $T_{\text{cros}}/N$ vs $\lambda$, where $N$ represents the number of qubits. From this plot, we estimated the value of $\lambda$ at which $T_{\text{cros}} = N$. The uncertainty in this estimate was propagated using standard error propagation techniques. Note that the procedure does not include the $2\sigma$ confidence passing criterion of standard Quantum Volume \cite{Math_monster_theoretic_fundations} due to the significant standard deviation resulting from the small number of sampled circuits and their realizations, which would impact the results.

\begin{table}[h]
\begin{tabular}{l|l|l}
Number of qubits $N$ & sampled circuits & number of simulations \\ \hline
4                    & 200        & 2000   \\
6                    & 100        & 1000      \\
8                    & 70         & 700    \\
10                    & 50         & 500     \\
12                    & 50         & 500     \\
14                   & 30          & 300
\end{tabular}
\caption{Summary of performed numerical experiments. The same simulations were performed for the Quantum Volume circuit, parity circuit, and double-parity circuit. Each sampled circuit was simulated a given number of times.}
\label{tab2}
\end{table}

\subsection{Discussion}

The presented simulations show agreement between the proposed measures with the Quantum Volume test. However, as we discussed in the \rd{main body of the work and above Appendices}, different tests have different sensitivities for various error models. Therefore discrepancies are inevitable in some scenarios. We argue that the proposed tests exhibit similar properties to Quantum Volume, agreeing with it in standard cases. Our primary objective was to develop benchmarks that replicate the scaling and qualitative behavior of the QV test, rather than achieving identical numerical results. The presented simulations confirm the reliability of the proposed tests and their agreement with QV in standard cases, making them suitable for benchmarking systems of larger sizes.

\section{\bl{Parity-preserving circuits as estimators for original QV test}}\label{app:est}

In this \bl{final Appendix we present the derivations of the formulas from Section \ref{sec:etimating}. Primarily, the} expressions for the heavy output frequency of the original quantum volume circuit and its modifications, which preserves parity on a subset of qubits. We start with the discussion of general noise restricted only by mild assumptions. Then, we gradually introduce various noise models to obtain explicit relations between heavy output frequencies in those particular examples which allows us to establish estimators for heavy output frequency in quantum volume circuits.

\subsection{Computing estimator $\widetilde{h_U}$ for standard QV circuit}\label{app:hu_est_original}

The equation for the heavy output probability \bl{in a circuit of  $N$ qubits and $T$ layers} with the 
noise \bl{modeled by a set of arbitrary quantum channels $(\Omega_z)_{z=1}^M$, each applied after some layer of two-qubit gates} is expressed as
\begin{equation}
    \widetilde{h_U} = \int_{C = (U_1,\ldots,U_M)} dC
\inner{\Omega_M(U_M\ldots\Omega_1(U_1\proj{0^{\otimes N} }U_1^\dagger)\ldots 
    U_M^\dagger)}{\Pi_C},
\end{equation}
where $C$ is a random quantum volume circuit of the size $(N, T)$, \bl{ composed} of 
random permutations and random two-qubit gates. The \bl{circuit $C$ is decomposed into $T$ sub-circuits, denoted as $U_z$}, each containing $T/M$ layers and 
$\Pi_C$ is the 
projector 
onto the corresponding heavy output subspace. Using the exponential 
distribution 
hypothesis~\cite{Math_monster_theoretic_fundations} for $N, T/M \gg 1$ we approximate one layer of the circuit by one large random unitary operator, and we estimate $h_U$ as
\begin{equation}
\begin{split}
    \widetilde{h_U} &= \sum_{\Pi} \mathbb{P}(\Pi) \int_{C \in \Pi} dC
\inner{\Omega_M(U_M\ldots\Omega_1(U_1\proj{0^{\otimes N} }U_1^\dagger)\ldots 
    U_M^\dagger)}{\Pi}\\
&= \sum_{\Pi} \mathbb{P}(\Pi) \int_{U_M \in \Pi,U_z} dU_1\ldots dU_M
    \inner{\Omega_M(U_M U_{M-1}^\dagger\ldots 
    U_1^\dagger\Omega_1(U_1\proj{0^{\otimes N} }U_1^\dagger)U_1\ldots 
    U_{M-1} U_M^\dagger)}{\Pi},
\end{split}
\end{equation}
where $\mathbb{P}(\Pi)$ is the 
probability that the heavy subspace projector $\Pi$ will be sampled and the 
integral is taken over independent Haar-random unitary operators $U_z$ with $z<M$ and ``$U_M \in \Pi$'' denotes \bl{unitary matrices} $U_M$ for which the heavy subspace projector of 
$U_M\ket{0}$ is 
$\Pi$. Assuming complete symmetry of the problem we obtain $\mathbb{P}(\Pi) = 
1/\binom{2^N}{2^{N-1}}$ and $\int_{U_z} dU_z U_z^\dagger \Omega_z(U_z \rho 
U_z^\dagger) U_z = a_z \rho + (1-a_z) 
\rho_*$, where
\begin{equation}
a_z = \frac{\inner{J_{\Omega_z}}{J_\Id}-1}{2^{2N}-1},
\end{equation}
with $J_{\Omega_z}$, $J_\Id$ representing the Choi-Jamio{\l}kowski isomorphism of noise and identity channels, and
$\rho_*$ is the 
maximally mixed state. Moreover, for the integral over unitary $U_M$ we have 
$\int_{U_M \in \Pi} dU_M U_M \proj{0^{\otimes N} } U_M^\dagger = p_* \Pi/\tr(\Pi) + 
(1-p_*)\Pi^\perp/\tr(\Pi^\perp) $, where $p_* = \frac{1+\ln 2}{2}\approx 0.84$ 
and $\Pi 
+ \Pi^\perp = \Id$. Let us introduce the following notation. For any 
bit-strings $j, i$ of length 
$N$ we define $w^{(z)}_{i,j} = \bra{i}\Omega_z(\proj{j})\ket{i}$ as the 
probability 
that the state $\ket{j}$ will be measured as $\ket{i}$ \bl{due to the influence of} $\Omega_z$. Moreover we denote $a = a_{M-1} \cdots a_1$, and an average probability of flipping $k$ bits as  
\begin{equation}\mathbb{P}^{(z)}_k = \sum_{i,j: \rho_H(i,j) 
= 
k} \frac{w^{(z)}_{i,j}}{2^N}\end{equation}
i.e.  the probability that uniformly 
sampled input 
basis state $\ket{j}$ is measured as $\ket{i}$ being under the effect of 
$\Omega_z$, where the Hamming distance between $i$ and $j$ is $k =0,\ldots,N$. 
Summarizing everything we get the value of the estimator $\widetilde{h_U}$:
\begin{equation}
\label{eq:hu}
    \begin{aligned}
        \widetilde{h_U}&= \sum_{\Pi} \mathbb{P}(\Pi) \int_{U_M \in \Pi,U_z} dU_1\ldots 
        dU_M
    \inner{\Omega_M(U_M U_{M-1}^\dagger\ldots 
    U_1^\dagger\Omega_1(U_1\proj{0^{\otimes N} }U_1^\dagger)U_1\ldots 
    U_{M-1} U_M^\dagger)}{\Pi}\\
&= \frac{1}{\binom{2^N}{2^{N-1}}} \sum_{\Pi} \int_{U_M \in \Pi} dU_M
    \inner{\Omega_M(U_M ( a \proj{0^{\otimes N} } + (1-a) \rho_* ) U_M^\dagger)}{\Pi}\\
&= \frac{1}{\binom{2^N}{2^{N-1}}} \sum_{\Pi} 
    \inner{\Omega_M\left( \left( \frac{p_* a}{\tr(\Pi)} + \frac{1-a}{2^N}  
    \right) \Pi + \left( \frac{(1 - p_*) a}{\tr(\Pi^\perp)} + \frac{1-a}{2^N}  
    \right) \Pi^\perp \right)}{\Pi}\\
&= \frac{1}{\binom{2^N}{2^{N-1}} 2^{N-1}} \sum_{\Pi} 
    \inner{\Omega_M\left( \underbrace{\left( p_*a + (1-a)/2\right)}_q \Pi + 
    \underbrace{\left( (1 - p_*) a 
    + (1-a)/2 \right)}_{1-q} \Pi^\perp \right)}{\Pi}\\
&= \frac{1}{\binom{2^N}{2^{N-1}}2^{N-1}}\sum_{\Pi,i,j} q w^{(M)}_{i,j} 
\delta_{i \in 
\Pi} \delta_{j \in \Pi} + (1-q) w^{(M)}_{i,j} \delta_{i \in \Pi} \delta_{j 
\not\in 
\Pi} 
\end{aligned}
\end{equation}
\begin{equation*}
\begin{aligned}
    & =\frac{1}{\binom{2^N}{2^{N-1}}2^{N-1}}\sum_{i,j}w^{(M)}_{i,j} 
        \left(q\sum_\Pi\delta_{i \in \Pi} \delta_{j \in \Pi} +  (1-q) 
        \sum_\Pi\delta_{i \in \Pi} \delta_{j \not\in \Pi}\right)  \\&=
        \frac{1}{\binom{2^N}{2^{N-1}}2^{N-1}}\sum_{i,j}w^{(M)}_{i,j} 
        \left(q(\delta_{i=j}\binom{2^N-1}{2^{N-1}-1}+\delta_{i\neq j} 
        \binom{2^N-2}{2^{N-1}-2} ) +  (1-q) \delta_{i\neq j} 
        \binom{2^N-2}{2^{N-1}-1} \right) \\&=
        \frac{q\binom{2^N-1}{2^{N-1}-1}\mathbb{P}^{(M)}_0 + 
        q\binom{2^N-2}{2^{N-1}-2}(1-\mathbb{P}^{(M)}_0)+(1-q)\binom{2^N-2}{2^{N-1}-1}
(1-\mathbb{P}^{(M)}_0)}{\binom{2^N}{2^{N-1}}2^{-1}}\\&=
        q\mathbb{P}^{(M)}_0 + 
        q\frac{2^{N-1}-1}{2^N-1}(1-\mathbb{P}^{(M)}_0)+(1-q)\frac{2^{N-1}}{2^N-1}
(1-\mathbb{P}^{(M)}_0)\\&= \frac{2^{N-1}-q}{2^N-1} +  \frac{2^N}{2^N-1} (q - 
1/2) 
\mathbb{P}^{(M)}_0 = \frac{1}{2} + \left(p_* - \frac{1}{2}\right)\frac{2^N 
\mathbb{P}^{(M)}_0 - 1}{2^N 
- 1} \prod_{z=1}^{M-1} a_z.
    \end{aligned}
\end{equation*}

\subsection{Computing the estimator $\widetilde{h_U^m}$ for parity-preserving circuits}\label{app:hu_est_m}

Using similar notation as in previous section, we can express
 the heavy-output probability $\widetilde{h_U^m}$ for a quantum volume circuit preserving parity on $m$ \bl{out of $N$} qubits as
\begin{equation}
    \widetilde{h_U^m} = \int_{C_m = (U_1,\ldots,U_M)} dC_m
\inner{\Omega_M(U_M\ldots\Omega_1(U_1\proj{0^{\otimes N} }U_1^\dagger)\ldots 
    U_M^\dagger)}{\Pi_{C_m}}.
\end{equation}
Based on the scheme description and exponential 
distribution 
hypothesis 
we may decompose unitaries representing layers of the circuit $U_1,\ldots,U_M$ with $N,T/M >\!> 1$ \bl{in the following way}. 
Each unitary operation mixes states on parity and non-parity subspaces 
separately and changes the position of \bl{qubits with preserved parity} $\mathcal{M}_t$. 
Moreover, \bl{as introduced in the main text,} the operation $U_1$ contains \bl{randomly placed $X$ gates to choose input parity}. 
\bl{Thus we apply following decomposition}
\begin{itemize}
\item $U_1 \to  \pi_1 B_m^\dagger V_1 B_m \pi_0^\dagger 
X_{m} \pi_0$,
\item $U_z \to  \pi_z B_m^\dagger V_z B_m 
\pi_{z-1}^\dagger $, where $z=2,\ldots,M$,
\end{itemize}
where $\pi_z$ are random subsystem permutations; $X_{m}$ is a tensor product of 
$X$ and $\Id$ operations on consecutive qubits with $X$ \bl{placed} randomly on at most $m$ qubits indexed 
by $\mathcal{M}_0$; $V_z = \proj{0}\otimes V_{z,0} + \proj{1}\otimes V_{z,1}$ 
are 
block-diagonal unitary operations with $V_{z,0}, V_{z,1}$ being independent 
Haar-random unitary operations defined on $N-1$ qubits. \bl{Finally} the $B_m$ being unitary 
transformation acts on first $m$ \bl{parity-preserved} qubits and connects parity subspace for 
each $m>1$ with the parity subspace for $m=1$ according to equations: 
$B_m\ket{0}\ket{p}=\ket{0}\ket{p},B_m\ket{0}\ket{p^\perp}=
\ket{1}\ket{p^\perp},B_m\ket{1}\ket{p}=
\ket{1}\ket{p},B_m\ket{1}\ket{p^\perp}=\ket{0}\ket{p^\perp}$ 
\bl{with $\ket{p}$ and $\ket{p^\perp}$ being arbitrary basis vectors from within and outside the parity subspace.} For $m=1$ we have simply $B_1=\Id$.
The projector on the heavy subspace $\Pi_{C_m}$ depends on two factors: $X_m$ 
which decide parity of the subspace the state belongs to and permutations $\pi_M$ 
which defines $\mathcal{M}_T$. To simplify computation we use the following 
identity
\begin{equation}
\Pi_{C_m} = \pi_M B_m^\dagger V_{M+1}^\dagger B_m (\Pi_{m,p_0} 
\otimes 
\Id_2^{\otimes N-m}) 
B_m^\dagger V_{M+1} B_m \pi_M^\dagger, ~\text{ where }~ \Pi_{m,p_0} = 
\sum_{b: 
\oplus_{k=1}^m b_k = 
p_0} \proj{b}
\end{equation}
and $p_0 \in \{0,1\}$ expresses the parity 
\bl{determined by the number of $X$ gates} used in $U_1$. Therefore, the integral over $C_m$ is \bl{an average} 
over the following independent and uniformly distributed variables: subsystem 
permutations $\pi_0,\ldots,\pi_M$, $X$ gates \bl{positions} in $X_m$ and mixing unitary 
operations $V_1,\ldots,V_{M+1}$. Additionally, the projector $\Pi_{m,p_0}$ is a 
dependent variable which varies with $X_m$ parity $p_0$. Let us fix $p_0 \in 
\{0,1\}$. Then, the effect of these $X$ gates can be summarized as
\begin{equation}
\int_{\pi_0, X_m} d \pi_0 dX_m 
\pi_0^\dagger 
X_{m} \pi_0 \proj{0^N} \pi_0^\dagger 
X_{m}^\dagger \pi_0 = \sigma_{m,p_0} \otimes \proj{0^{M-m}}, ~\text{ where }~ 
\sigma_{m,p_0} = 
\frac{1}{2^{m-1}} \Pi_{m,p_0}.\end{equation} Going further, $\int_{V_z}dV_z V_z 
\rho 
V_z^\dagger = (\Delta 
\otimes \Phi_*^{\otimes N-1})(\rho)$, where $\Delta$ is maximally dephasing 
qubit channel and $\Phi_*$ is maximally depolarizing qubit channel. Combining 
with $B_m$ we obtain averaged action of one layer as a channel
\begin{equation*}
\Psi(\rho) = \int_{V_z}dV_z B_m^\dagger V_z B_m 
\rho 
B_m^\dagger V_z^\dagger B_m = \tr\left((\Pi_{m,0} \otimes \Id_2^{\otimes N-m} 
)\rho\right) \sigma_{m,0} \otimes \rho_*^{\otimes N - m} + \tr\left((\Pi_{m,1} 
\otimes \Id_2^{\otimes N-m} 
)\rho\right) \sigma_{m,1} \otimes \rho_*^{\otimes N - m} .    
\end{equation*}
Finally let us define 
symmetric transformation of the noise channels $\Omega_z$ as
\begin{equation}
\Omega_z^{(s)}(\rho) = \frac{1}{N!} \sum_{\pi_z}  \pi_z^\dagger \Omega_z(\pi_z 
\rho \pi_z^\dagger) \pi_z.\end{equation}
Summarizing everything we get the following expression for heavy output frequency estimator in considered circuits
\begin{equation}\label{eq:hum}
\begin{split}
    \widetilde{h_U^m} &= \int_{C_m = (U_1,\ldots,U_M)} dC_m
    \inner{\Omega_M(U_M\ldots\Omega_1(U_1\proj{0}U_1^\dagger)\ldots 
    U_M^\dagger)}{\Pi_{C_m}}\\
&= \frac12 \sum_{p_0}
\inner{\Psi(\Omega_M^{(s)}(\ldots\Psi(\Omega_1^{(s)}(\Psi(\sigma_{m,p_0} 
    \otimes \proj{0})\cdots)}{\Pi_{m,p_0} \otimes 
    \Id_2^{\otimes N-m}}\\
&= \frac12 \sum_{p_0,\ldots,p_M \in \{0,1\}} \prod_{z=1}^M 
\tr\left((\Pi_{m,p_z} \otimes 
\Id_2^{\otimes N-m})\Omega_z^{(s)}(\sigma_{m,p_{z-1}}\otimes \rho_*^{\otimes N 
- m})\right) \delta_{p_M=p_0}.
\end{split}
\end{equation}

\subsection{Estimating $h_U$ for a given noise model type $(\Omega_z)_{z=1}^M$}\label{App:estimation}

Having established concise expressions for heavy output frequency in different circuits we may leverage them, in particular scenarios, to obtain an estimator of original heavy output frequency.

\subsubsection{No estimation for general noise channels $\Omega_z$}\label{App:estimation-z}

Let us fix $\Omega_1 = Z_\lambda^{\otimes N}$ with $Z_\lambda = \proj{0} + e^{i\lambda} \proj{1}$, $\lambda\in \mathbb{R}$, 
and \bl{set all remaining noise channels to identity}  $\Omega_2,\ldots,\Omega_M = 
\Id$. 
\bl{Applying} this noise model to Eq.~\eqref{eq:hu} we get $\widetilde{h_U} = 
\frac12 +  (p_* - \frac12)a_1 = \frac{1}{2} + (p_* - 
\frac12)(|1+e^{i\lambda}|^{2N}-1)/(4^N-1)$, which varies 
with $\lambda$ as $\widetilde{h_U} \in [1/2 - \mathcal{O}(4^{-N}),p_*]$. Meanwhile, based 
on Eq.~\eqref{eq:hum} we get $\widetilde{h_U^m} = \frac12 \sum_{p} 
\tr\left( (\Pi_{m,p} \otimes 
\Id_2^{\otimes N-m}) \Omega_1^{(s)} (\sigma_{m,p}\otimes \rho_*^{\otimes N 
- m})\right)=1$ for any $m = 1,\ldots, N$ and any $\lambda$. Thus, as mentioned in the main text, one may find peculiar error models which, on average, do not affect the parity-preserving quantum volume circuits. We stress however that this result follows from the fact that in above calculations parity-preserving two-qubit gates were ``exposed'' to the noise. Introduction of random single qubit gates immersed in two-qubit ones as discussed in Appendix \ref{app:appl} resolves this problem.

\subsubsection{Upper bound for the estimator $\widetilde{h_U}$}\label{App:estimation-upper}
\bl{There is a simple way to upper bound heavy output frequency using above derivations.}
One can show by the Cauchy-Schwarz inequality that for any $\Omega_z$ it holds 
$\inner{J_{\Omega_z}}{J_\Id} \le 4^{N} \mathbb{P}_0^{(z)}$. Therefore, we may 
estimate $|a_z| \le \mathbb{P}_0^{(z)}+\frac{1}{4^N-1}$ and provide the 
following upper bound for the estimator $\widetilde{h_U}$
\begin{equation}
\begin{split}
\widetilde{h_U} &\le \frac{1}{2}+\left(p_*-\frac12\right)\left| 
\mathbb{P}_0^{(M)} + \frac{1}{2^N-1}\right|\prod_{z=1}^{M-1}|a_z|\\
&\le \frac{1}{2}+\left(p_*-\frac12\right)\left[
\prod_{z=1}^{M}\mathbb{P}_0^{(z)} + 
\left(1+\frac{1}{2^N-1}\right)\left(1+\frac{1}{4^N-1}\right)^{M-1} -1\right] \\
&\le \frac{1}{2}+\left(p_*-\frac12\right)
\prod_{z=1}^{M}\mathbb{P}_0^{(z)} + \mathcal{O}(4^{-N})
\end{split}
\end{equation}
for $M \le 2^N$.

\subsubsection{Precise estimation for the measurement noise $\Omega_M$}\label{App:estimation-m}

Let us assume noiseless execution of all quantum circuit layers, except the last one $\Omega_z = \Id$ for $z<N$. In that noise model, all errors 
occur at the measurement stage. The value of the original heavy output 
probability estimator Eq.~\eqref{eq:hu} equals then $\widetilde{h_U} = \frac12 + 
(p_*-\frac12)\frac{2^N \mathbb{P}_0^{(M)}-1}{2^N-1}$. Meanwhile, from the 
Eq.~\eqref{eq:hum} we get
\begin{equation}
\begin{split}
\widetilde{h_U^m} &=  \frac12 \sum_{p} 
\tr\left( (\Pi_{m,p} \otimes 
\Id_2^{\otimes N-m}) \Omega_M^{(s)} (\sigma_{m,p}\otimes \rho_*^{\otimes N 
- m})\right)\\
&=  \frac{1}{2\binom{N}{m}} \sum_{p, \{q_1,\ldots,q_m\} \subset \{1,\ldots,N\}} 
\tr\left( \pi_{q_1,\ldots,q_m}(\Pi_{m,p} \otimes 
\Id_2^{\otimes N-m})\pi_{q_1,\ldots,q_m}^\dagger \Omega_M 
(\pi_{q_1,\ldots,q_m}(\sigma_{m,p}\otimes 
\rho_*^{\otimes N 
- m})\pi_{q_1,\ldots,q_m}^\dagger)\right)\\
&=  \frac{1}{2^N\binom{N}{m}} \sum_{p, q_1,\ldots,q_m} \,\,\, \sum_{i,j: 
\oplus_{k=1}^m 
i_{q_k} = p, \oplus_{k=1}^m j_{q_k} = p}
\tr\left( \proj{i} \Omega_M 
(\proj{j})\right)\\
&=  \frac{1}{2^N\binom{N}{m}} \sum_{p, q_1,\ldots,q_m} \sum_{i,j}
w_{i,j}^{(M)} \delta_{\oplus_{k=1}^m 
i_{q_k} = p} \delta_{\oplus_{k=1}^m j_{q_k} = p}\\
&=  \frac{1}{2^N\binom{N}{m}} \sum_{d=0}^N \sum_{i,j: \rho_H(i,j)=d} 
w_{i,j}^{(M)} \sum_{p, q_1,\ldots,q_m} 
 \delta_{\oplus_{k=1}^m 
i_{q_k} = p} \delta_{\oplus_{k=1}^m j_{q_k} = p}\\
&=  \frac{1}{\binom{N}{m}} \sum_{d=0}^N \sum_{i,j: \rho_H(i,j)=d} 
\frac{w_{i,j}^{(M)}}{2^N} \sum_{l=0}^{m} \binom{d}{l} \binom{N-d}{m-l} 
\delta_{l \in 2\mathbb{N}}=  \frac{1}{\binom{N}{m}} \sum_{d=0}^N 
\mathbb{P}_d^{(M)} \sum_{l=0}^{m} \binom{d}{l} \binom{N-d}{m-l} 
\delta_{l \in 2\mathbb{N}}\\
&= \sum_{d=0}^N \mathbb{P}_d^{(M)} f_{m,d},
\end{split}
\end{equation}
where we introduced shorthand notation $f_{m,d} = \sum_{l=0}^m \frac{\binom{d}{l} \binom{N-d}{m-l}}{\binom{N}{m}}  
 \delta_{l \in 2 \mathbb{N}}$. 

Let us extend the definition of $\widetilde{h_U^m}$ for $m=0$, $h_U^0 = 1$ and define  
 $\ket{\widetilde{h_U^*}} = 
 \sum_{m=0}^N \widetilde{h_U^m} \ket{m}$. To estimate $\widetilde{h_U}$ we should estimate 
 $\mathbb{P}_0^{(M)}$. To do so, we can experimentally evaluate the values of  
 $\widetilde{h_U^m}$, define the vector $\ket{h_U^*}$ and compute the following inner 
 product of $\ket{\widetilde{h_U^*}}$ and $\ket{v} = \sum_{m=0}^N v_m \ket{m}$, such 
 that $v_m = \frac{\binom{N}{m}}{2^{N-1}}-\delta_{m=0}$. Then, one obtains
\begin{equation}
\begin{split}
 \inner{v}{\widetilde{h_U^*}} &= \sum_{m=0}^N v_m \widetilde{h_U^m} = \sum_{m=0}^N \left( 
 \frac{\binom{N}{m}}{2^{N-1}}-\delta_{m=0} \right)\sum_{d=0}^N 
 \mathbb{P}_d^{(M)} 
 \sum_{l=0}^m
         \frac{\binom{d}{l} \binom{N-d}{m-l}}{\binom{N}{m}}\delta_{l \in 2 
         \mathbb{N}}  \\
 &= \sum_{m,d=0}^N \sum_{l \in 2 \mathbb{N}}  \left( 
 \frac{\binom{N}{m}}{2^{N-1}}-\delta_{m=0} \right) \mathbb{P}_d^{(M)} 
         \frac{\binom{d}{l} \binom{N-d}{m-l}}{\binom{N}{m}}  = 
         \sum_{m,d=0}^N \sum_{l \in 2 \mathbb{N}}
 \frac{\binom{N}{m}}{2^{N-1}} \mathbb{P}_d^{(M)} 
         \frac{\binom{d}{l} \binom{N-d}{m-l}}{\binom{N}{m}} - \sum_{d=0}^N 
         \sum_{l \in 2 \mathbb{N}}  \mathbb{P}_d^{(M)} 
         \frac{\binom{d}{l} \binom{N-d}{-l}}{\binom{N}{0}}  \\
&= \frac{1}{2^{N-1}}\sum_{d=0}^N \sum_{l \in 2 \mathbb{N}}
  \mathbb{P}_d^{(M)} 
         \binom{d}{l} \sum_{m=0}^N \binom{N-d}{m-l} -1= 
         \sum_{d=0}^N\frac{1}{2^{d-1}} \sum_{l \in 2 \mathbb{N}}
  \mathbb{P}_d^{(M)} 
         \binom{d}{l}  -1\\
&= 2\mathbb{P}_0^{(M)}  -1 + \sum_{d=1}^N \frac{1}{2^{d-1}} \mathbb{P}_d^{(M)} 
2^{d-1}
= 2\mathbb{P}_0^{(M)}  -1 + (1 - \mathbb{P}_0^{(M)}) = \mathbb{P}_0^{(M)}.
\end{split}
\end{equation}
Substituting above equation into \eqref{eq:hu} we obtain the desired 
result
\begin{equation}
    \widetilde{h_U} = \frac12 + 
    \left(p_*-\frac12\right)\frac{2^{N}\inner{v}{\widetilde{h_U^*}}-1}{2^N-1}.
\end{equation}

\subsubsection{Precise estimation for the maximally depolarizing noise}\label{App:estimation-d}

To calculate the estimator $\widetilde{h_U}$ for depolarizing noise \bl{with strength $\epsilon$} we notice that $\inner{J_{\Phi_\epsilon}}{J_\Id} = 
4-3\epsilon$ which gives $a_z = \frac{(4-3\epsilon)^N-1}{4^{N}-1}$. \bl{Moreover the no bit-flip probability is equal
$\mathbb{P}_0^{(M)} = (1-\epsilon/2)^N$ which} gives
\begin{equation}
\widetilde{h_U} = \frac12 + \left(p_* - \frac12\right)\frac{(2-\epsilon)^N-1}{2^N-1}
\left(\frac{(4-3\epsilon)^N-1}{4^{N}-1}\right)^{M-1}.
\end{equation}

To calculate $\widetilde{h_U^m}$ we notice that \bl{for this particular noise model}
\begin{equation}
\begin{split}
&\tr\left((\Pi_{m,p_z} \otimes 
\Id_2^{\otimes N-m})\Omega_z^{(s)}(\sigma_{m,p_{z-1}}\otimes \rho_*^{\otimes N 
- m})\right) = \tr\left(\Pi_{m,p_z} \Phi_\epsilon^{\otimes 
m}(\sigma_{m,p_{z-1}})\right)\\
=&\frac{1}{2^{m-1}} \sum_{i,j} \tr(\proj{i} \Phi_\epsilon^{\otimes 
m}(\proj{j}) ) \delta_{\oplus j = p_{z-1}} \delta_{\oplus i = 
p_z}=\frac{1}{2^{m-1}} \sum_{d=0}^m \sum_{i,j: \rho_H(i,j) = d} (\epsilon/2)^d 
(1-\epsilon/2)^{m-d}\delta_{\oplus j = p_{z-1}} \delta_{\oplus i = p_z}\\
=& \sum_{d=0}^m \binom{m}{d} (\epsilon/2)^d 
(1-\epsilon/2)^{m-d} \delta_{p_z \oplus d = p_{z-1}} = \frac{1+(-1)^{p_z - 
p_{z-1}}(1-\epsilon)^m}{2}.
\end{split}
\end{equation}
\bl{Applying this observation to} Eq.~\eqref{eq:hum} we obtain \rd{the final estimate}
\begin{equation}
\begin{split}
\widetilde{h_U^m} &= \frac12 \sum_{p_0,\ldots,p_M \in \{0,1\}} \prod_{z=1}^M 
\frac{1+(-1)^{p_z - 
p_{z-1}}(1-\epsilon)^m}{2} \delta_{p_M=p_0} = \sum_{p_1\ldots,p_{M-1} \in 
\{0,1\}, p_0=p_M=0} \prod_{z=1}^M 
\frac{1+(-1)^{p_z - 
p_{z-1}}(1-\epsilon)^m}{2}\\
&= \sum_{k=0}^M \delta_{k \in 2\mathbb{N}} \binom{M}{k}
\left[\frac{1-(1-\epsilon)^m}{2}\right]^{k}
\left[\frac{1+(1-\epsilon)^m}{2}\right]^{M-k} = \frac{1+(1-\epsilon)^{mM}}{2}.
\end{split}
\end{equation}
\end{document}